\documentclass[useAMS,usenatbib]{mn2e}
\usepackage{graphicx}
\usepackage{times}
\usepackage{subfigure}
\newcommand{\zsun}{Z$_{\odot}$}
\newcommand{\msun}{M$_{\odot}$}
\newcommand{\ha}{H$\alpha$}
\newcommand{\hb}{H$\beta$}
\newcommand{\gq}{NGC\,4900}
\newcommand{\mg}{Mg$_2$}
\newcommand{\fei}{Fe\,{\sc{i}}}
\newcommand{\oiii}{[O\,{\sc{iii}}]}
\newcommand{\oi}{[O\,{\sc{i}}]}
\newcommand{\nni}{[N\,{\sc{i}}]}
\newcommand{\nii}{[N\,{\sc{ii}}]}
\newcommand{\sii}{[S\,{\sc{ii}}]}
\newcommand{\hii}{H\,{\sc{ii}}}

\title[Stellar Populations in NGC 4900]{Stellar Populations in the
Central Region of the Barred Spiral Galaxy NGC 4900} 
\author[Cantin, Robert, Moll\'a, and Pellerin]{Simon Cantin$^1$, Carmelle Robert$^1$, Mercedes Moll\'a$^2$,
and Anne Pellerin$^3$ \\ 
$^1$ Universit\'e Laval, D\'epartement de physique, de g\'enie physique et 
d'optique, and Centre de recherche \\ en astrophysique du Qu\'{e}bec,  
Qu\'ebec, QC, CANADA, G1V 0A6 (simon.cantin.1@ulaval.ca and carobert@phy.ulaval.ca) \\ 
$^2$ Departamento de Investigaci\'on B\'asica, CIEMAT Avda, Complutense 22,
28040 Madrid, Spain\\ 
$^3$  George P. and Cynthia W. Mitchell Institute for Fundamental Physics and Astronomy,\\ Department of Physics and Astronomy, Texas A\&M University, College Station, TX 77843, USA
}

\def\jnl@style{\it}
\def\aaref@jnl#1{{\jnl@style#1}}
\def\aaref@jnl#1{{\jnl@style#1}}

\def\aj{\aaref@jnl{AJ}}                   % Astronomical Journal
\def\araa{\aaref@jnl{ARA\&A}}             % Annual Review of Astron and Astrophys
\def\apj{\aaref@jnl{ApJ}}                 % Astrophysical Journal
\def\apjl{\aaref@jnl{ApJ}}                % Astrophysical Journal, Letters
\def\apjs{\aaref@jnl{ApJS}}               % Astrophysical Journal, Supplement
\def\ao{\aaref@jnl{Appl.~Opt.}}           % Applied Optics
\def\apss{\aaref@jnl{Ap\&SS}}             % Astrophysics and Space Science
\def\aap{\aaref@jnl{A\&A}}                % Astronomy and Astrophysics
\def\aapr{\aaref@jnl{A\&A~Rev.}}          % Astronomy and Astrophysics Reviews
\def\aaps{\aaref@jnl{A\&AS}}              % Astronomy and Astrophysics, Supplement
\def\azh{\aaref@jnl{AZh}}                 % Astronomicheskii Zhurnal
\def\baas{\aaref@jnl{BAAS}}               % Bulletin of the AAS
\def\jrasc{\aaref@jnl{JRASC}}             % Journal of the RAS of Canada
\def\memras{\aaref@jnl{MmRAS}}            % Memoirs of the RAS
\def\mnras{\aaref@jnl{MNRAS}}             % Monthly Notices of the RAS
\def\pra{\aaref@jnl{Phys.~Rev.~A}}        % Physical Review A: General Physics
\def\prb{\aaref@jnl{Phys.~Rev.~B}}        % Physical Review B: Solid State
\def\prc{\aaref@jnl{Phys.~Rev.~C}}        % Physical Review C
\def\prd{\aaref@jnl{Phys.~Rev.~D}}        % Physical Review D
\def\pre{\aaref@jnl{Phys.~Rev.~E}}        % Physical Review E
\def\prl{\aaref@jnl{Phys.~Rev.~Lett.}}    % Physical Review Letters
\def\pasp{\aaref@jnl{PASP}}               % Publications of the ASP
\def\pasj{\aaref@jnl{PASJ}}               % Publications of the ASJ
\def\qjras{\aaref@jnl{QJRAS}}             % Quarterly Journal of the RAS
\def\skytel{\aaref@jnl{S\&T}}             % Sky and Telescope
\def\solphys{\aaref@jnl{Sol.~Phys.}}      % Solar Physics
\def\sovast{\aaref@jnl{Soviet~Ast.}}      % Soviet Astronomy
\def\ssr{\aaref@jnl{Space~Sci.~Rev.}}     % Space Science Reviews
\def\zap{\aaref@jnl{ZAp}}                 % Zeitschrift fuer Astrophysik
\def\nat{\aaref@jnl{Nature}}              % Nature
\def\iaucirc{\aaref@jnl{IAU~Circ.}}       % IAU Cirulars
\def\aplett{\aaref@jnl{Astrophys.~Lett.}} % Astrophysics Letters
\def\apspr{\aaref@jnl{Astrophys.~Space~Phys.~Res.}}
\def\bain{\aaref@jnl{Bull.~Astron.~Inst.~Netherlands}} 
\def\fcp{\aaref@jnl{Fund.~Cosmic~Phys.}}  % Fundamental Cosmic Physics
\def\gca{\aaref@jnl{Geochim.~Cosmochim.~Acta}}   % Geochimica Cosmochimica Acta
\def\grl{\aaref@jnl{Geophys.~Res.~Lett.}} % Geophysics Research Letters
\def\jcp{\aaref@jnl{J.~Chem.~Phys.}}      % Journal of Chemical Physics
\def\jgr{\aaref@jnl{J.~Geophys.~Res.}}    % Journal of Geophysics Research
\def\jqsrt{\aaref@jnl{J.~Quant.~Spec.~Radiat.~Transf.}}
\def\memsai{\aaref@jnl{Mem.~Soc.~Astron.~Italiana}}
\def\nphysa{\aaref@jnl{Nucl.~Phys.~A}}   % Nuclear Physics A
\def\physrep{\aaref@jnl{Phys.~Rep.}}   % Physics Reports
\def\physscr{\aaref@jnl{Phys.~Scr}}   % Physica Scripta
\def\planss{\aaref@jnl{Planet.~Space~Sci.}}   % Planetary Space Science
\def\procspie{\aaref@jnl{Proc.~SPIE}}   % Proceedings of the SPIE

\begin{document}

\date{}

\pagerange{\pageref{firstpage}--\pageref{lastpage}} \pubyear{2010}

\maketitle

\label{firstpage}

\begin{abstract}
We present OASIS observations obtained at the Canada-France-Hawaii Telescope for the SB(rs)c galaxy \gq. About 800 spectra in the wavelength range 4700-5500\,\AA\ and 6270-7000\,\AA\ have been collected with a spatial resolution of $\sim$\,50 pc. This galaxy is part of a sample to study the stellar populations and their history in the central region of galaxies. In this paper, we present our iterative technique developed to describe consistently the different stellar components seen through emission and absorption lines.  In \gq\ we find many young bursts of star formation distributed along the galaxy large scale bar on each side of the nucleus.  They represent nearly 40 per cent of the actual stellar mass in the field of view. The age for these bursts ranges from 5.5 to 8\,Myr with a metallicity near and above 2\,\zsun. The extinction map gives {\it E(B$-$V)} values from 0.19$\pm$0.01 near the youngest bursts to 0.62$\pm$0.06 in a dusty internal bar perpendicular to the large scale bar. The \mg\ and \fei\ absorption lines indicate the superposition of a background stellar population with an age between 100\,Myr to 3\,Gyr and a sub-solar metallicity on average. We propose that all these episodes of star formation are the consequence of a secular evolution. In this scenario, the galactic large scale bar plays an important role with respect to the recent bursts and the dusty nuclear bar observed.  The iterative technique allows us to improve the determination of the stellar population parameters, mainly an older age is obtained for the old component and more reliable stellar population masses are found.  A composite/transition type activity in the galaxy nucleus is also revealed with this technique. 

\end{abstract}


%%%%%%%%%%%%%%%%%%%%%%%%%%%%%%%%%%%%%%%%%%%%%%%%
\section{Introduction}

The central region of spiral galaxies hides a large variety of activities and stellar populations.  Observations of nearby galaxies and theoretical models for the formation of galaxies indicate the superposition of disk stars, old and metal-poor objects from the bulge, and often additional populations related to peculiar phenomenon near the galaxy center \citep[see for example the study of the central region of the Milky Way by][]{2003MNRAS.338..857V}. The assemblage and evolution of these components is still puzzling. The presence of a large scale bar in the galaxy is believed to be an efficient way to drive gas toward the central kiloparsecs of the galaxy \citep{1996ApJ...462..114N}, which may trigger star formation or turn on the central engine \citep{2009ApJ...692.1623H}. However, \citet{2004ApJ...600..595R} showed that a bar alone will only bring gas to the central region, but further structures and perturbations such as nuclear bars, rings, or spirals are needed to drive the gas deeper into the galaxy. Nuclear rings are found in about 20 per cent of spirals \citep{2005A&A...429..141K} and these are very often, if not always, barred spirals \citep{1996FCPh...17...95B}. Interactions, mergers or cosmic filaments are considered to play a role in the accretion of gas and its flow process as well \citep{2008IAUS..245..151C}. Secular evolution is proposed to be important for the formation of the bulge, called a pseudobulge, for late-type spirals \citep{2004ARA&A..42..603K}.  This slow process may be internal -- due to  non-axisymetry from a bar, oval disk, global spirals, or a three axial dark halo -- or environmental, due to galaxy harassment for example.  It rearranges the material from the disk, creating inner and outer rings and a dense concentration of gas in the galaxy center which may in turn form new stars, nuclear structures, and becomes the pseudobulge.  As opposed to true (quickly-formed) bulges, which are believed to assemble in less than 1~Gyr early in the galaxy history \citep[$\geq$10~Gyr ago;][]{1999Ap&SS.267..357R,2006A&A...457L...1Z,2006ApJ...651..491C,2008A&A...478..335B}, pseudobulges build up on time scale of 1 to 3~Gyr,  are rapidly rotating,  and have luminosity profile closer to a disk with a S\'ersic index between 1 and 2 \citep[for a summary of pseudobulges properties, see][]{2008IAUS..245..107K}.

One way to deepen our understanding of galaxy formation and evolution is to obtain a detailed description of their stellar and gaseous contents. In distant galaxies, where individual stars cannot be resolved, information about the stellar populations (i.e. their age, metallicity, initial mass function [IMF], stellar formation rate, mass, etc.) is gathered using evolutionary synthesis techniques on the integrated galaxy spectrum. This type of study is somewhat difficult and limited. For instance, it is highly dependent on the wavelength range studied since specific generations of stars will dominate the integrated flux while the emission from others will be washed out.  For example, OB stars, even in small number, may account for most of the UV flux and hide older generations of stars \citep[e.g.][]{2007MNRAS.381..228P}. In the optical, young and old populations may contribute to the spectrum, but often one is ignored for a quick analysis of the dominant component.  Furthermore when considering the addition of several stellar populations, a unique solution is difficult to obtain depending on the number of stellar indicators available. This is mainly caused by degeneracy effects between the age, metallicity, and reddening, as can be seen in models for a single population \citep[e.g.][]{2009MNRAS.397..695C}. Nevertheless, new instrumentation for the observation of distant galaxies encourage us to pursue the development of detailed synthesis techniques for the reconstruction of the history of galaxies.  In particular, integral field units, with high spatial resolution spectroscopic capabilities, are very promising if one wants to isolate and characterize individual stellar populations in crowded regions of distant galaxies.

In this paper, we present a study of the stellar populations and gas in the central region of the barred galaxy \gq\ (RA: 13h00m39.1s, DEC: +02$^\circ$30m05s, J2000) using visible data obtained with the integral field unit OASIS (Optically Adaptive System for Imaging Spectroscopy). \gq\ is a SB(rs)c galaxy located in the Virgo cluster at a distance of 13.3\,Mpc (using a redshift of 960$\pm$3\,km\,s$^{-1}$ from NED and H$_0$=72\,km\,s$^{-1}$\,Mpc$^{-1}$).  It is located at about 12$^\circ$ South-East of M87. It does not show any evidence of a recent interaction with other cluster members and \citet{1997A&A...326..449M} did not find any particular nuclear activity\footnote{In 1999, \gq\ hosted a supernova (SN\,1999br) which was located on its outskirts, far from the region observed with OASIS.}. \gq\ is part of a larger sample of galaxies observed with OASIS and represents a typical object for the presentation of our analysis technique;  papers by Cantin et al. will follow to  present the results for the other galaxies.  

Using the high resolution data from OASIS, the spectral synthesis code {\sc lavalsb} from \citet{2006ApJ...641..252D}, and models from \citet{2005MNRAS.357..945G}, we characterize the galaxy components in a consistent way and propose scenarios for the evolution of the galaxy central region. We have developed an iterative technique which has the advantage to separate a very young stellar population from an older one by taking into account the emission lines from the gas associated with the young population, the nebular and the stellar continuum from this young population, along with the stellar absorption lines and the continuum from an old population. This technique is different from the one used in other studies, which often do not intend to study in detail all the stellar populations \citep[e.g.][]{2006MNRAS.371.1087A,2007MNRAS.380..506G,2009arXiv0905.3349S} but where, for example, the contribution from a young population is neglected even if emission lines are seen.

This paper is organized as follow.  The OASIS data and their reduction are presented in \S2 along with the description of the method applied to measure spectral indicators  and the synthesis models used to gather the information for the gas and the stellar populations.  A demonstration on how the indicators are applied is done in \S3 along with a description of the integrated spectrum.  In \S4 we present spatial maps of the continuum flux, emission lines, and extinction. A detailed analysis for all the spectra is done in \S5 where we first describe our iterative technique to separate in an accurate way the contribution from the different stellar populations.  A discussion of the iterative technique and an evolutionary scenario for \gq\ are presented in \S6. Finally, our conclusions are summarized in \S7.

%%%%%%%%%%%%%%%%%%%%%%%%%%%%%%%%%%%%%%%%%%%%%%%%
\section{OASIS Data}
\subsection{Observations and Data Reduction}

\begin{figure}
\resizebox{\hsize}{!}{\includegraphics{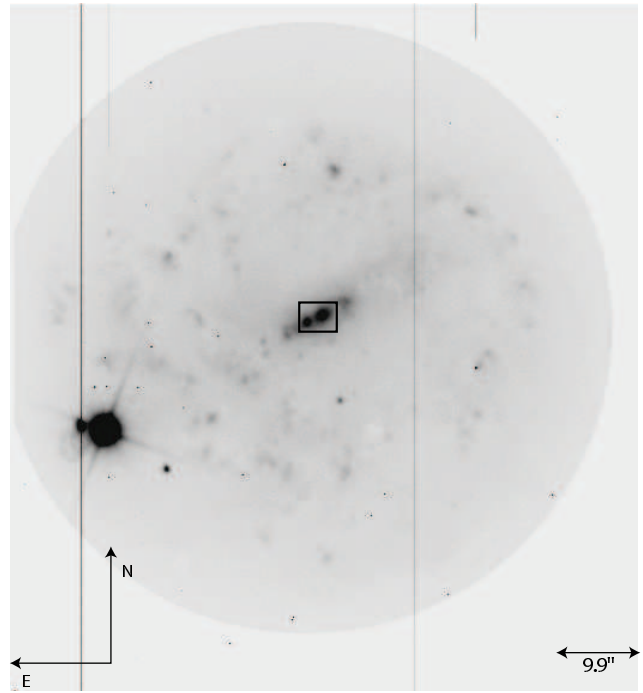}}
\caption{OASIS acquisition image of \gq. The image was obtained without a filter.  The rectangle indicates the field covered by the lenslet array used in the OASIS spectroscopic mode. It samples a region 790\,pc$\times$690\,pc centered on the galaxy coordinates.}
\label{acs}
\end{figure}

The data for \gq\ has been collected with the instrument OASIS at the 3.6\,m Canada-France-Hawaii Telescope on March 18 and 19, 2001.  OASIS was used in the Cassegrain f/8 mode with the EEV detector and without adaptive optics. OASIS is an integral field unit which is composed of a matrix of lenslets. Documentation for OASIS may be found on the web page of the William Herschel Telescope, where it is now located. The spatial resolution of 0.41\arcsec/lenslet was selected to allow a large field of view of 15\arcsec$\times$12\arcsec.  Fig.~\ref{acs} shows the OASIS acquisition image along with the location of the lenslet array used in the spectroscopic mode.  Two wavelength ranges have been covered using the grating MR1 (4760 to 5558\,\AA) and MR2 (6210 to 7008\,\AA). These configurations offer an instrumental spectral resolution of 2.17 and 2.23\,\AA\,pixel$^{-1}$, respectively.

\begin{figure*}
\subfigure{\includegraphics[width=0.49\textwidth]{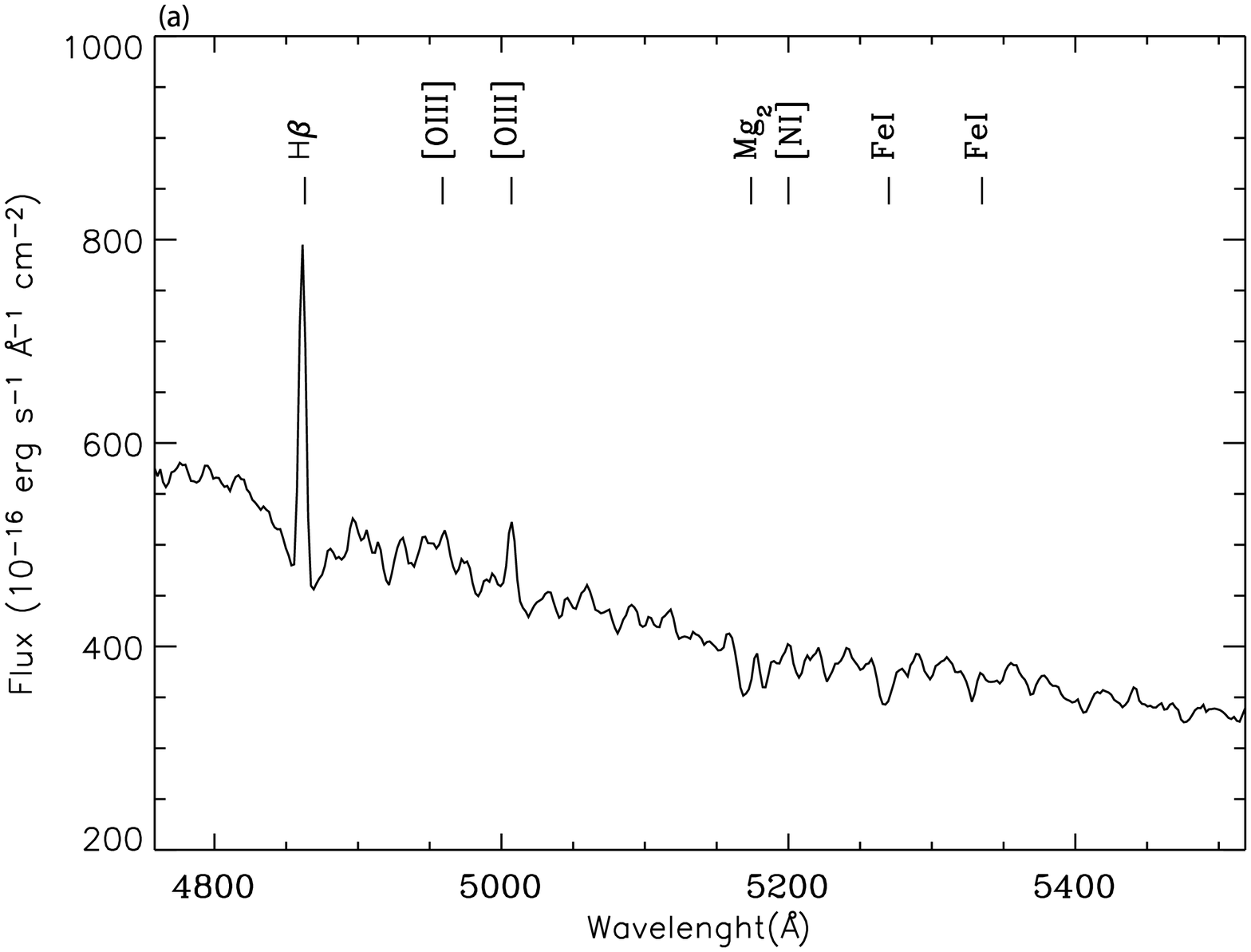}}
\subfigure{\includegraphics[width=0.47\textwidth]{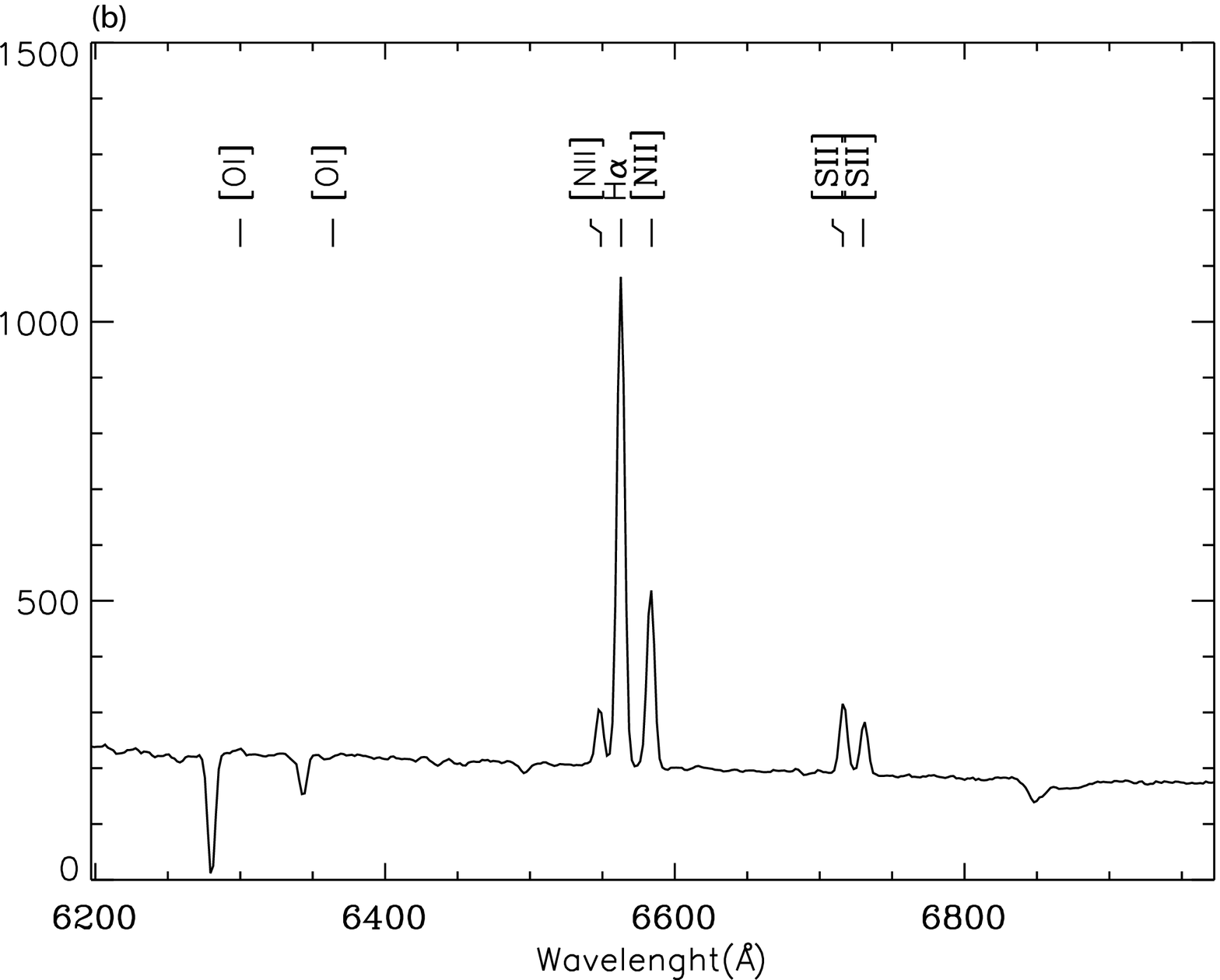}}
\caption{The OASIS integrated spectrum of \gq. The MR1 configuration is shown in (a) and the MR2 configuration in (b).  The features near 6280, 6345, 6495, and 6850\,\AA\ are caused by sky line subtractions and telluric absorptions.}
\label{intspec}
\end{figure*}

One exposure of 2400\,s was obtained for each of the gratings. Corresponding sky exposures were limited to 900\,s due to time constraints. A set of calibration files (bias, dark, and micro-pupils) along with spectra of Neon and Mercury lamps and a spectrophotometric standard were collected to perform the calibrations.

The data reduction was mainly done with the {\sc xoasis} program (version 6.0) developed by the TIGER team at the CRAL-Observatoire de Lyon. With {\sc xoasis}, we performed the usual operations, i.e overscan removal, and dark current and bias sustraction.  We also corrected for weak fringing in the detector using dome flat exposures.

The next step in the data reduction, the mask extraction, was more crucial. The mask was created using {\sc  xoasis} where a model of the telescope and instrument optical path was considered for each of the lenslets.  The mask was applied to each image to produce a datacube where the spectra were automatically (but roughly) wavelength calibrated.  The refinement of the wavelength calibration was done using the neon lamp image obtained in the same telescope position and instrument configuration as the science files to be calibrated. This gave a final wavelength resolution in MR1 and MR2 of 2.3\,\AA\,pixel$^{-1}$.  The datacubes were corrected for the instrument transmission using the flat field task. This was followed by the {\sc xoasis} routine to remove comic rays.  The sky background was subtracted using an image of the sky in the case of the galaxy or using the image's surrounding sky information in the case of the standard star.

The last steps done with the {\sc  xoasis} software consisted in the flux calibration and the merging of the datacubes.  We used the {\sc  xoasis} flux calibration task. Considering the noise level and the uncertainty in the reduction process, we estimate the flux level to be accurate to within 10 per cent.  Finally, we merged each calibrated datacube with itself. While this operation did not change the signal-to-noise ratio or any characteristics of the spectra, it simplified the analysis by reordering the lenslet identification number from a concentric pattern to a rectangular system starting with lenslet \#1 in the lower left South-East corner, and moving up North then West, column by column.

For the analysis, the datacubes were easily manipulated with {\sc iraf}\footnote{IRAF is distributed by the National Optical Astronomy Observatories, which are operated by the Association of Universities for Research in Astronomy, Inc., under cooperative agreement with the National Science Foundation.}. We corrected for the galactic reddening using the law of \citet{1989ApJ...345..245C} and an extinction {\it E(B$-$V)}=0.024 (NED) and then deredshifted the spectra using the galaxy velocity of 960\,km\,s$^{-1}$ (NED). To take into account the 0.8\arcsec\ seeing limit at the time of the observation, we spatially smoothed over 4 adjoining lenslets for a final spatial resolution of $\sim$\,53\,pc (one pixel on the images will cover $\sim$\,26\,pc). Rejecting the low signal spectra on the edge of the image, this leaves us with an almost rectangular matrix of dimension 30$\times$26 lenslets, or spectra, where the signal-to-noise ratio per pixel reaches 40 in MR1 (measured between 5050 and 5150\,\AA) and 50 in MR2 (between 6400 and 6500\,\AA).

\begin{table}
\begin{minipage}{85mm}
\centering
\caption{Line parameters}
\label{mesurelist}
\begin{tabular}{@{}lccc@{}}
\hline Line & Center & Blue continuum & Red continuum \\ &(\AA) &band 
(\AA) &band (\AA) \\ \hline \hb & 4861 & 4841 & 4881 \\ \oiii & 4959
& 4939 & 5027 \\ \oiii & 5007 & 4987 & 5027 \\ Mg$_2$ & 5159 - 5195 &
5082 - 5140 & 5210 - 5248 \\ \fei & 5270 & 5250 & 5290 \\ \fei & 5335
& 5315 & 5355 \\\hline \oi & 6300 & 6280 & 6320 \\ \oi & 6364 & 6344&
6384 \\ \nii & 6548 & 6523 & 6603\\ \ha & 6563 & 6523 & 6603 \\ \nii &
6583 & 6523 & 6603 \\ \sii & 6716 & 6696 & 6750 \\ \sii & 6730 & 6696
& 6750 \\ \hline
\end{tabular}
\end{minipage}
\end{table}

%%%%%%%%%%%%%%%%%%%%%%%%%%%%%%%%%%%%%%%%%%%%%%%%
\subsection{Measurement of Spectral Lines and the Synthesis Codes}
\label{sec_mes_lin_ext}
The selected instrument configurations allow us to see a number of emission lines which will be used to characterize the stellar populations, the extinction along with other gas properties, and nuclear activity, if any. These emission lines are: \hb, \oiii\,$\lambda\lambda$\,4959,\,5007, \nni\,$\lambda$$\lambda$\,5200, \oi\,$\lambda\lambda$\,6300,\,6364, \ha, \nii\,$\lambda\lambda$\,6548,\,6583, and \sii\,$\lambda\lambda$\,6716,\,6730.  In this wavelength range, we can also identify a number of stellar absorption lines which will be useful. These lines are \mg\ around 5177\,\AA\ (which is composed of the magnesium molecular band MgH and the magnesium absorption triplet Mg$_b$), \fei\,$\lambda$\,5270, \fei\,$\lambda$\,5335, and \hb\ (when not contaminated by nebular emission). The list of the spectral lines measured is given in Table~\ref{mesurelist} along with the position of continuum bands on each side of the lines.  An example of the spectral features observed may be seen in the integrated spectrum of Fig.~\ref{intspec}.

We estimated that for each invidivual spectrum, the galaxy velocity dispersion is small compared to the spectral resolution since each lenslet covers a relatively small spatial region. Therefore, we did not apply any correction to the spectra for this effect. Spectral line measurements were performed with an {\sc iraf} script to assure the uniformity of the result over the large number of spectra. The automatic measurements required that we first normalize the spectrum continuum.  In this routine, two continuum bands (see Table~\ref{mesurelist}) on each side of a line must be specified. The lines were not measured with the standard indicator definitions from the Lick system chiefly because our data has a higher spectral resolution than the one used in this system.  A Gaussian was fitted to the line using the theoretical line center as a first reference for the Gaussian centroid. Most lines were fitted using a single Gaussian, but some (\ha\ and \nii\,$\lambda\lambda\,$6548,\,6583 or \sii\,$\lambda\lambda$\,6716,\,6730) needed a deblending routine. For each line, we extracted a center, an equivalent width (EQW), and a full width at half maximum. In the case of the \mg\ line, an index was calculated according to the definition of \citet{1994ApJS...95..107W}. A flux value was also calculated as the product of the Gaussian flux with the continuum value under the line as given by the non-normalized spectrum.  Less than 30 per cent of the automatic measurements failed (giving peculiar values for the line parameters or extreme numbers compared to those obtained for the surrounding spectra) mainly due to a poorer signal-to-noise ratio in the continuum and a weaker line intensity. These have been redone manually.

The \hb\ emission was carefully measured at the base of the absorption profile when observed. No absorption component underneath the \ha\ emission was obvious and no correction was done.  
Prior to all calculation, special care was given to properly align the MR1 and MR2 images. It was accomplished by superposing the continuum peaks observed in the two configurations. A small shift, 1 pixel to the North and 2 pixels to the East for MR1 relative to MR2, was applied. A total of 631 spectra are found to be overlapping in the two fields of view. 
The extinction is then measured considering the ratio of the \ha\ and \hb\ emission lines for a model assuming the recombination case B of \citet{2006agna.book.....O}. The gas abundance 12+$\log$(O/H) and the ionization parameter $q$ are estimated using a method similar to the one proposed by \citet{2002ApJS..142...35K} with the flux ratios: \nii\,$\lambda$\,6583/\sii\,$\lambda$\,6716+6730, \nii\,$\lambda$\,6583/\ha\, \nii\,$\lambda$\,6583/\oiii\,$\lambda$\,5007, and \oiii\,$\lambda$\,4959+5007/\hb. The metallicity determination depends strongly on the calibrators used \citep[see][]{2008ApJ...681.1183K}. Other calibrators,  such as those of \citet{2004MNRAS.348L..59P}, available in the wavelength range considered  here, would give a metallicity estimate shifted downward by up $\sim$0.2 dex compared to our result. In our work, we assume that solar metallicity is defined by 12+$\log$(O/H)=8.7 \citep{1998SSRv...85..161G}.  AGN activity is looked for based on the line ratios: \nii\,$\lambda$\,6583/\ha\ and \sii\,$\lambda$\,6716+6730/\ha\ as proposed by \citet{2006ApJS..167..177D} and \citet[][hereafter BPT]{1981PASP...93....5B}.  The \oi\,$\lambda$\,6300/\ha\ diagnostic is omitted since we cannot accurately measure the weak \oi\ line in our spectra.

The synthesis code {\sc lavalsb} \citep{2003ApJS..144...21R,2006ApJ...641..252D} is used to reproduce the emission lines EQW associated with very young star forming regions ($\leq$20\,Myr). {\sc lavalsb} was developed in parallel with {\sc starbust99} \citep{1999ApJS..123....3L}.  It uses the evolutionary tracks of the Geneva group, without rotation \citep{1992A&AS...96..269S,1993A&AS..102..339S,1993A&AS...98..523S,1993A&AS..101..415C,1994A&AS..103...97M}, at four metallicities\footnote{Here we adopt for the solar metallicity: Z$_\odot$=0.02 %\citep[but see the lower revised value by][]{2007SSRv..130..105G,2008ASPC..384...39A}.}  
\citep[][]{2009ARA&A..47..481A}.} Z=0.002, 0.005, 0.02, and 0.04.  For the young stellar populations we assume a metallicity identical to the one found for the gas. For this paper, the IMF used is a power law with a Salpeter slope $\alpha$=2.35 and cutoff masses of 1 and 100\,M$_\odot$. An instantaneous burst of star formation (where all the stars are formed at the same initial time) is considered.  Nebular continuum flux, from the free-free and bound-free emission in hydrogen and helium, is also taken into account.  The equivalent width of the hydrogen emission lines is calculated directly from the number of ionizing H\,{\sc i} photons. These values are to be compared to the \ha\ and \hb\ equivalent widths observed in order to obtain the age of the young stellar population. The flux level from the young population, including its nebular contribution to the continuum, is to be used to estimate the mass of the young population.

Measurements of the observed absorption lines are compared with models from \citet{2005MNRAS.357..945G} to study the stellar populations directly.  \citeauthor{2005MNRAS.357..945G} give synthesized spectra for metallicities of Z=0.004, 0.008, 0.019, and 0.030.  These models allow the study of stellar populations with an age between 4\,Myr and 17\,Gyr (or 60\,Myr and 17\,Gyr for the highest metallicity 1.5\,Z$_\odot$). These models have been computed for a single burst population, i.e. an instantaneous burst of star formation with no chemical evolution in the population beside normal stellar evolution. It considers an IMF with a Salpeter slope and masses between 0.1 and 120\,M$_\odot$. This smaller lower mass cut-off, compared to the one for young stellar populations, is required to properly replicate the oldest ($>$1\,Gyr) population optical proprieties \citep{2002MNRAS.334..883Z}. For populations younger than 20\,Myr, the contribution to the optical flux from stars smaller than 1\,M$_\odot$ is negligible compared to the flux of%completely dwarfed by
 B and A type stars. The spectral resolution of the models of \citeauthor{2005MNRAS.357..945G} 
is high (0.3\,\AA\ pixel$^{-1}$) since they used the theoretical spectra for individual stars from the library of \citet{2005MNRAS.358...49M}.  Therefore we degraded these models to match our observed spectra. Then, we measured the spectral indices over the synthetic spectra using the same method as for the observational data. From the absorption lines \mg, \fei\,$\lambda$\,5270, \fei\,$\lambda$\,5335, and \hb\ we obtain simultaneously the metallicity and age of a stellar population.  The absolute flux from the population found is also used to estimate its contribution to the total mass. Other IMFs like the one of \citet{1993MNRAS.262..545K} and \citet{2003PASP..115..763C}  would reduce the number of low mass stars in the models. This would slightly change the shape of the countinuum flux  for the oldest populations and consequently modify the stellar mass of these populations.  On the other hand, the age and metallicity values would probably remain unchanged with these IMFs since these parameters are determined through spectral indices that are mostly influenced by  more massive stars. Testing other IMFs is beyond the scope of this paper. 

%%%%%%%%%%%%%%%%%%%%%%%%%%%%%%%%%%%%%%%%%%%%%
\section{Integrated Spectrum}

\begin{table}
\begin{minipage}{95mm}
\centering
\caption{Line measurements in the integrated spectrum}
\label{intsp}
\begin{tabular}{@{}llll@{}}
\hline &Indicator & Value &\\ \hline &EQW(\ha) &$-38.5\pm0.1$ \AA \\
&EQW(\hb) &$-8.2\pm0.2$ \AA \\ &EQW(\fei\,$\lambda$5270) &
$+0.9\pm0.1$ \AA \\ &EQW(\fei\,$\lambda$5335) & $+0.4\pm0.2$ \AA \\
&I(\mg\,$\lambda$5177) & $+0.065\pm0.005$ mag \\
&Log(\nii\,$\lambda$6583/\ha) &$-0.46\pm0.01$ \\
&Log(\oiii\,$\lambda$5007/\hb) &$-0.40\pm0.01$ \\
&Log(\sii\,$\lambda$6716+6730/\ha) &$-0.60\pm0.02$ \\
&Log(\oi\,$\lambda$6303/\ha) &$-1.5\pm0.2$ \\ \hline
\end{tabular}
\end{minipage}
\end{table}

In order to guide us with the detailed analysis, to compare our data with previous studies from the literature, and to show how we use the stellar indicators and the models, we first present the integrated spectrum and its analysis.

\begin{figure}
\centering \includegraphics[width=0.47\textwidth]{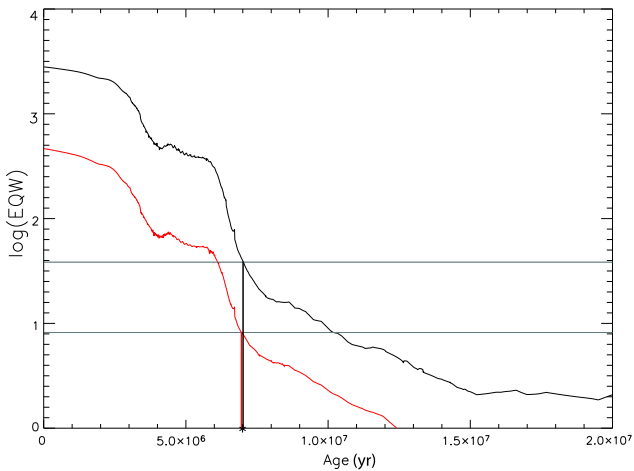}
\caption{Equivalent widths of \ha\ and \hb\ based on the synthesis code {\sc lavalsb}. The values for \ha\ (black curve) and \hb\ (red curve) have been calculated for an instantaneous burst, at solar metallicity, and using a standard IMF. The equivalent widths observed for the integrated spectrum are indicated as horizontal lines (their uncertainties are also considered, but are too small to be noticed on the plot).  Both emission lines are consistent with a stellar population of $\sim$\,7\,Myr.}
\label{haint}
\end{figure}

The integrated spectrum was created by adding together all the OASIS individual spectra.  This spectrum covers a region with an average diameter of $\sim$\,720~pc centered on the galaxy coordinates.  The Balmer decrement reveals a global internal extinction {\it E(B$-$V)}=0.25$\pm$0.02 which fits very well with the published result of \citet{1997ApJS..112..315H}. The reddening corrected spectrum is shown in Fig.~\ref{intspec}.  The strong and narrow emission lines clearly indicate the presence of hot stars. The  \mg\ and \fei\ absorption features are a more direct evidence for a stellar population, although without modeling, it is not easy to say if they belong to the same young population responsible for the emission lines or to an older underlying population.

\begin{figure}
\centering
\subfigure{\includegraphics[width=0.46\textwidth]{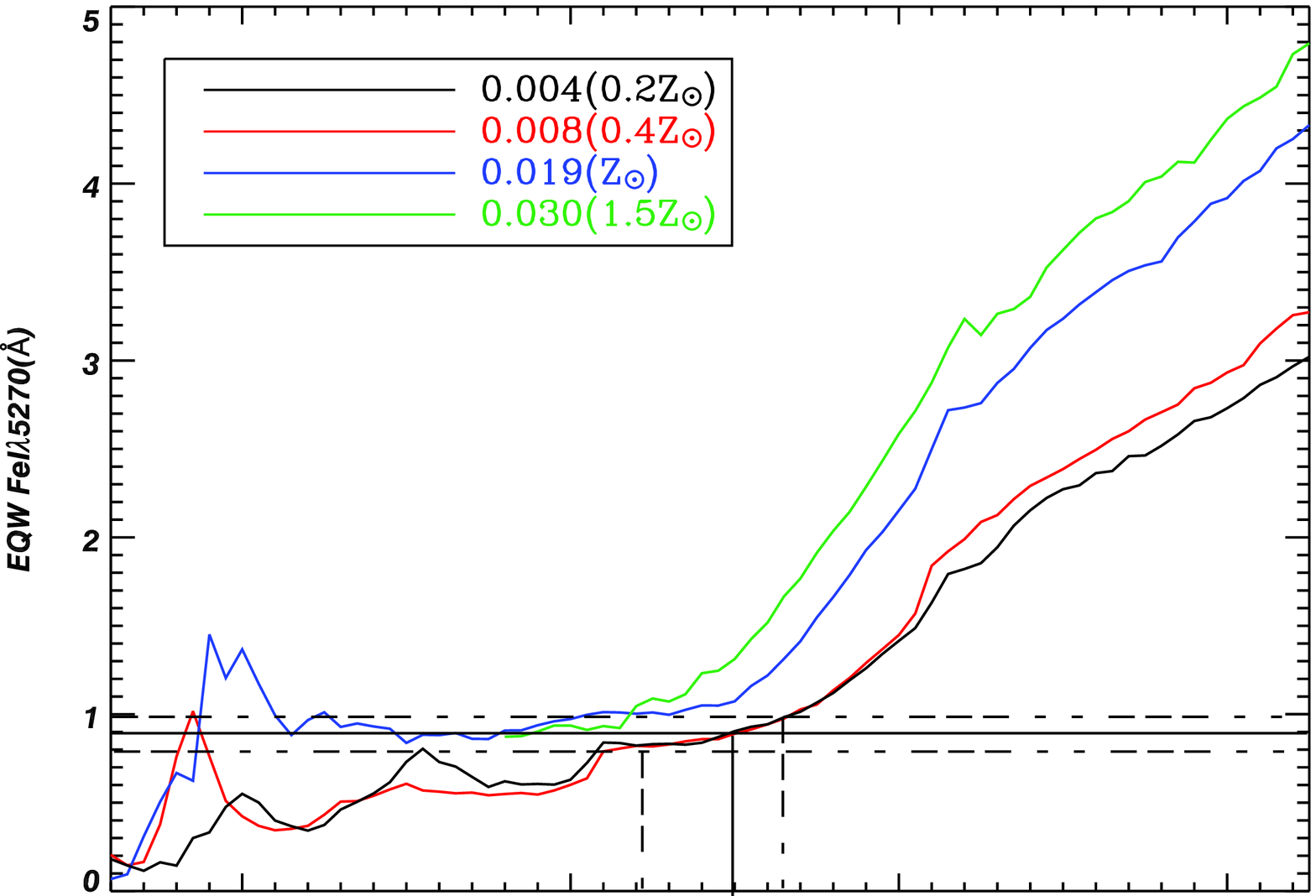}}
\subfigure{\includegraphics[width=0.46\textwidth]{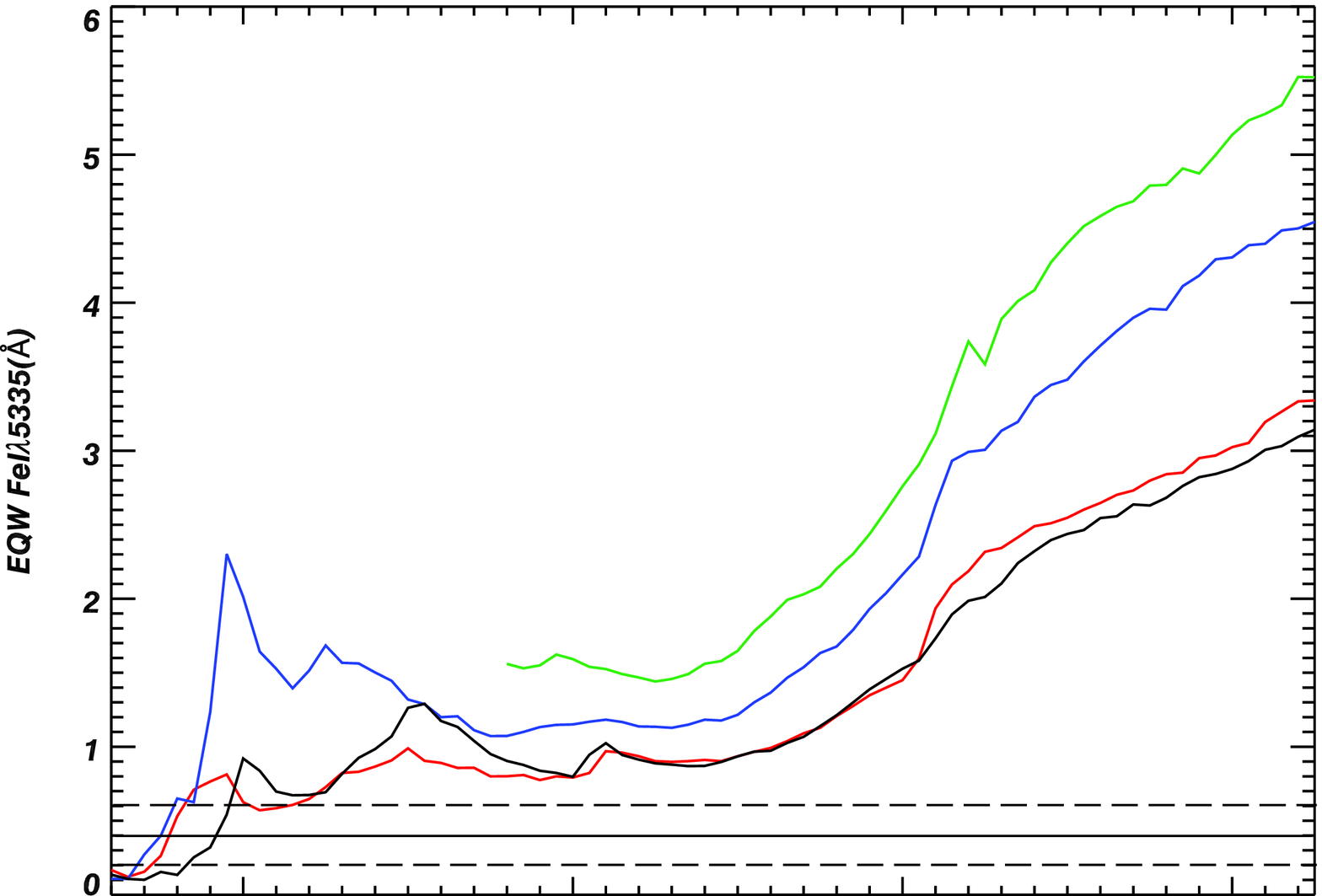}}
\subfigure{\includegraphics[width=0.46\textwidth]{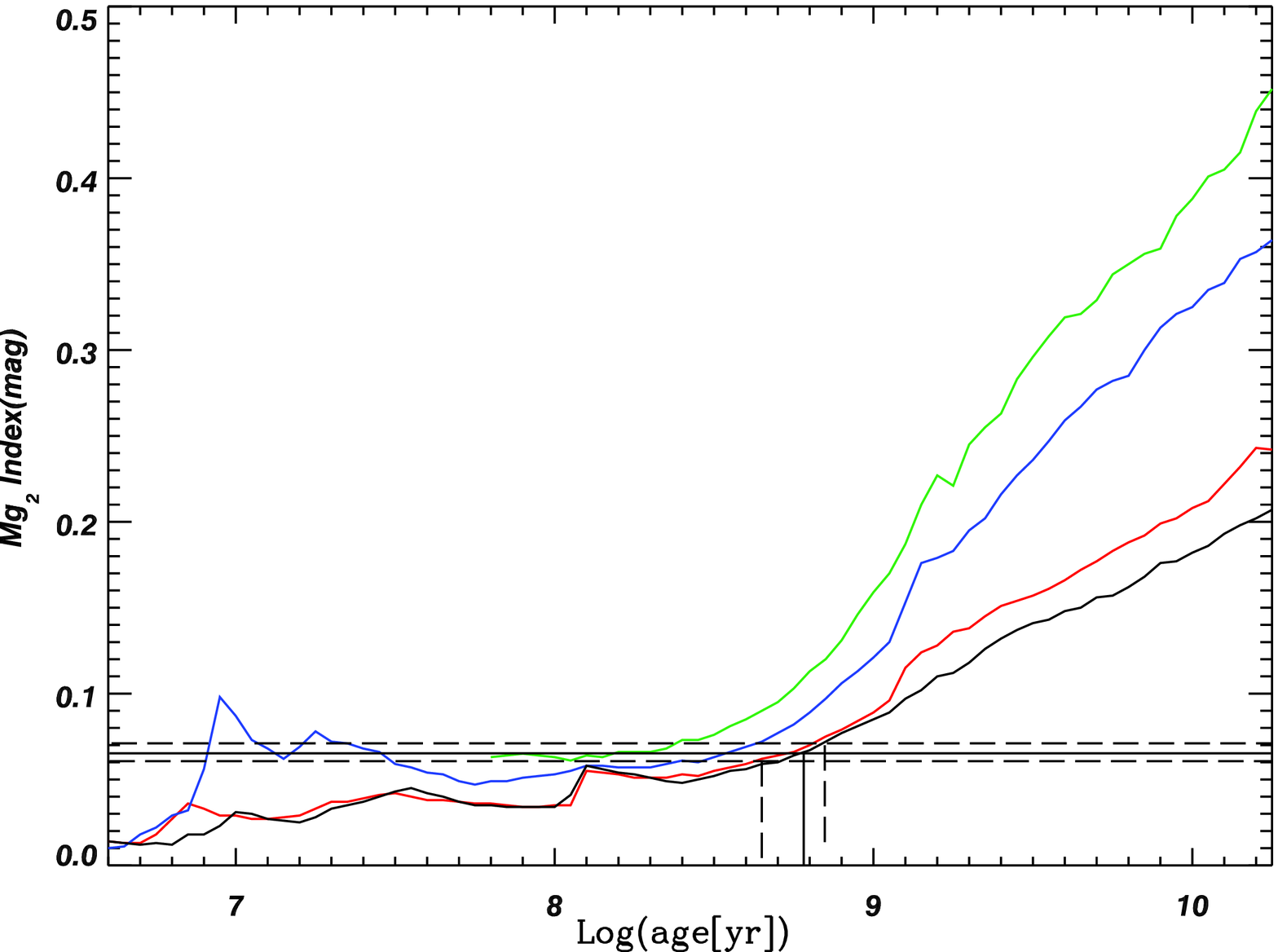}}
\caption{The evolution of absorption lines with the age and metallicity of the stellar population extracted from the models of \citet{2005MNRAS.357..945G}. The two top panels show the equivalent width of the iron lines and the bottom panel presents the Mg$_2$ index.  Different metallicities have been used as indicated in the top panel.  The values observed for the integrated spectrum are indicated as horizontal lines (uncertainties are represented using dash lines). }
\label{evindsp}
\end{figure}

Line equivalent widths and the spectral indice measured in the integrated spectrum are given in Table~\ref{intsp}. In this analysis, all line measurements are used to obtain the stellar population characteristics independently of a possible contamination by other
populations. We notice that the nebular emission line ratios (see Table~\ref{intsp}) do not indicate any peculiar activity for the integrated spectrum. These ratios are also in good agreement with those of \citet{1997ApJS..112..315H}. 

\begin{figure*}
\centering
\subfigure{\label{conti_a}\includegraphics[width=0.49\textwidth]{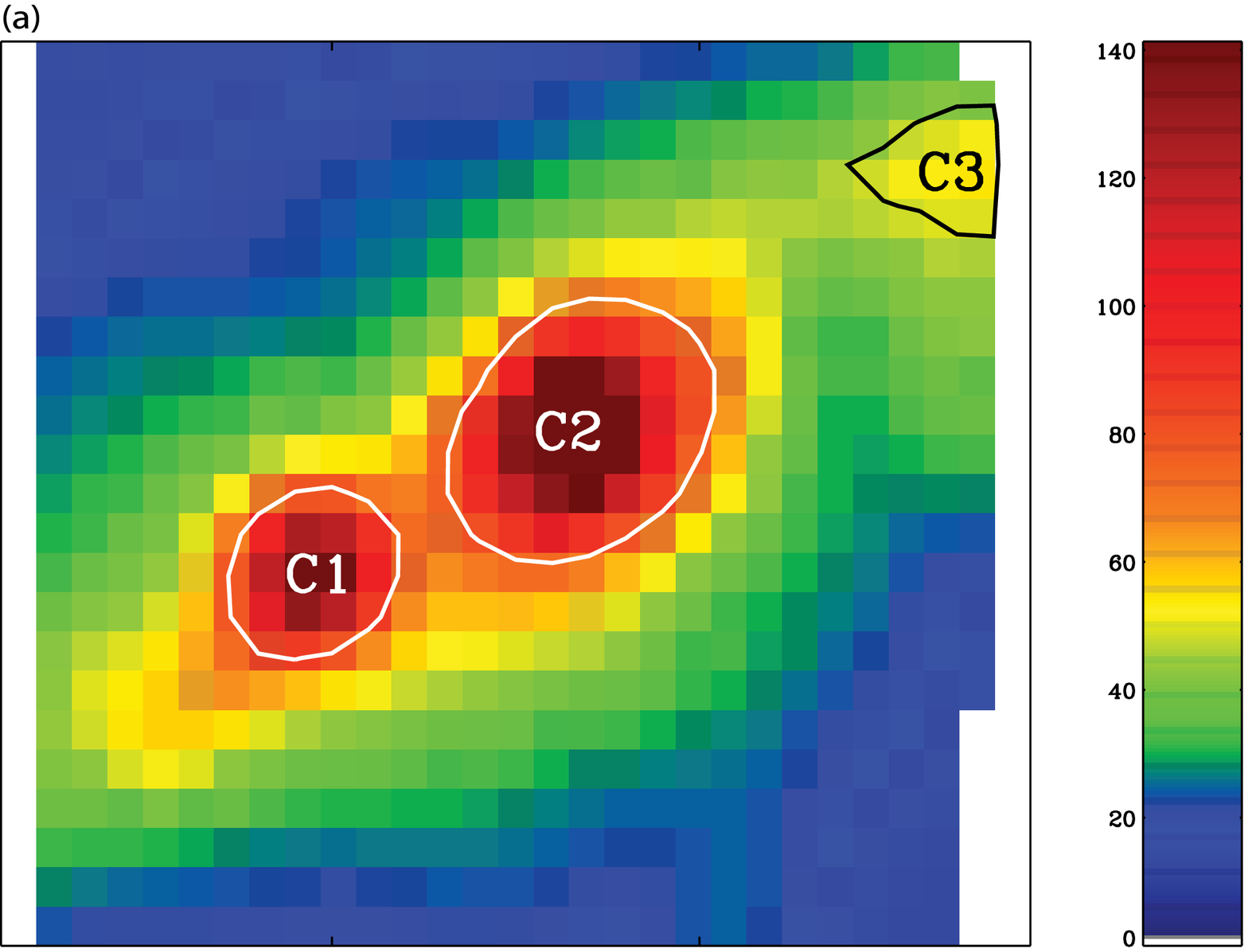}}
\subfigure{\label{conti_b}\includegraphics[width=0.49\textwidth]{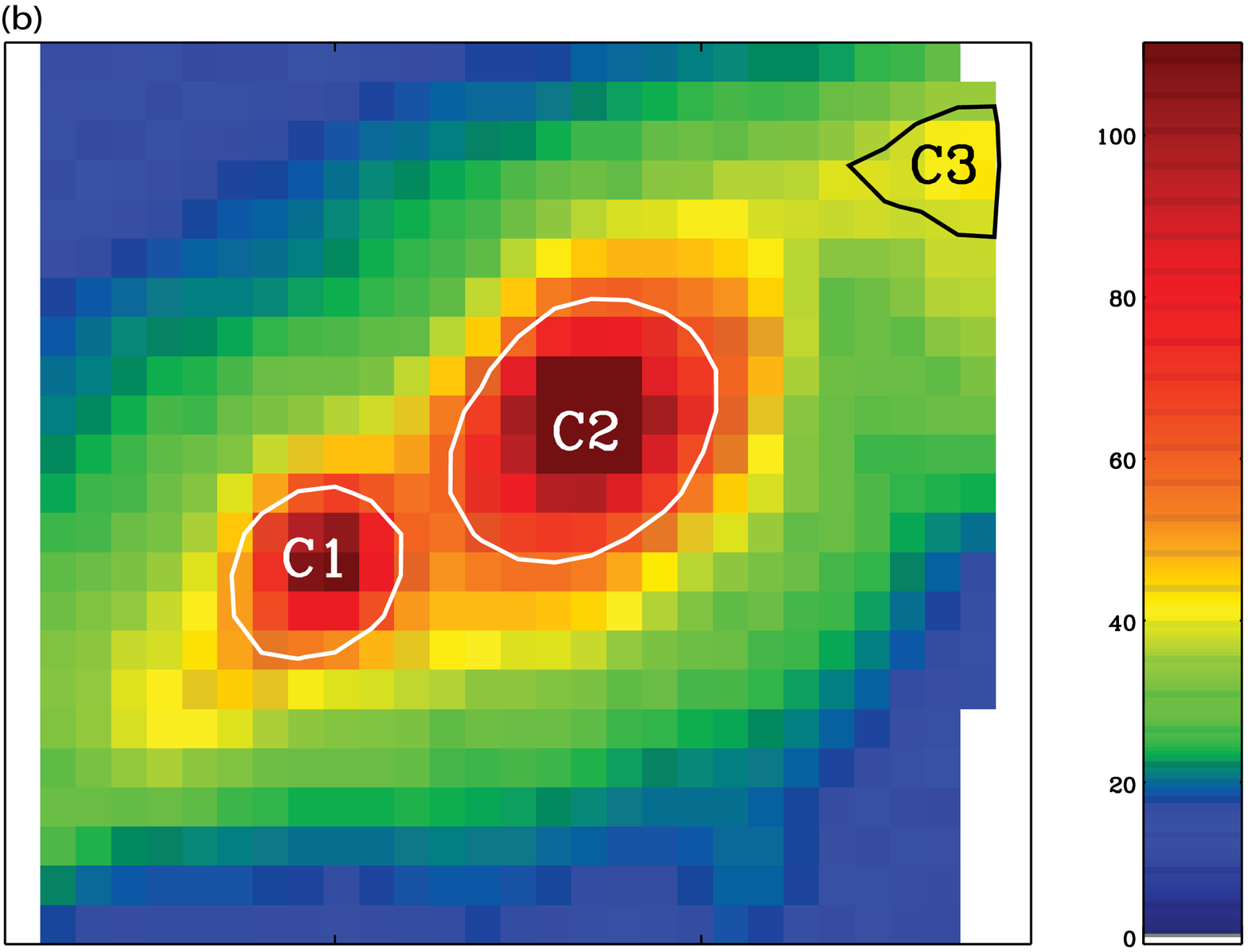}}
\subfigure{\label{conti_c}\includegraphics[width=0.49\textwidth]{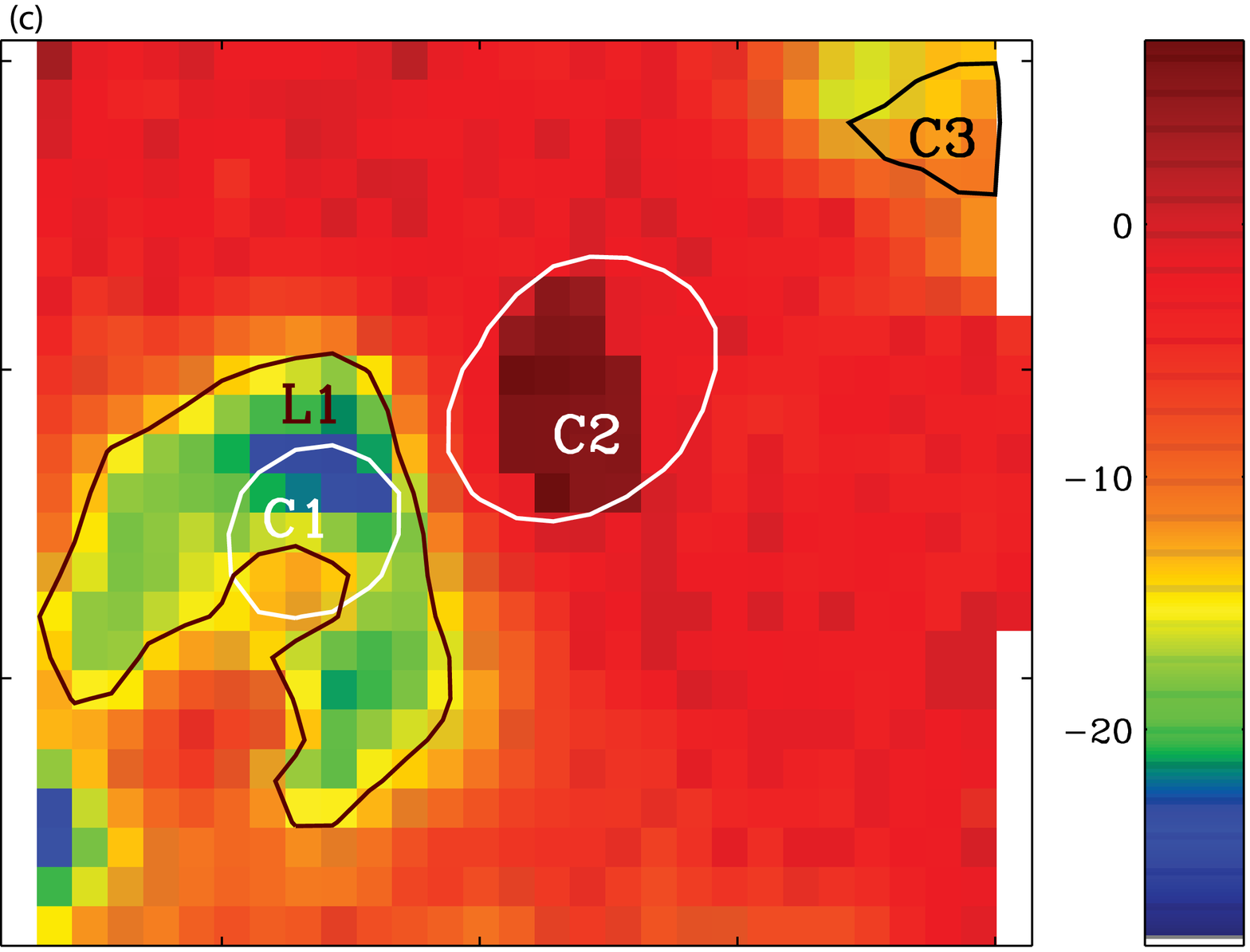}}
\subfigure{\label{conti_d}\includegraphics[width=0.49\textwidth]{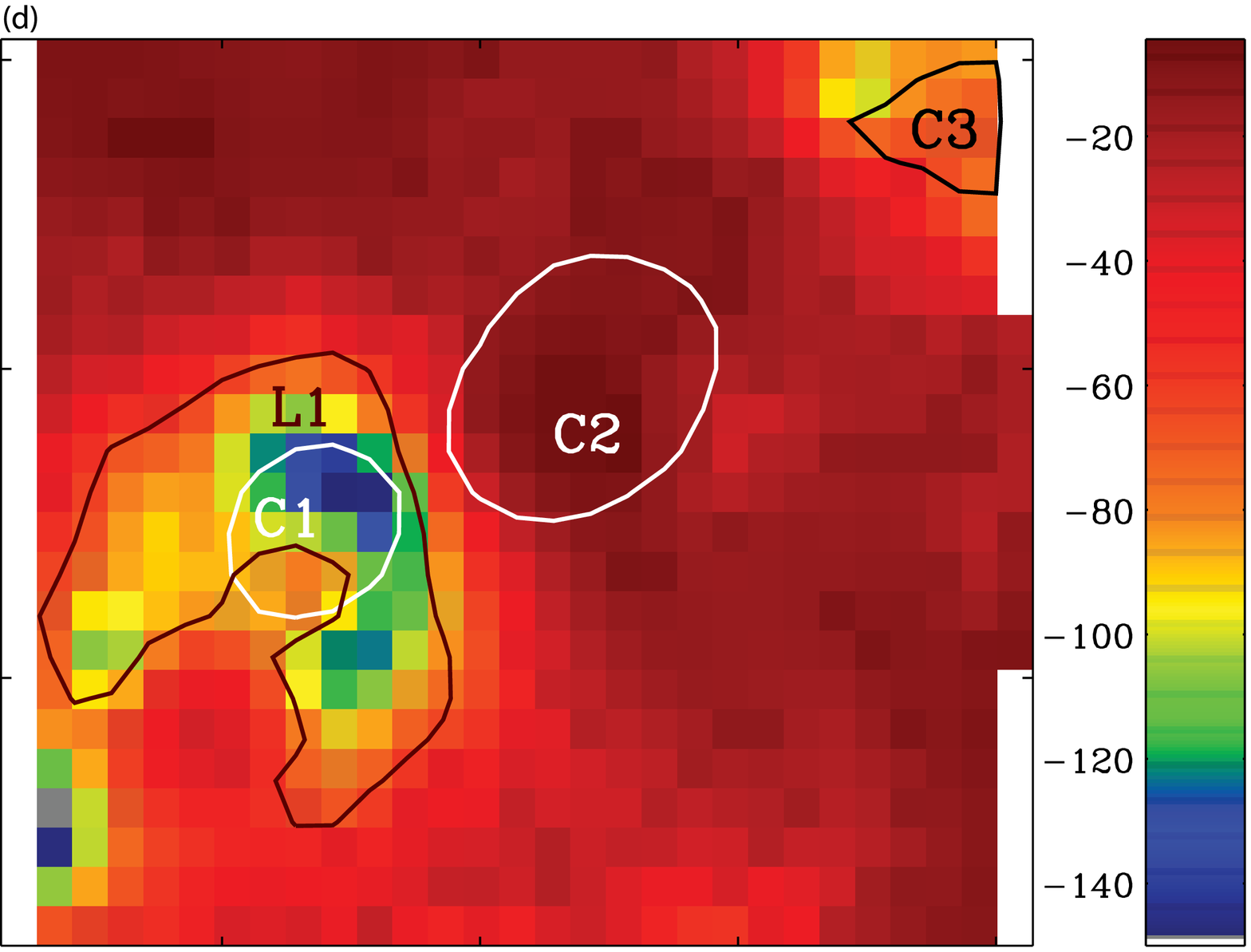}}
\caption{Continuum and nebular emission maps. The continuum fluxes (10$^{-16}$ erg s$^{-1}$ cm$^{-2}$ \AA$^{-1}$) from the MR1 and MR2 configurations are shown in (a) and (b), respectively.   The \hb\ and \ha\  equivalent widths (\AA)  are in (c) and (d), respectively. An equivalent width is given for the dominant component only, either the emission (negative value) or the absorption (positive value) component. Regions C1, C2, and C3 are delimited by contour lines extracted on the average continuum map. The region L1 is defined by a contour line on the \hb\ emission map. Only the area in which spectra are available for the two configurations is shown. On all maps North is up and East is to the left.}
\label{conti}
\end{figure*}

An oxygen abundance 12+$\log$(O/H)=8.70$\pm$0.05, i.e. Z$_\odot$, is found for the gas and it is assumed to correspond to the abundance of the young stellar population responsible for the emission lines. The equivalent widths of the \ha\ and \hb\ emission lines are then compared with models from {\sc lavalsb} at this metallicity, as shown in Fig.~\ref{haint}. This comparison gives an age of 7.0$\pm$0.2\,Myr for a single young ionizing stellar population.  The uncertainty is based on the equivalent width measurements using only a burst model at solar metallicity. For example, a small decrease in the model abundance to 0.4\,Z$_\odot$ would only slightly increase the age of the young population and its uncertainty. 

Fig.~\ref{evindsp} presents the comparison of the observed measurements for the iron and magnesium absorption lines with the model values.  These indicators should be equally used to get an age and a metallicity. First, if we ignore the larger uncertainty for the \fei\,$\lambda$5335 line (middle panel in Fig.~\ref{evindsp}), the three lines suggest a young population with an age close to 7-8\,Myr and a metallicity near solar  (it is difficult to give a better estimate of the metallicity at these young ages as models with a metallicity higher that solar are missing). This is close to the population found based on the emission lines. On the other hand, if a lower weight is given to the uncertain  \fei\,$\lambda$5335 line a much older population, between 120 to 630\,Myr with a lower metallicity of 0.2-0.4\,Z$_\odot$, seems more likely. We did not use the absorption component of \hb\ here because it is strongly contaminated by the nebular emission. In conclusion, the integrated spectrum clearly indicates that  the galaxy central region is dominated by a metal rich 7\,Myr  population, but the characteristics of the population responsible for the absorption lines remain uncertain.

%%%%%%%%%%%%%%%%%%%%%%%%%%%%%%%%%%%%%%%%%%%%%%%
\section{Maps of  Continuum Fluxes, Line Equivalent Widths, and Extinction}

In this section we take into account the spatial information to localize the stellar populations and estimate the extinction in each spectrum. In the next sections, we will present a  rigorous analysis of the stellar population parameters considering their relative contribution to the spectrum.

Fig.~\ref{conti_a} and \ref{conti_b} show maps of the continuum fluxes in the MR1 and MR2 configurations, respectively. The flux was calculated over the whole spectral range in each configuration, excluding all features which would deviate from the average by 3 sigma or more.  The maps display three peaks in the continuum flux. Contour lines around these peaks define the regions C1, C2, and C3. These are oriented from South-East to North-West  along the large scale bar of the galaxy.  The brighter peaks correspond to the two bright regions seen on the OASIS acquisition image (Fig.~\ref{acs}). The region C2 corresponds to the location of the galaxy center. The importance of the flux in the region C3 is difficult to establish while it seems only partially included in the OASIS field of view.

Fig.~\ref{conti_c} and \ref{conti_d} are maps of the equivalent widths of the \hb\ and \ha\ lines.  In these maps, the equivalent width of the emission component {\it or} the equivalent width of the absorption component is shown, depending on the strongest component
in each spectrum. For example, while \ha\ is always in emission (a  negative value for the equivalent width), 
an absorption component (positive value) is stronger in \hb\ inside region C2. It is stricking to see the morphological differences between the continuum and the emission line maps. Strong emission lines are seen near C1 and C3. The emission line region L1 is much more extended than C1, while its maximum is located $\sim$80\,pc to the North of the C1 continuum peak. Another emission region is also seen in the image South-East corner, right at the edge the field of view. The continuum region C2 barely displays any emission line; at this location, an \hb\ absorption component is clearly found underneath a weak emission while \ha\ shows only a weak emission (see for example the spectrum in Fig.~\ref{spec_young}). Maps created for the other emission lines
observed (\oiii, \nii, and \sii) are not presented here because they display the same morphology as observed for \ha. The emission lines are most likely  the consequence of the interstellar gas ionized by the ultraviolet flux from OB stars present in young stellar populations. The continuum and emission line maps (not like the integrated spectrum) offer the best evidence for the presence and the superposition of both old and young stellar populations in the galaxy center.

From the ratio of the emission lines \ha\ and \hb\ (with a careful measurement of the line emission component only) important spatial variations of the extinction are found, as shown in Fig.~\ref{ebv}. It indicates  {\it E(B$-$V)} values between 0.02 and 1.3, with an average extinction of 0.47$\pm$0.08. Larger values are found in isolated pixels, where the signal is low or the \hb\ emission line is very weak. For a few spectra (mainly in the North-East corner of the image), a clear \hb\ emission component could not be observed over the noise and we adopted an extinction value computed from the average of four adjoining spectra, assuming that the extinction changes slowly from one spectrum (26\,pc) to the next. The extinction map shows a peculiar morphology; a clear dust lane is seen superimposed to the continuum region C2, and oriented perpendicular to the galaxy large scale bar. This heavily dust-obscured zone, identified X1 on the map, displays an average extinction of 0.62$\pm$0.06. Over the South-East boundary of this zone, near  C1 and L1, the region X2 identifies a much lower extinction zone with an average {\it E(B$-$V)} of 0.19$\pm$0.01. The region X2 is surrounded by a background with an average extinction of 0.31$\pm$0.01. Over the North-West boundary of the dust lane, near C3, the extinction has an average value of 0.39$\pm$0.01.

%%%%%%%%%%%%%%%%%%%%%%%%%%%%%%%%%%%%%%%
\section{Characteristics of the Stellar Populations}

\subsection{The Iteration Technique}

\begin{figure*}
\centering \includegraphics[width=\textwidth]{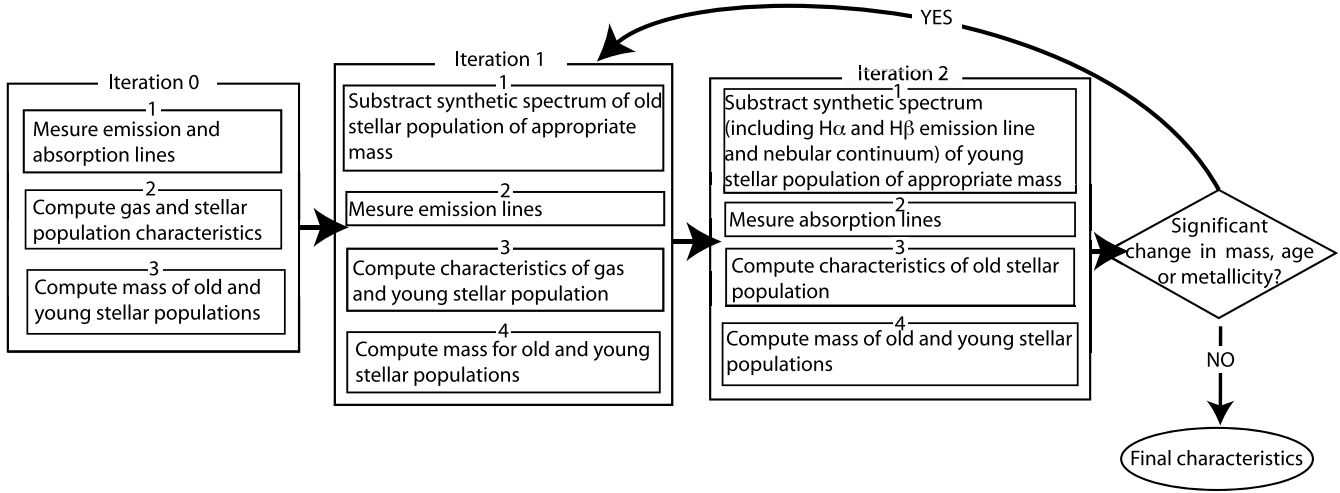}
\caption{Schematic representation of the iterative technique.}
\label{iter_proc}
\end{figure*}

For a consistent analysis of the stellar populations, we aim at disentangling the fluxes associated to each population before determining their characteristics. An iterative technique is proposed in order to isolate the different stellar populations present in a spectrum. The emission lines of \hb\ and \ha, when present, are used to characterize a young population ($<$20\,Myr) ionizing the nebular gaz, while absorption lines are used independently to describe directly a stellar population. The absorption lines may show the same age as the one described by the emission lines but, based on our experience with this galaxy and the others in our sample, the absorption lines mostly describe an older generation of stars than the one found with the emission lines. So for simplicity's sake, we say through the paper that we find a young population based on the emission lines and an old population based on the absorption lines. In any case, if only a young population is present the two sets of indicators will provide comparable characteristics for the population. %For simplicity's sake, we consider the possibility of only two different populations in each spectrum: a young one based on the \hb\ and \ha\ emission lines and a second population based on the absorption lines. Note that the possibility that both type of lines describe the same stellar population exists. However, in the galaxy region studied here, given the history of the disk, bulge, and nuclear region formation and the morphology of the continuum flux and emission line maps obtained,  it is very probable that the absorption lines will be dominated by an underlying older population different from the one responsible for the emission lines.  To simplify the following text we therefore say that the absorption line are describing an old population. 
Our iterative technique is summarized in Fig.~\ref{iter_proc}. We use a program based on Bayesian statistics to compare line measurements between observations and models. This program looks at a given set of indicators in order to identify a single age-metallicity pair solution or a single mass solution, with its uncertainty, as described in the Appendix.  Our iterative technique begins with a preliminary analysis, named ``iteration~0'', which gives a starting point for the characteristics of the stellar populations. Iteration~0 looks at the emission lines and  the absorption lines without taking into account the contribution of the other population to the spectral lines studied (as done for the integrated spectrum).

\begin{figure}
\centering \includegraphics[width=0.47\textwidth]{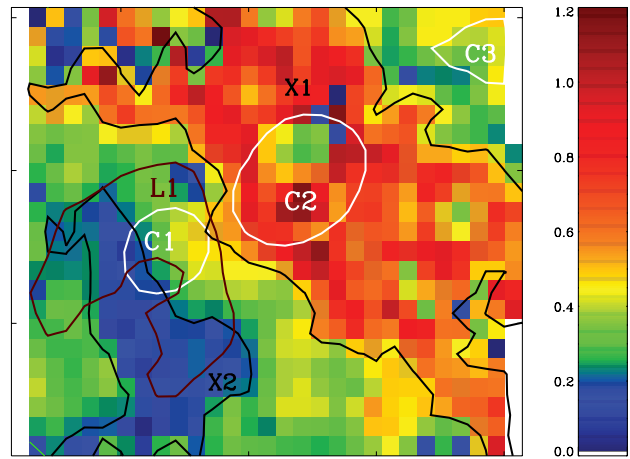}
\caption{Extinction map. Only the area which overlap between the two configurations is shown (i.e. the \hb\ line is part of the MR1 configuration while \ha\ is in MR2). Areas of high (X1) and low (X2) extinction are identified.
Regions C1, C2, C3, and L1 are as in Fig.~\ref{conti}.}
\label{ebv}
\end{figure}

Iteration 1 consists in subtracting a theoretical model for the old population in order to study the emission lines alone. This model is adapted from the models of  \citet{2005MNRAS.357..945G}, as described in \S2.2,  and is selected using the age and metallicity found after iteration 0 for the old population. The continuum flux level of this theoretical model (which is related to the mass of the old stellar population on the line of sight for a specific spectrum) is set such that the continuum flux left after its substraction from the observed spectrum is as close as possible to zero without being negative. The spectrum resulting from the subtraction is then reprocessed as if from a single young burst alone.  Bayesian statistics are used to compare line measurements between observations and {\sc lavalsb} models. To complete iteration 1, the model which represents the old stellar population and the best model for the young population are summed.   A nebular continuum  calculated using {\sc lavalsb} based on the ionizing flux of the young population is also included. The sum of these models are then scaled so that when compared with the observed spectrum (using again Bayesian statistics), it reveals a new mass for both, the old and the young, stellar populations.

Iteration 2 consists of subtracting the theoretical model for the young population (including the nebular continuum) found in the previous iteration from the observed spectrum in order to study the absorption lines.  Prior to the subtraction, synthetic emission line profiles of \ha\ and \hb\ only (no synthetic lines are used for \oiii, \nii, or \sii) are added to the spectral energy distribution of the young population spectrum. It is assumed that the \ha\ line profile is the same as the one observed (we consider that the absorption component underneath \ha\ is negligible), while the \hb\ line is a copy of the \ha\ profile scaled with respect to the theoretical 2.86 line intensity ratio used for the extinction correction (see \S\ref{sec_mes_lin_ext}). Once this theoretical spectrum is subtracted, revised values for the age and metallicity of the old stellar population are obtained. At this step, the \hb\ line might still present some emission and cannot be used for determining the characteristics of the old population. As the iteration proceeds,  the \hb\ profile should present a pure absorption. The masses of both stellar populations are recalculated by comparing the combined model for the two populations with the observed spectrum.

The next steps consist in repeating iterations 1 and 2 until there is no significant change, from the previous step, in the age, metallicity, and mass found for both populations, or until the difference for the stellar indicator measurements are of the order of their uncertainties. In general for most spectra, we need to repeat  iterations 1 and 2 three times. In the following subsections, we present the results from iteration 0 and, in more details, the characteristics obtained after the final iteration.

%%%%%%%%%%%%%%%%%%%%%%%%%%%%%%%%%%%%%%%%%%%%%
\subsection{Iteration 0}

We corrected the individual spectra with their respective extinction value (as given in Fig.~\ref{ebv}) using the reddening law of Cardelli, Clayton, \& Mathis (1989) within {\sc iraf}.  As done for the integrated spectrum, we estimated the age and metallicity of a young population, based on the emission lines, and an old population, based on the absorptions. This analysis is done again without taking into account the contribution to the continuum from one population when studying the other. It will be used as a starting point for the iteration process.

\begin{figure*}
\centering
\subfigure{\label{ages_a}\includegraphics[width=0.47\textwidth]{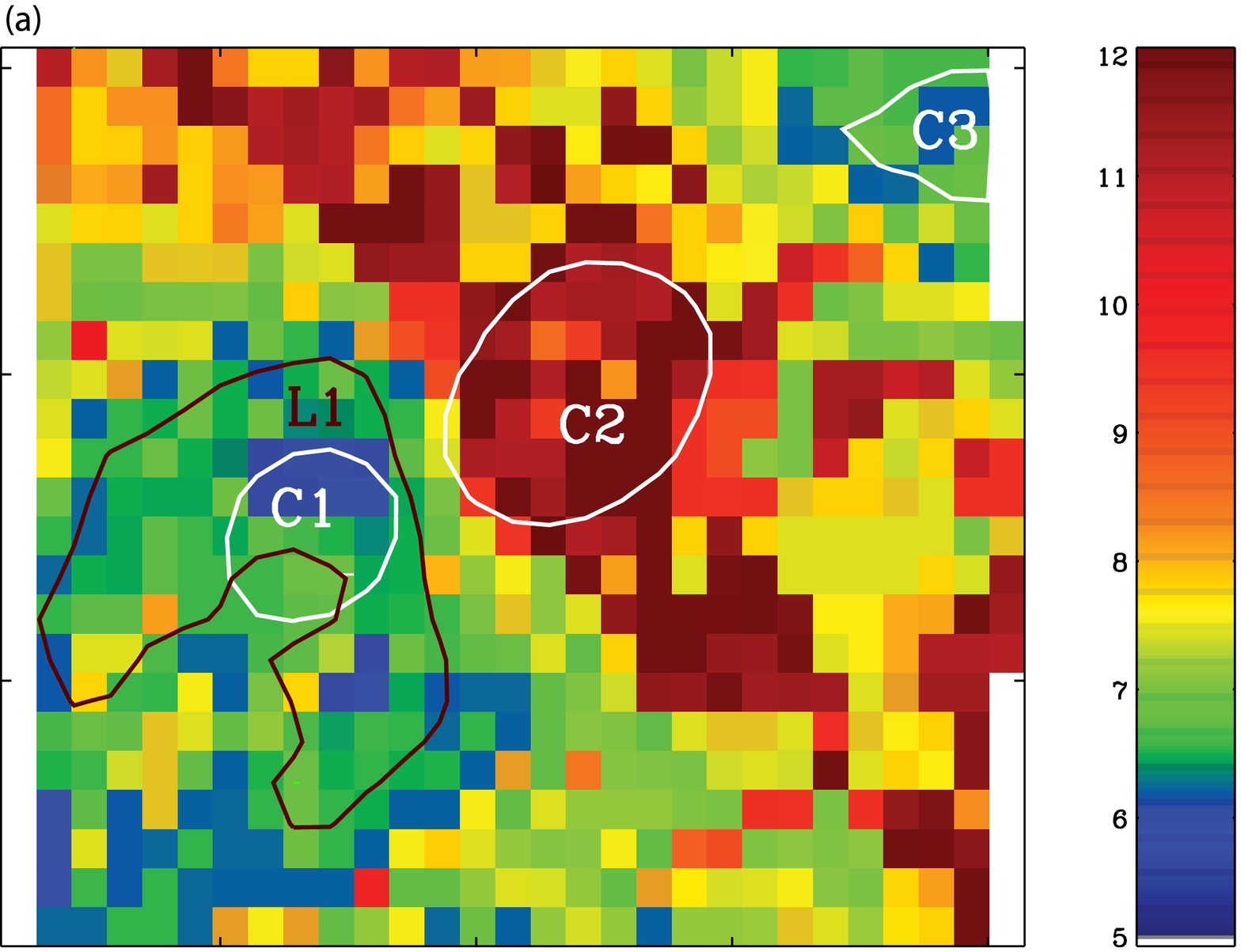}}
\subfigure{\label{ages_b}\includegraphics[width=0.47\textwidth]{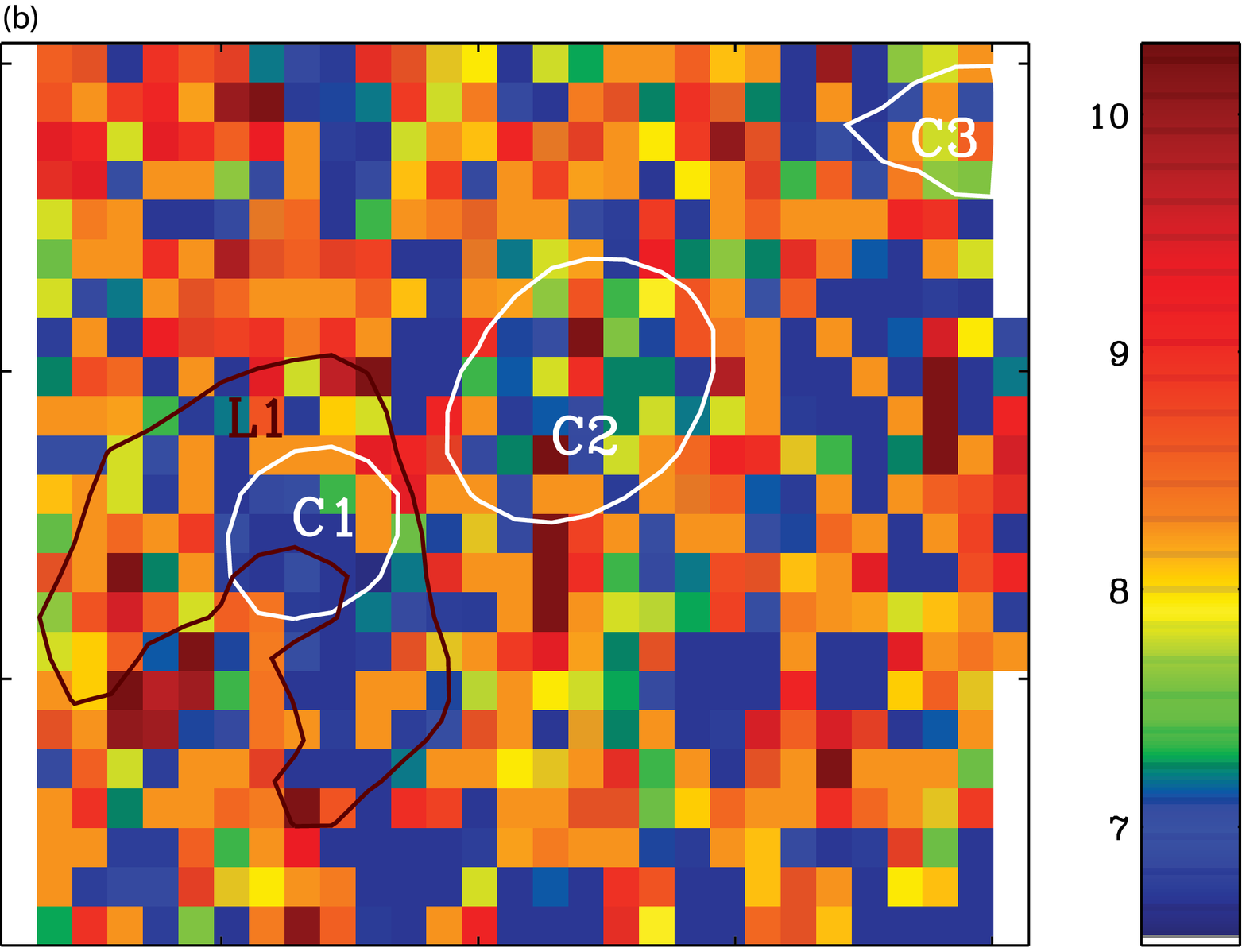}}
\subfigure{\label{ages_c}\includegraphics[width=0.47\textwidth]{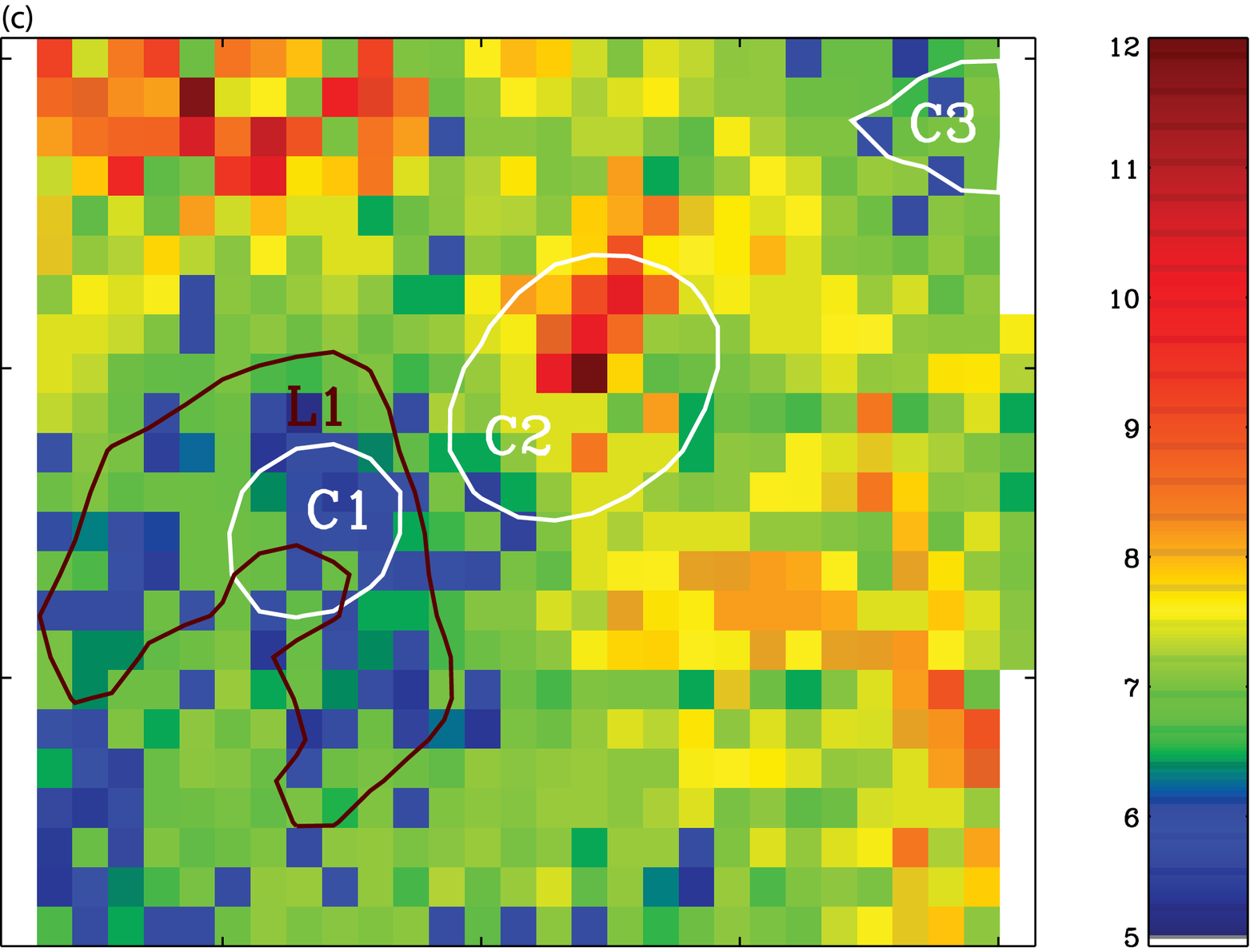}}
\subfigure{\label{ages_d}\includegraphics[width=0.47\textwidth]{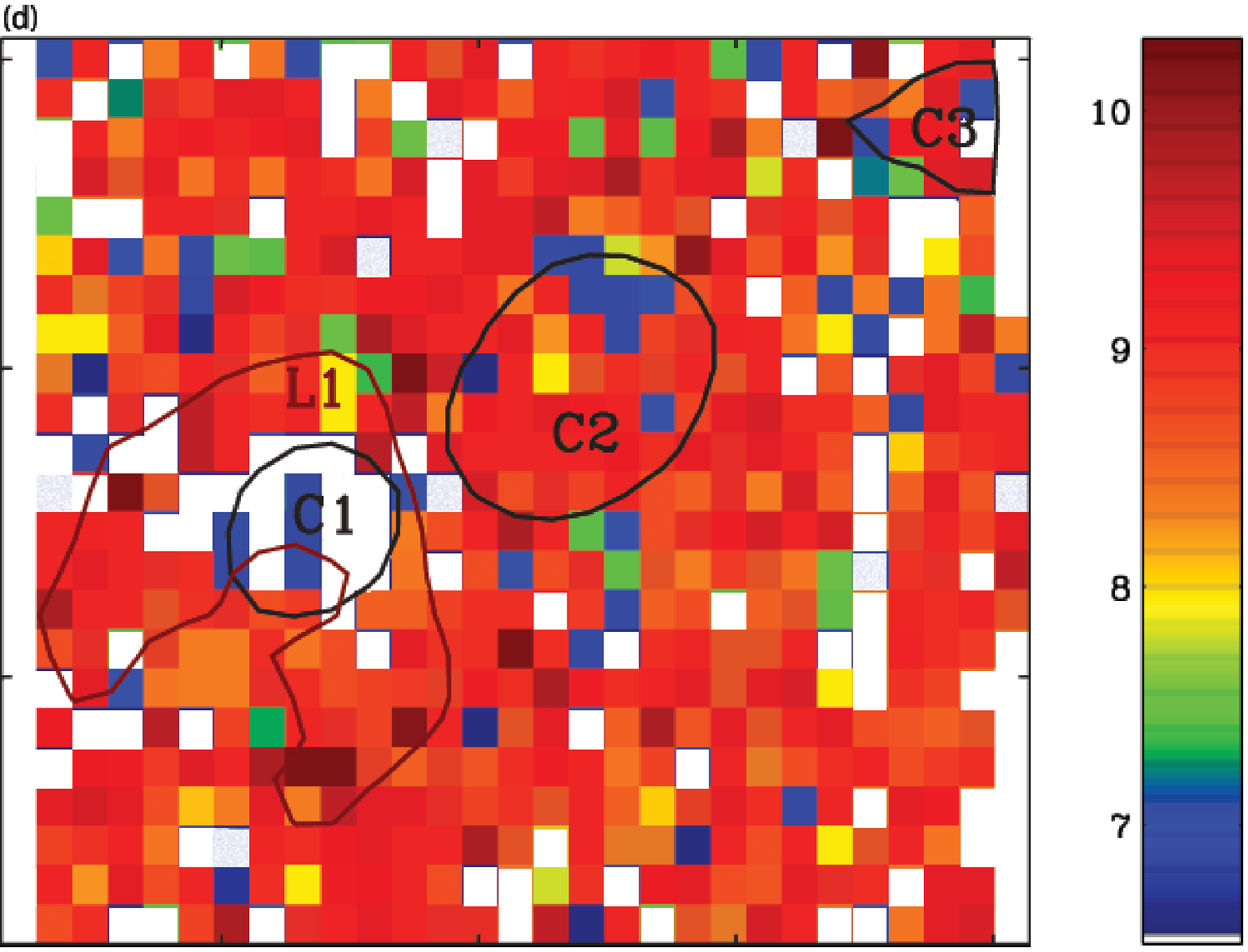}}
\caption{Age distributions. At the top are maps obtained after the iteration 0. At the bottom, after the final interation. 
To the left, (a) and (c), the age is shown, in Myr, based on the emission lines. To the right, (b) and (d), the age is given, using a logarithm scale, by the absorption lines. Contour lines are as in Fig.~\ref{conti}. }
\label{ages}
\end{figure*}

%At this step, maps of the ratio of the emission lines indicate that some region, particularly in C2 and on the periphery reach AGN-like level, still most of the observed region show line ratio typical of \hii\ regions.
 At this step in the iteration process, most of the lenses in the observed region show emission line ratio typical of \hii\ regions (only a few spectra, notably in C2 and on the periphery, display emission line ratio values as found in AGNs). The emission lines reveal a metallicity for the gas, measured in term of 12+$\log$(O/H), which is slightly above the solar value along the C1-C2-C3 axis, i.e. along the galaxy large scale bar. The gas metallicity remains closer to the solar value away from this axis (but in these regions it is also more difficult to make good estimates because of the lower signal in \oiii\,$\lambda$\,4959).  Adopting the gas metallicity for the stars responsible for the ionizing flux, we find that the youngest regions are located in the South-East image corner (see Fig.~\ref{ages_a}). We obtain an age of about 5.7$\pm$0.3\,Myr at the location of the stronger emission peak in L1, an average age of 6.5$\pm$0.3\,Myr in regions C1 and C3, and 12$\pm$1\,Myr in C2.  The age  given by the \mg\ and \fei\ lines  (see Fig.~\ref{ages_b}), is on average over the whole map, 750$\pm$250\,Myr. There is no clear spatial region with a different and specific age except for a few patches which may look younger (for example near C1). The most probable metallicity given by the absorption lines  is Z=0.2\,\zsun.

%%%%%%%%%%%%%%%%%%%%%%%%%%%%%%%%%%%%%%%%%%%
\subsection{Final Iteration}

Using the results from iteration 0, we follow the steps described for the iteration technique (Fig.~\ref{iter_proc}).  The final results for the age, metallicity, and mass of the young and old populations are presented here along with examples of the final model for selected spectra. As described at the end of this section, the iteration technique also helps us to reveal a low level of peculiar activity in the galaxy center.

\begin{figure*}
\centering
\subfigure{\includegraphics[width=0.49\textwidth,height=6cm]{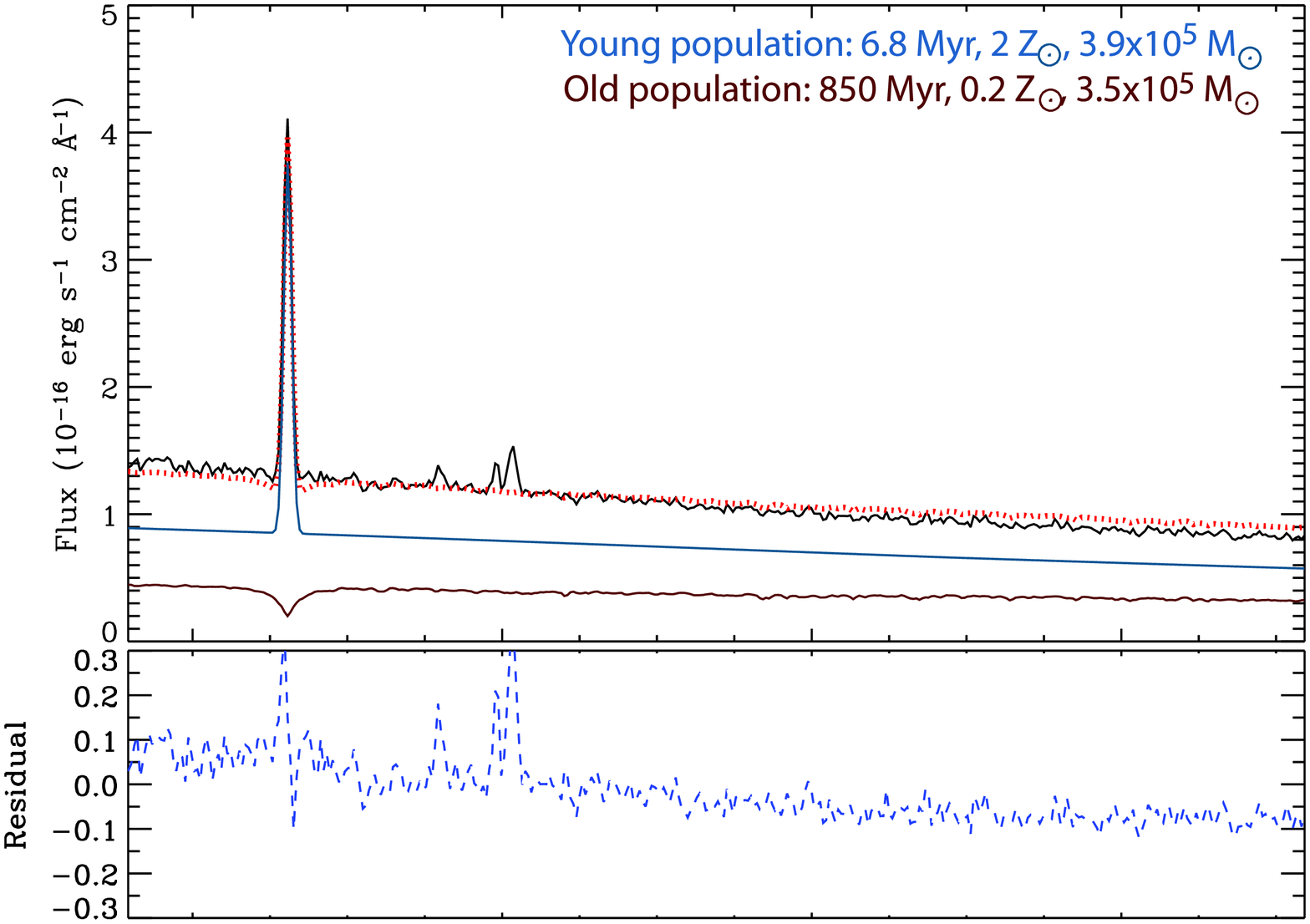}}
\subfigure{\includegraphics[width=0.49\textwidth,height=6cm]{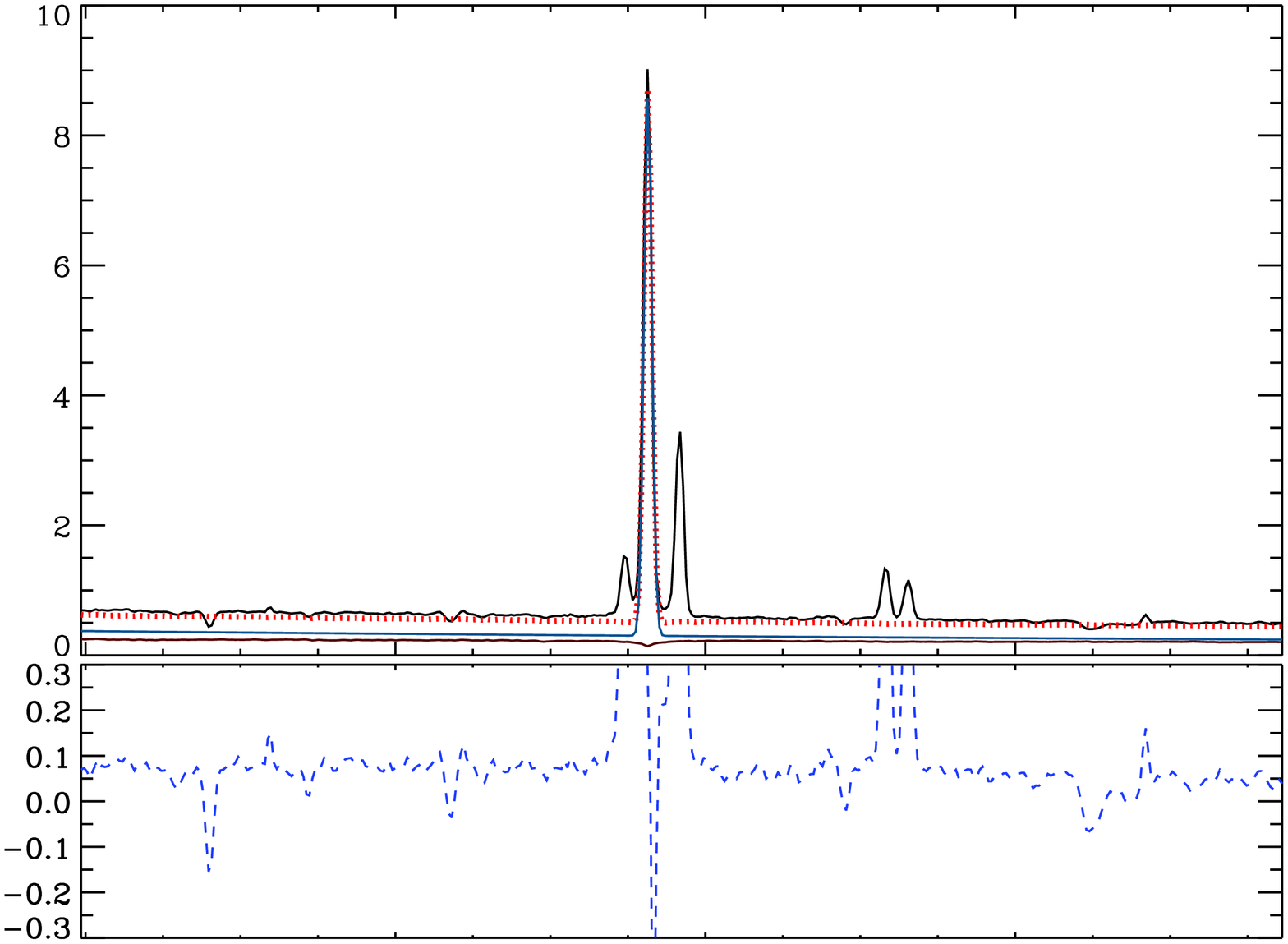}}
\subfigure{\includegraphics[width=0.49\textwidth,height=6cm]{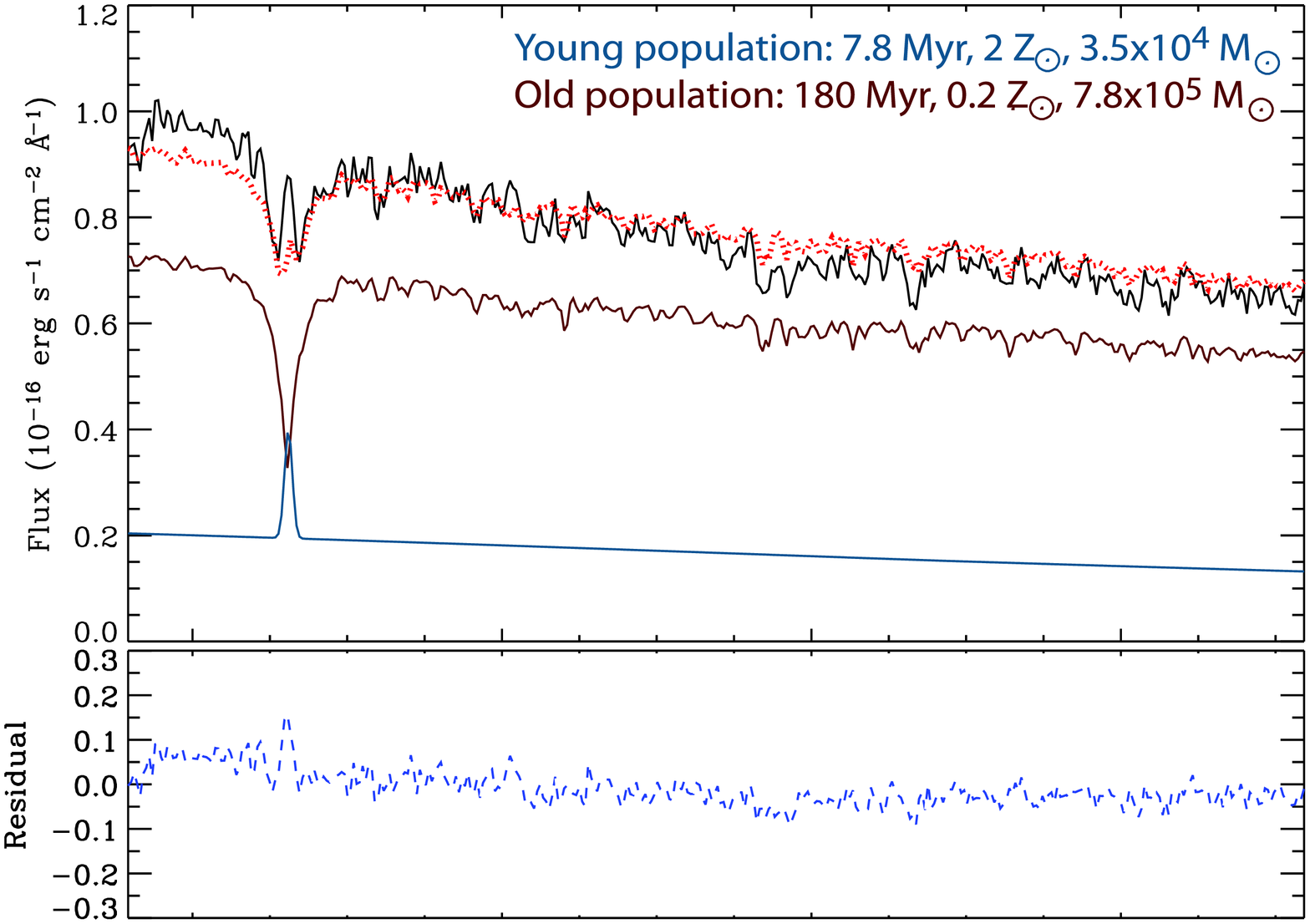}}
\subfigure{\includegraphics[width=0.49\textwidth,height=6cm]{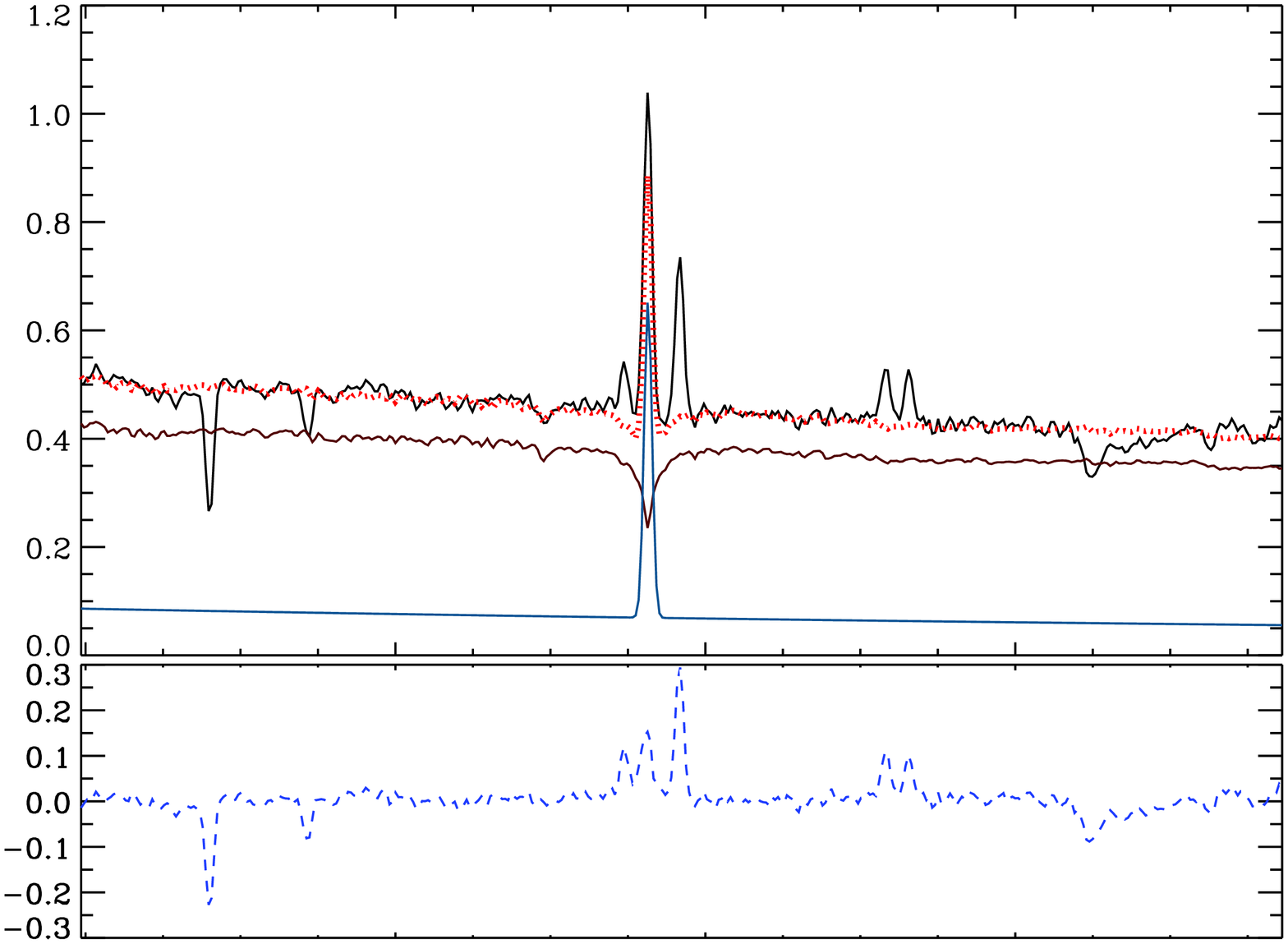}}
\subfigure{\includegraphics[width=0.49\textwidth,height=6cm]{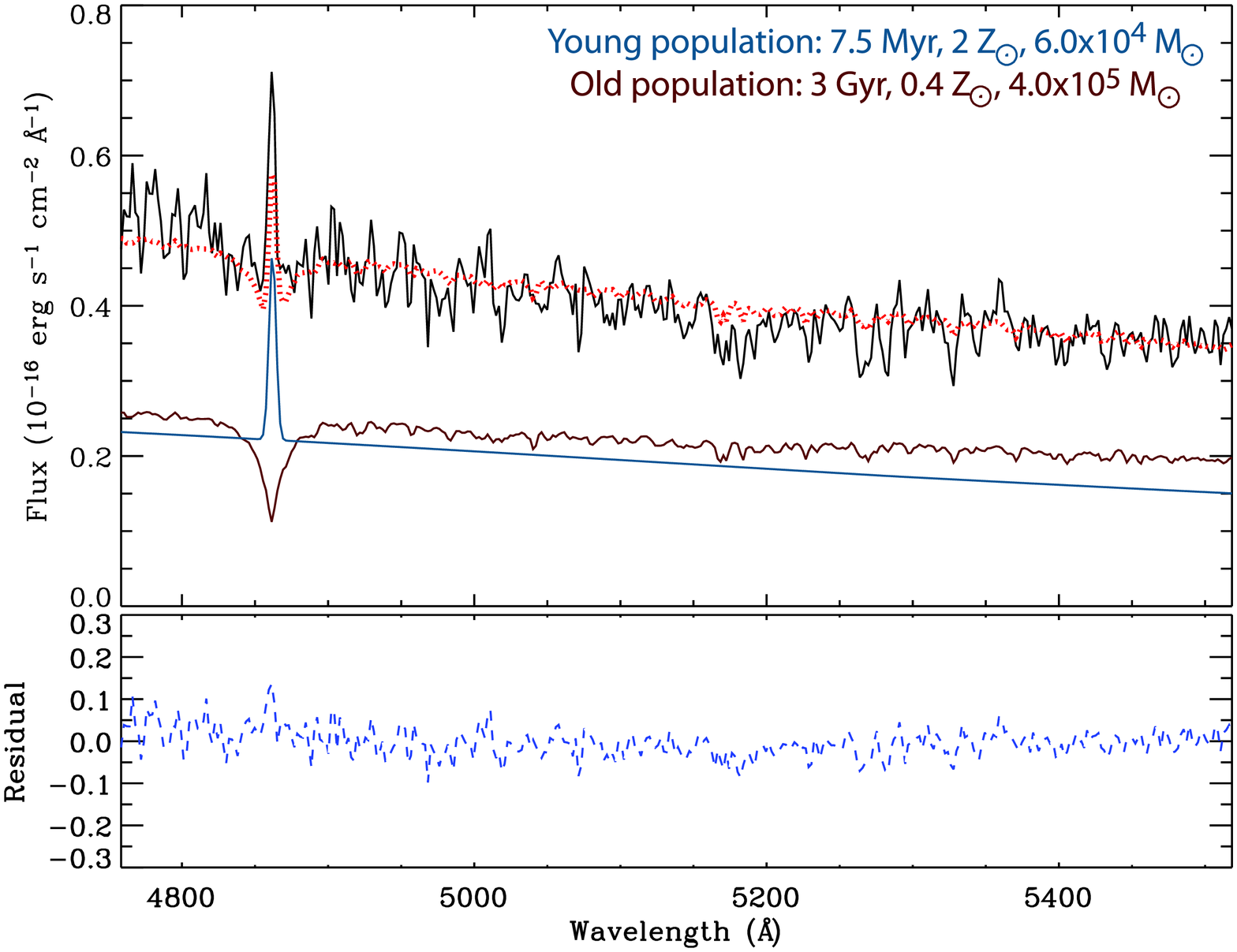}}
\subfigure{\includegraphics[width=0.49\textwidth,height=6cm]{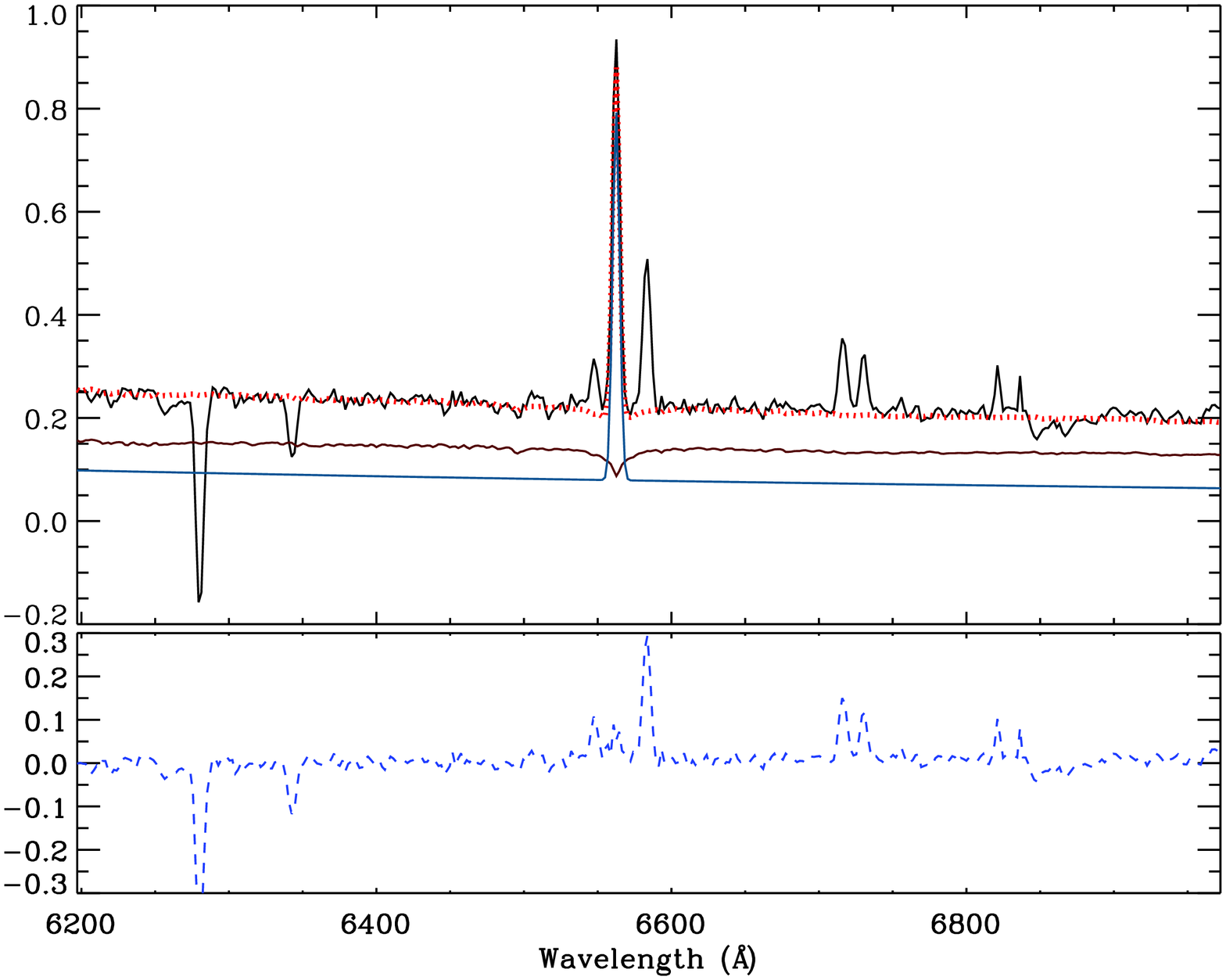}}
\caption{Examples of best models for three different regions. Top panels: spectrum \# 322, in C1, dominated by strong emission lines.  Middle panels: spectrum \# 496, in C2, dominated by an old stellar populations. Bottom panels: spectrum \# 330, in the high extinction zone X1 at the North of C1, where the flux from the two stellar populations is nearly equals. The black line represents the observed spectrum in the MR1 configuration (left) and the MR2 configuration (right). The contribution of a young stellar population characterized using the emission line is shown by the blue curve while the old stellar population studied using the absorption lines is represented by the red curve. The best model obtained by combining the contribution from the two stellar populations is shown by the red dotted curve. The blue dotted curve is the residual obtained after subtraction of the best model from the observed spectrum. Characteristics of the stellar populations used for the models are indicated on the right panels.}
\label{spec_young}
\end{figure*}

Fig.~\ref{spec_young} shows the best model obtained after the final iteration for three spectra as examples. One of them is located in C1, where the young population is dominating the flux, another one is located near the galaxy center, in C2, where weak emission lines are seen, and the last spectrum was taken North of C1, in X1, where both a young and an old population are more or less equally contributing to the flux. This figure presents the individual contributions to the final model from the young and the old components, along with the residual between the observed spectrum and the final model.  Since the best models are found using measurements in many  lines and in the continuum, a perfect superposition is not easy to achieve. The overall residual is generally close to zero (and will never be equal to zero in the emission lines of \oiii, \nii, and \sii\ since they are not taken into account in the young population models) and indicate a good fit of the final models. The spectrum in C1 (top panels in Fig.~\ref{spec_young}) is best reproduced by a young population of 6.8$\pm$0.2\,Myr with a metallicity of 2\,\zsun\   (based on an oxygen abundance of 9.1$\pm$0.3) and a mass of 3.9$\pm$0.4\,$\times$10$^{5}$\,\msun, superposed with an averaged old stellar population of 850$\pm$150\,Myr with a metallicity  of 0.2\,\zsun\  and a mass of 3$\pm$1\,$\times$10$^{5}$\,\msun. In this case, the flux from the young stellar population is so important that it is not possible to constrain the characteristics of the old component. Therefore, an averaged old population spectrum was simply adopted. In   C2 (middle panels in Fig.~\ref{spec_young}), the best model was found by co-adding a young population of 7.8$\pm$0.2\,Myr at 2\,\zsun\ (based on an oxygen abundance of 9.0$\pm$0.2) and a small stellar mass of 3.5$\pm$0.4\,$\times$10$^{4}$\,\msun,  with an old stellar population of 180$\pm$50\,Myr at  0.2\,\zsun\ and a mass of 7.8$\pm$0.5\,$\times$10$^{5}$\,\msun. The best model for the spectrum in X1 (bottom panels in Fig.~\ref{spec_young}) was obtained by adding a young population of 7.5$\pm$0.3\,Myr at  2\,\zsun\ (based on an oxygen abundance of 8.9$\pm$0.4) and a very low mass of 6.0$\pm$0.5\,$\times$10$^{4}$\,\msun, with an old stellar population of 3$\pm$1\,Gyr at 0.4\,\zsun\ and a mass of 4.0$\pm$0.3\,$\times$10$^{5}$\,\msun.

\subsubsection{Age and Metallicity}

Fig.~\ref{ages} also shows two maps of the distribution of the age for the stellar populations after the final iteration, one based on the emission lines (Fig.~\ref{ages_c}) which is limited to younger populations and the other based on the absorption lines (Fig.~\ref{ages_d}). As found after iteration 0, younger populations are located near regions C1 and C3. The age at the location of the emission peak  in L1 is 5.5$\pm$0.1\,Myr and the average age is 6.0$\pm$0.1\,Myr in C1 and 6.4$\pm$0.1\,Myr in C3. In  C2, a few spectra display ratios of the emission lines that indicate some peculiar activity in the nucleus (see \S\,5.3.4). Excluding these spectra, an age of 7.9$\pm$0.3\,Myr is found in C2 based on the emission lines. Surrounding  C2, along the dust lane, the emission lines give and average value for the age of 8.2$\pm$0.4\,Myr.  Note that in the upper North-East corner of this map, the age is often given by the \ha\ line only as an emission in \hb\ is not always easy to measure.  

The absorption lines reveal a more or less homogeneous old population of 850$\pm$150\,Myr, on average, distributed all over the central galaxy region observed (Fig.~\ref{ages_d}).  Within the region C2, the average age is somewhat older, 940$\pm$120\,Myr, but no other specific pattern or gradient is seen for the old population age.  It is difficult to study the absorption lines near C1 because the contribution to the continuum from the young population is very strong. After its subtraction, the signal-to-noise ratio in the absorption lines is very low and models give unreliable values (often the lower age limit from the models with an uncertainty equals to zero or to an extremely large value). Therefore, in the map based on the absorption lines, we have replaced unreliable values (due directly or indirectly to a low signal-to-noise ratio) by white pixels. 

The final iteration reveals values for the age of the stellar populations which are significantly different from those obtained after iteration 0. A comparison of the age maps in Fig.~\ref{ages} clearly shows this. Numbers for the young population age are slightly but systematically smaller after the final iteration mainly because the stellar absorption component underneath the emission lines is subtracted in the iteration process. The old population age is, on the other hand, increased after the final iteration. 
This increase  is brought forth mostly by two effects, first because the subtraction of the young population continuum flux has a tendency  to increase the equivalent width of the absorption lines and second, because the addition of the \hb\ absorption feature to the age determination process results, in this case, in a larger probability for an older age value.

\begin{figure}
\centering
\subfigure{\includegraphics[width=0.47\textwidth]{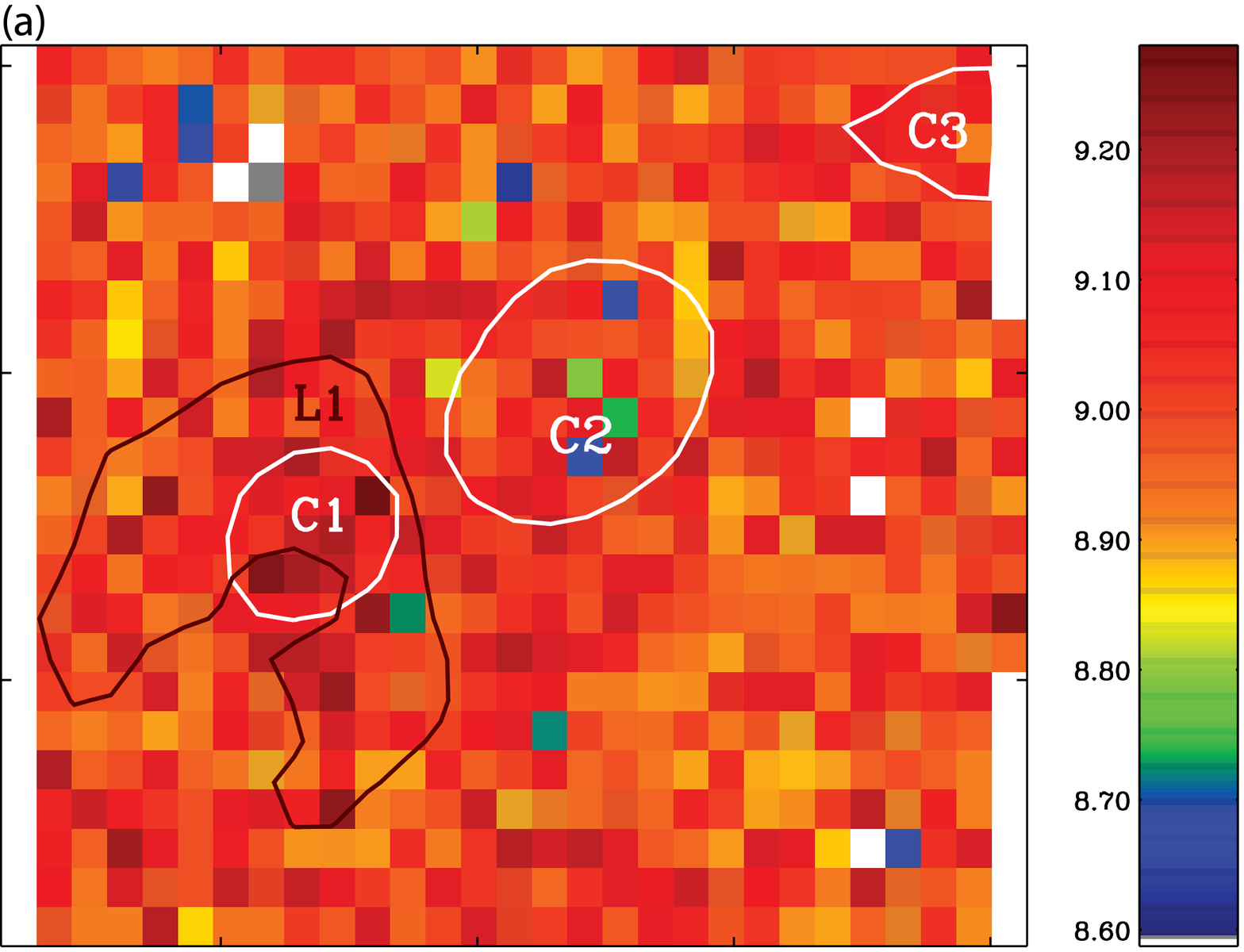}}
\subfigure{\includegraphics[width=0.47\textwidth]{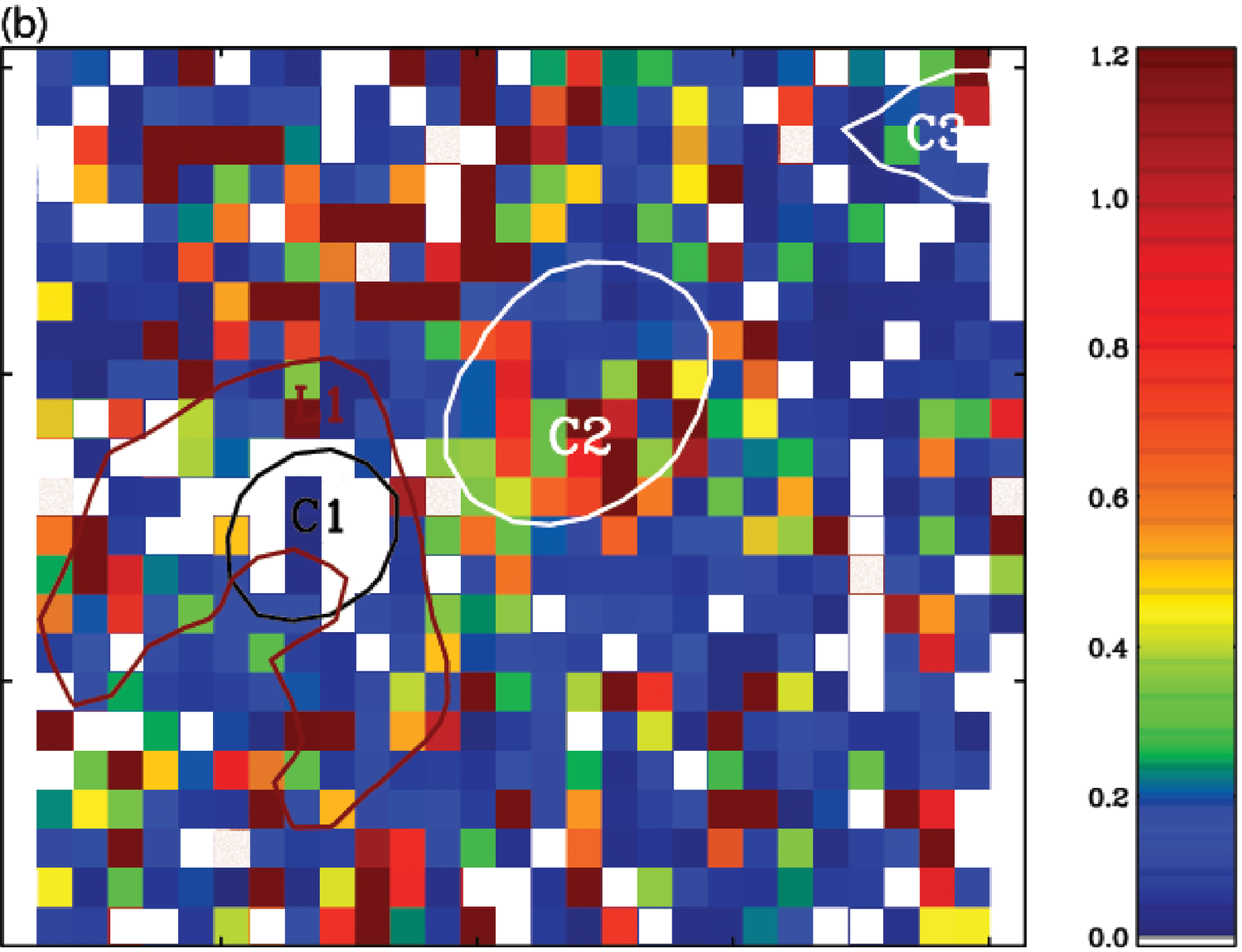}}
\caption{Metallicity distributions after the final iteration. In (a), the gas metallicity 12+log(O/H) is shown. In (b), the stellar metallicity Z is given. Isolated pixels with a lower metallicity value in (a) display very week emission lines. White pixels in (b) indicate poor line measurements because of a low signal-to-noise ratio for the absorption lines studied. Contour lines are as in Fig.~\ref{conti}.}
\label{met_rev}
\end{figure}

There is not a significant difference between the metallicity calculated after the final iteration and iteration 0. The gas and stellar metallicity maps are almost featureless (Fig.~\ref{met_rev}). After the final iteration, the average gas metallicity is supersolar at 12+log(O/H)$= 8.9\pm$0.1 (i.e. 2\,\zsun) while the stellar metallicity is lower with an average value of Z=0.2$\pm$0.2\,\zsun. Inside the region  C1, the gas metallicity increases to an average value 12+log(O/H)=9.1$\pm$0.1, i.e. 3\,\zsun. For the same region, the estimated stellar metallicity  is unreliable because of the very strong flux from the young population which reduces the signal-to-noise ratio in the absorption lines. In C2, if we assume that the peculiar activity seen in a few spectra has a negligible effect on the continuum and absorption lines, we find on average a larger stellar metallicity, Z$\simeq$1.0$\pm$0.2\,\zsun. A larger value for the stellar metallicity in the galaxy center is not surprising. The difference between the stellar and gas metallicity is, on the other hand, quite important.

\begin{figure}
\centering
\subfigure{\label{age_histo_a}\includegraphics[width=0.47\textwidth]{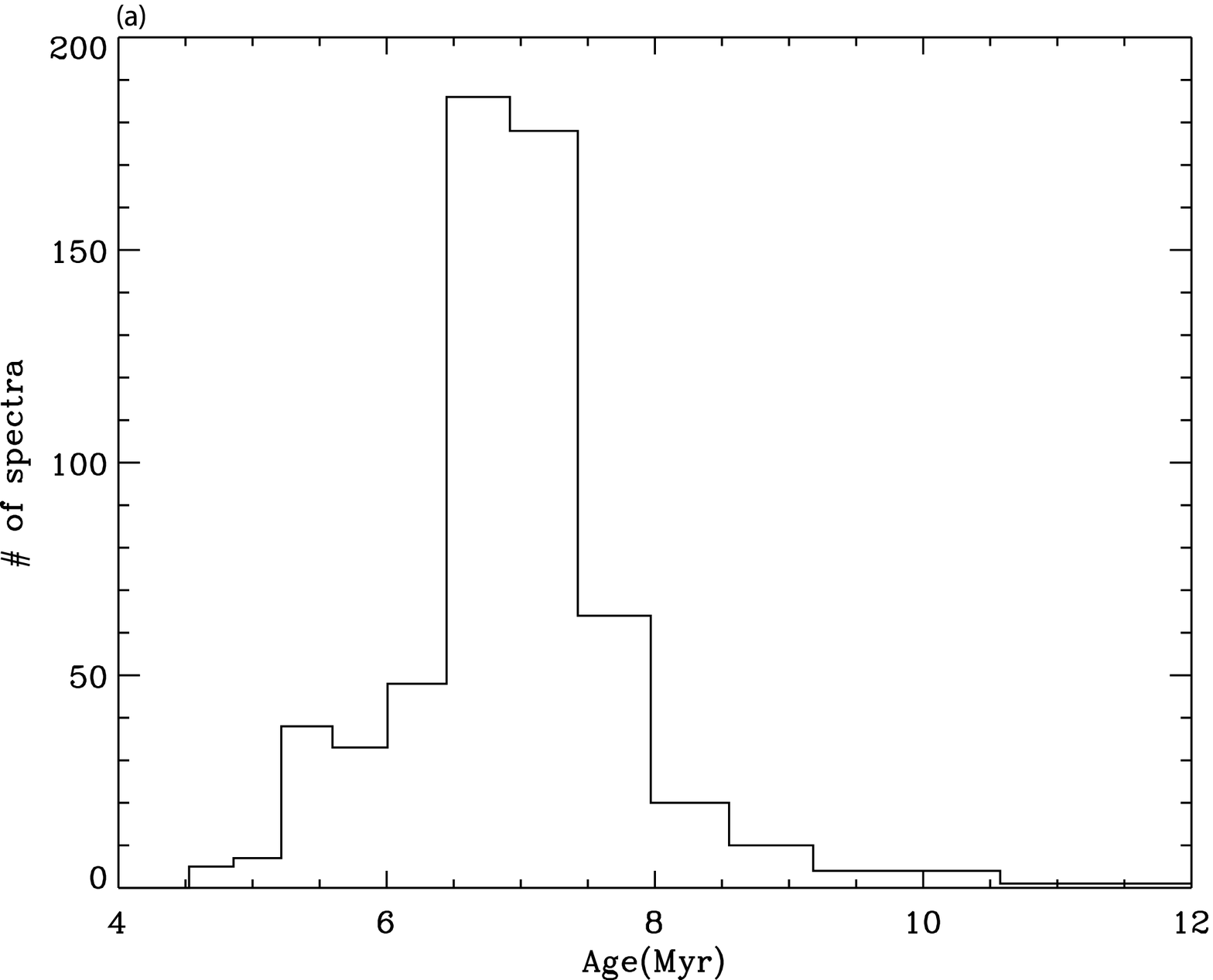}}
\subfigure{\label{age_histo_b}\includegraphics[width=0.47\textwidth]{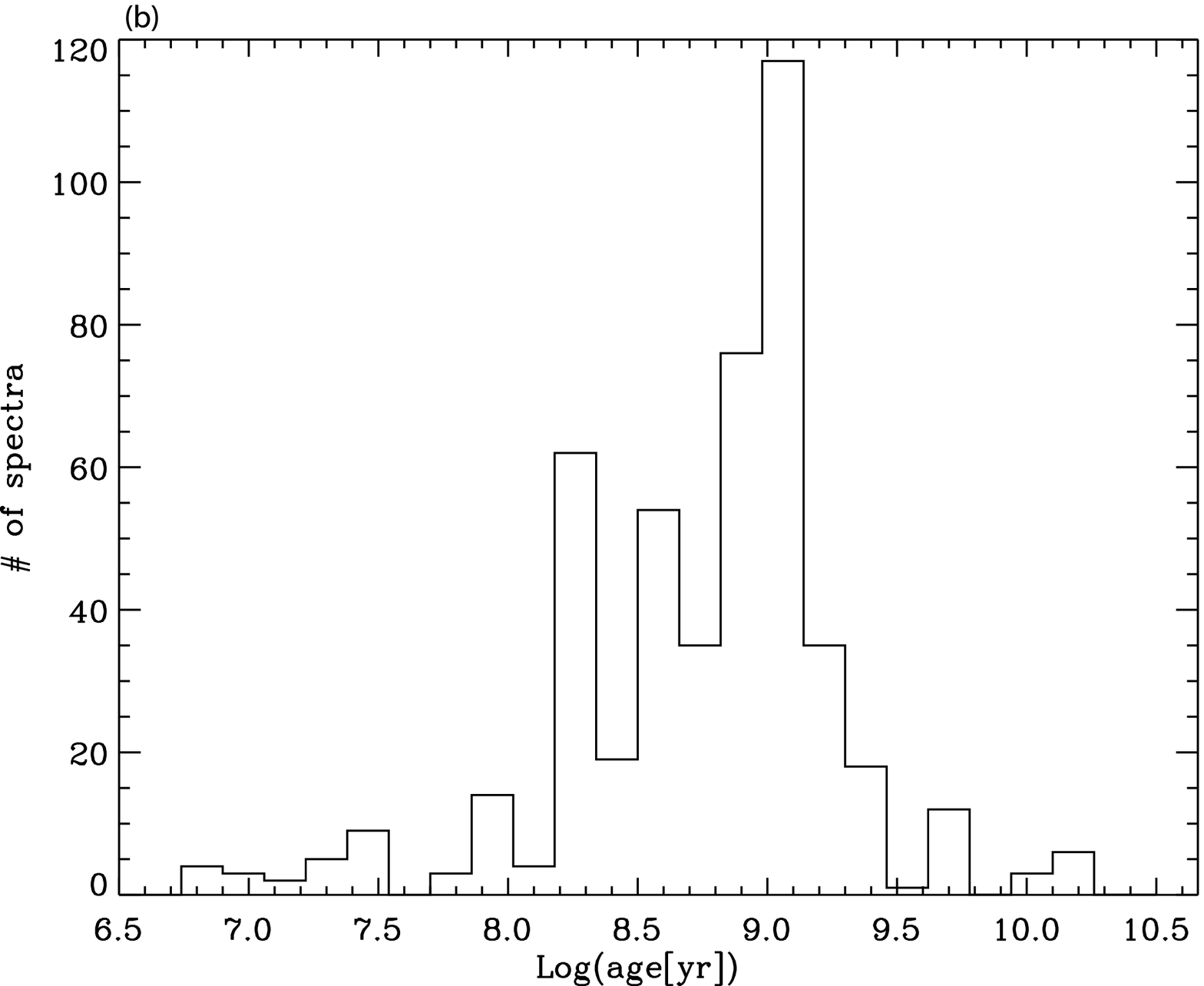}}
\caption{Histograms of the number of spectra as a function of the age after the final iteration. In (a) the age is shown based on the emission lines. In (b) the age is given by the absorption lines (as for Fig.~\ref{ages}). }
\label{age_histo}
\end{figure}

To get a better idea of how the various populations found are distributed, we plot in Fig. \ref{age_histo} two histograms showing the number of spectra as a function of the age. The younger populations, found using the emission lines, are distributed in a narrow spike, centered at about 7\,Myr (Fig.~\ref{age_histo_a}). An asymmetry indicated by a second spike, much weaker, near 6\,Myr, corresponding to the average population in C1. The histogram based on the age extracted from the absorption lines shows an old and long episode of star formation in the galaxy central region (Fig.~\ref{age_histo_b}). It started about 3\,Gyr ago, was very active around 1\,Gyr, and lasted until about 100\,Myr. There is also an indication of a star-forming activity between 6 and 30\,Myr and around 10\,Gyr in this histogram, but it is associated with sporadic pixels on the age map which are often at the limit of our rejection criteria due to a lower signal in the absorption lines.

%%%%%%%%%%%%%%%%%%%%%%%%%%%
\subsubsection{Mass}

\begin{figure}
\centering
\subfigure{\label{mass_dis_a}\includegraphics[width=0.47\textwidth]{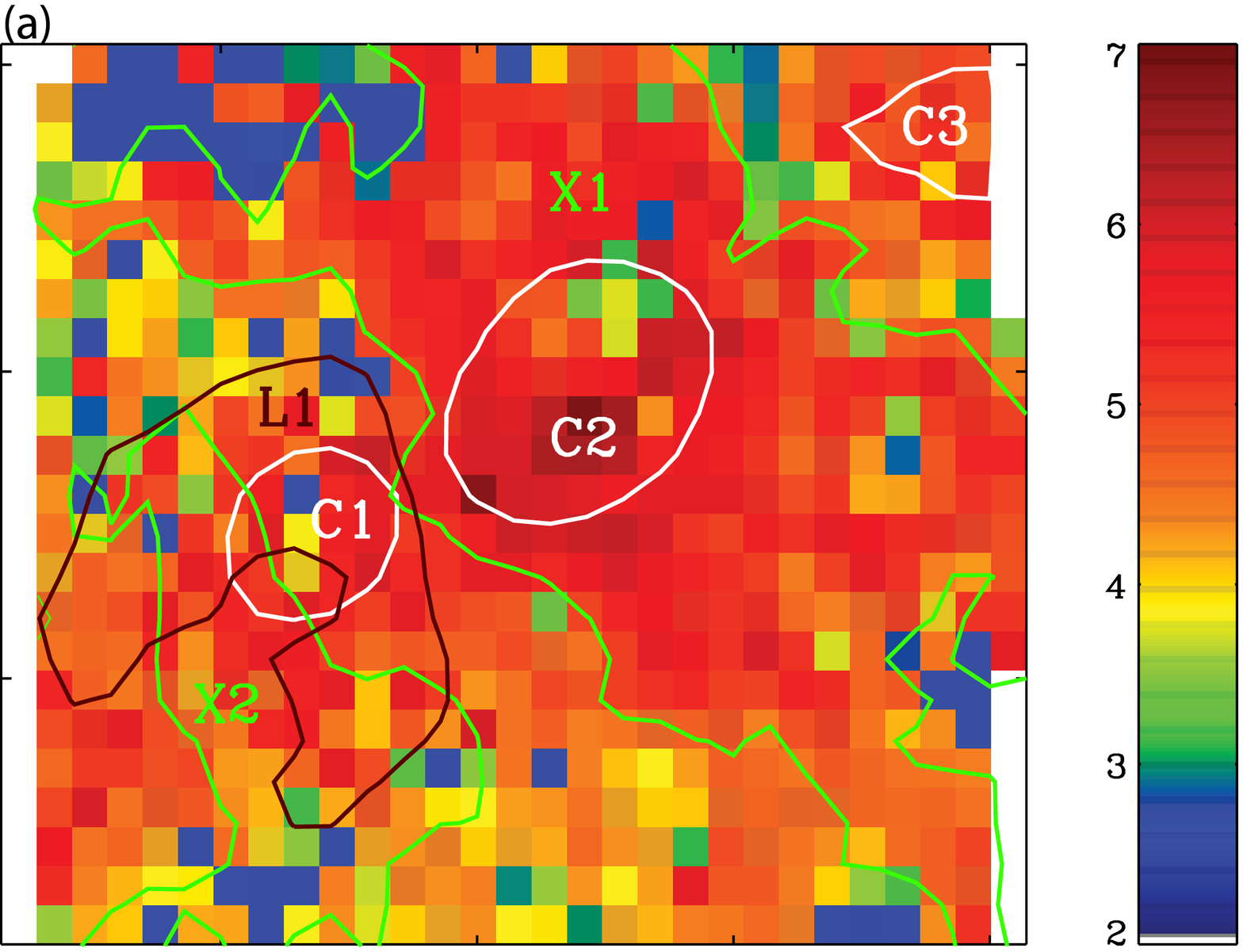}}
\subfigure{\label{mass_dis_b}\includegraphics[width=0.47\textwidth]{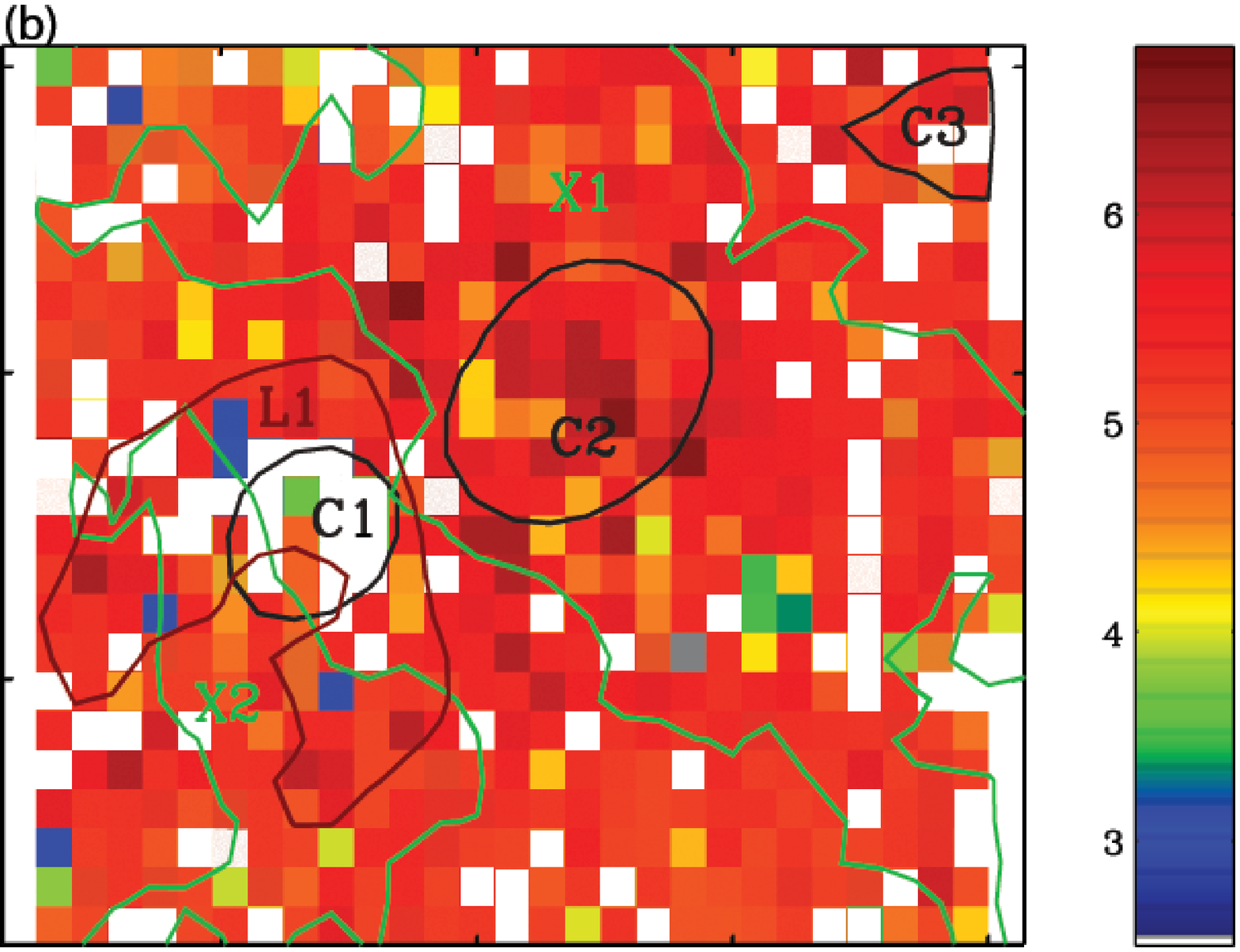}}
\subfigure{\label{mass_dis_c}\includegraphics[width=0.47\textwidth]{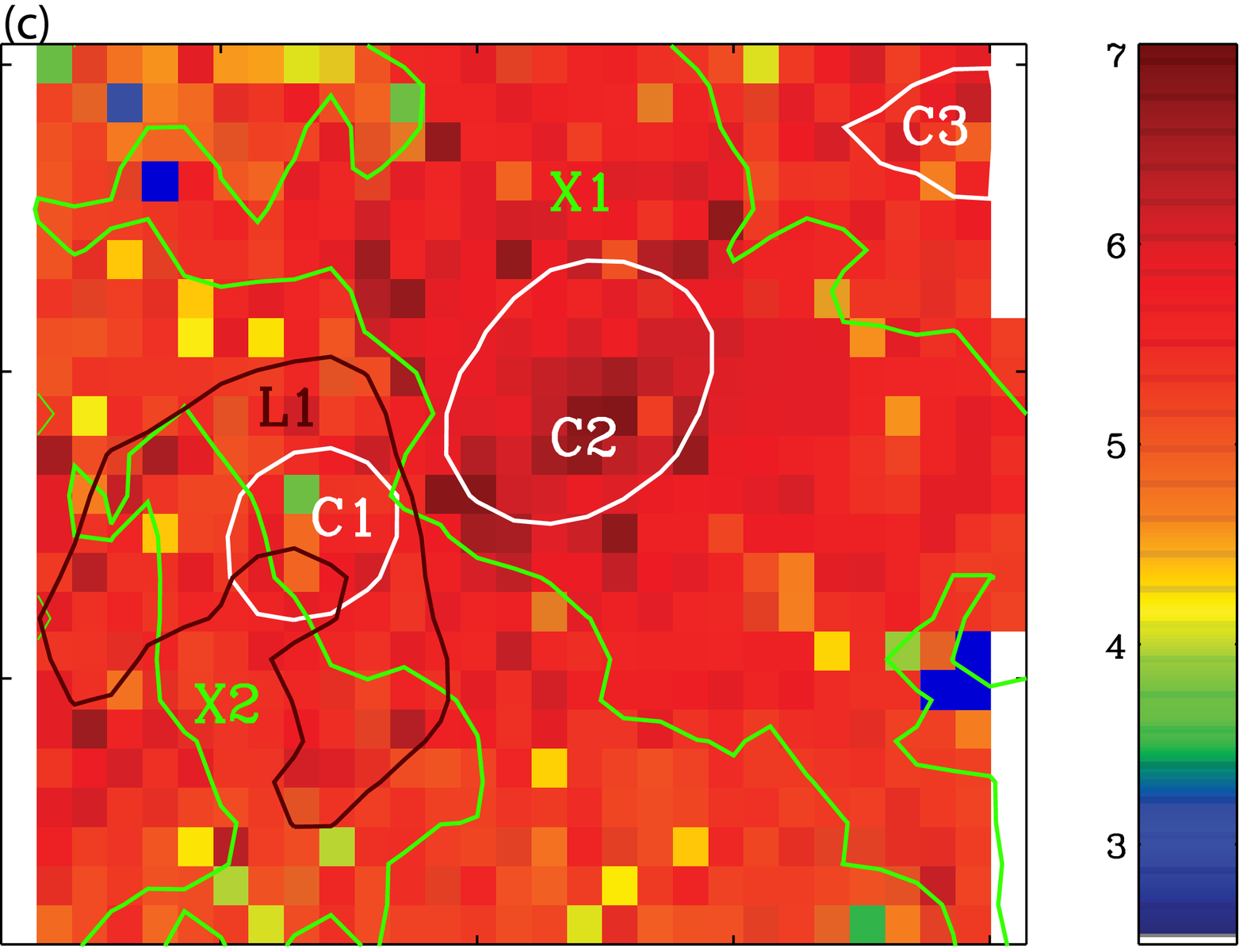}}
\caption{Mass distributions after the final iteration. The maps show the current stellar population mass, in log(mass~[\msun]). In (a), the mass of the young population based on the emission lines. In (b), the mass of the old population based on the absorption lines. The total population mass is shown in (c). White pixels in (b) indicate missing measurements because of a low signal-to-noise ratio for the absorption lines studied. Contour lines are as in Fig.~\ref{conti} and \ref{ebv}.}
\label{mass_dis}
\end{figure}

Fig.~\ref{mass_dis} shows the mass distributions for the young and the old stellar populations, individually, as well as for the total stellar mass derived in each spectrum. The mass of the stellar populations is obtained during the iteration process by comparing the continuum level of the observed spectrum with the best model selected based on the line indicators after each time step. The mass estimated here are current values. The current mass of a stellar population is smaller than the initial mass as a consequence of stellar evolution; the most massive stars evolve and die more quickly and do not contribute to the flux and mass observed now.  We estimate, based on models for an instantaneous burst with a standard Salpeter IMF, that the current mass of a 1\,Gyr population would be $\sim$26 per cent smaller then the initial mass, and for a 10\,Gyr population it would be $\sim$40 per cent smaller.  We find the highest mass concentration of the young populations along the axis C1-C2-C3 and underneath the dust lane (Fig.~\ref{mass_dis_a}). In  C1, where the young population is very important, we have considered the average 850\,Myr old population. If this population was younger, 10\,Myr for example, then the mass of the young population in C1 would not change significantly, i.e. by less than a factor of 10. Typically in a spectrum in C1 and C2, there is about 5$\pm$1\,$\times$10$^5$\,\msun, corresponding to an average density of about 50\,\msun\,pc$^{-2}$ and 40\,\msun\,pc$^{-2}$, respectively. Outside these regions there is an average mass of 8$\pm$1$\times$10$^4$\,\msun\ per spectrum and an average density of 30\,\msun\,pc$^{-2}$ (except to the North-East where it is smaller by a factor $\sim$\,10). Summing for all pixels, the total mass in new stars is 1.14$\pm$0.05\,$\times$10$^8$\,\msun.

The old stellar population is more massive at the center than outside (Fig.~\ref{mass_dis_b}), with values from 2$\pm$1\,$\times$10$^{5}$\,\msun\ per spectrum in the region C2 ($\sim$\,20\,\msun\,pc$^{-2}$) to 8$\pm$3\,$\times$10$^4$\,\msun\ per spectrum in the surrounding. The mass of the old population near C1 is often not shown on the map. If we compute the mass in C1 using an old population with the average age of 850\,Myr, we find the same value as in the surrounding region. (If we consider instead a younger age, 10\,Myr for exemple, the mass found would be reduced by a factor of $\sim$10). Summing for all pixels available, we find that the old stellar population represents 2.07$\pm$0.06\,$\times$10$^8$\,\msun,  about 60 per cent of the total stellar mass. It clearly shows the importance of the most recent star forming episode that occurred about 7\,Myr ago.

Fig.~\ref{mass_dis_c} shows the map for the total stellar mass obtained after adding together the two stellar populations characterized with the iterative technique. The mass is strongly concentrated in the galaxy center in C2. The region C1, where we are considering the average underlying older population, is also important. This is similar to the continuum maps (see Fig.~\ref{conti_a} and \ref{conti_b}). We also distinguish the galaxy large scale bar in the total stellar mass map.  Within the section of the bar observed (supposing a width of 260\,pc), the average density is 40$\pm$5\,\msun\,pc$^{-2}$ and it includes 2.2$\pm$0.1\,$\times$10$^8$\,\msun, i.e. almost 2/3 of the mass present in the galaxy central area. The region C2 contains about half of the mass of the bar observed.

%%%%%%%%%%%%%%%%%%%%%%%%%%%%%%%%%%%%%%%%%%%%%
\subsubsection{Peculiar Activity}

\begin{figure}
\centering
\subfigure{\label{gas_param_a}\includegraphics[width=0.47\textwidth]{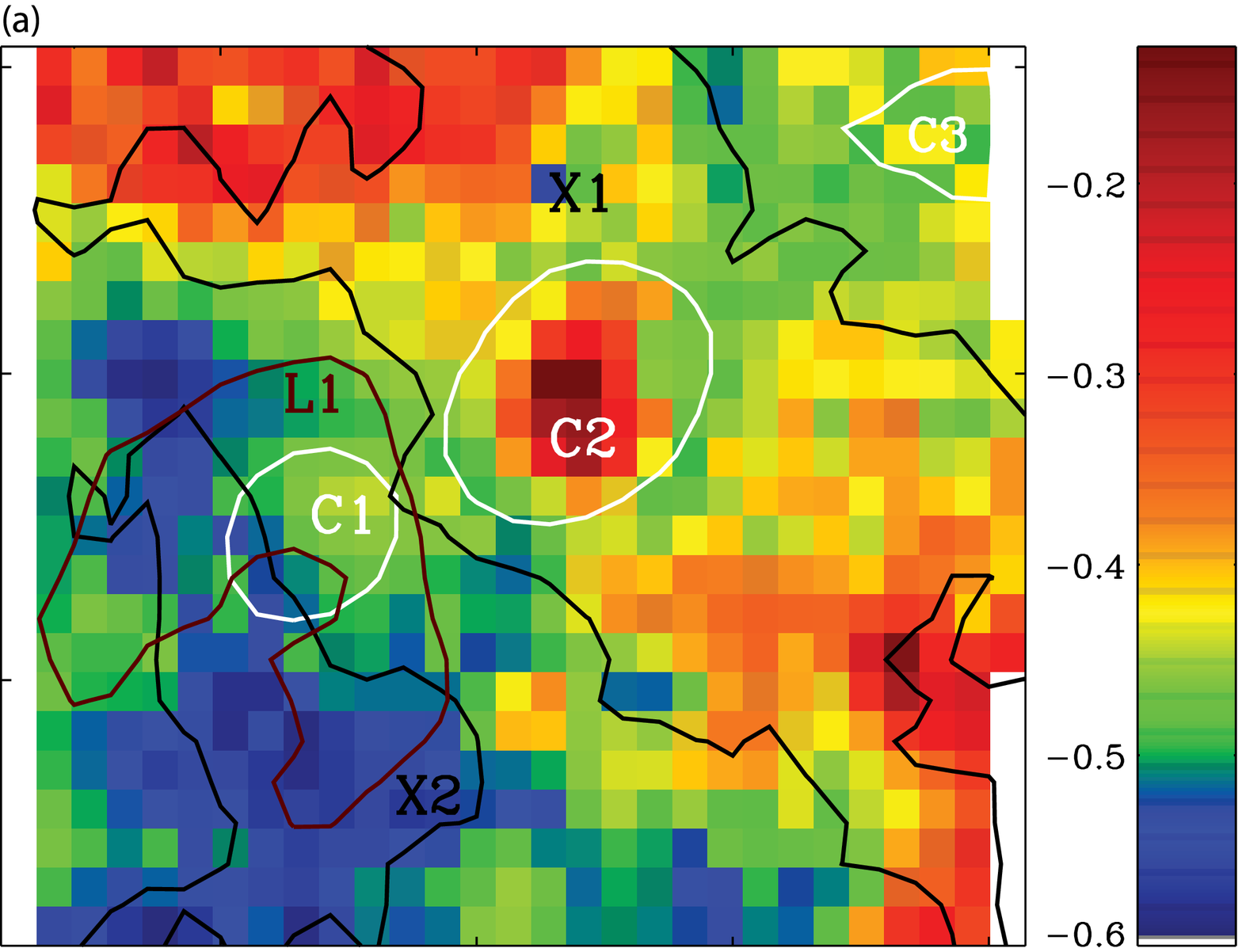}}
\subfigure{\label{gas_param_b}\includegraphics[width=0.47\textwidth]{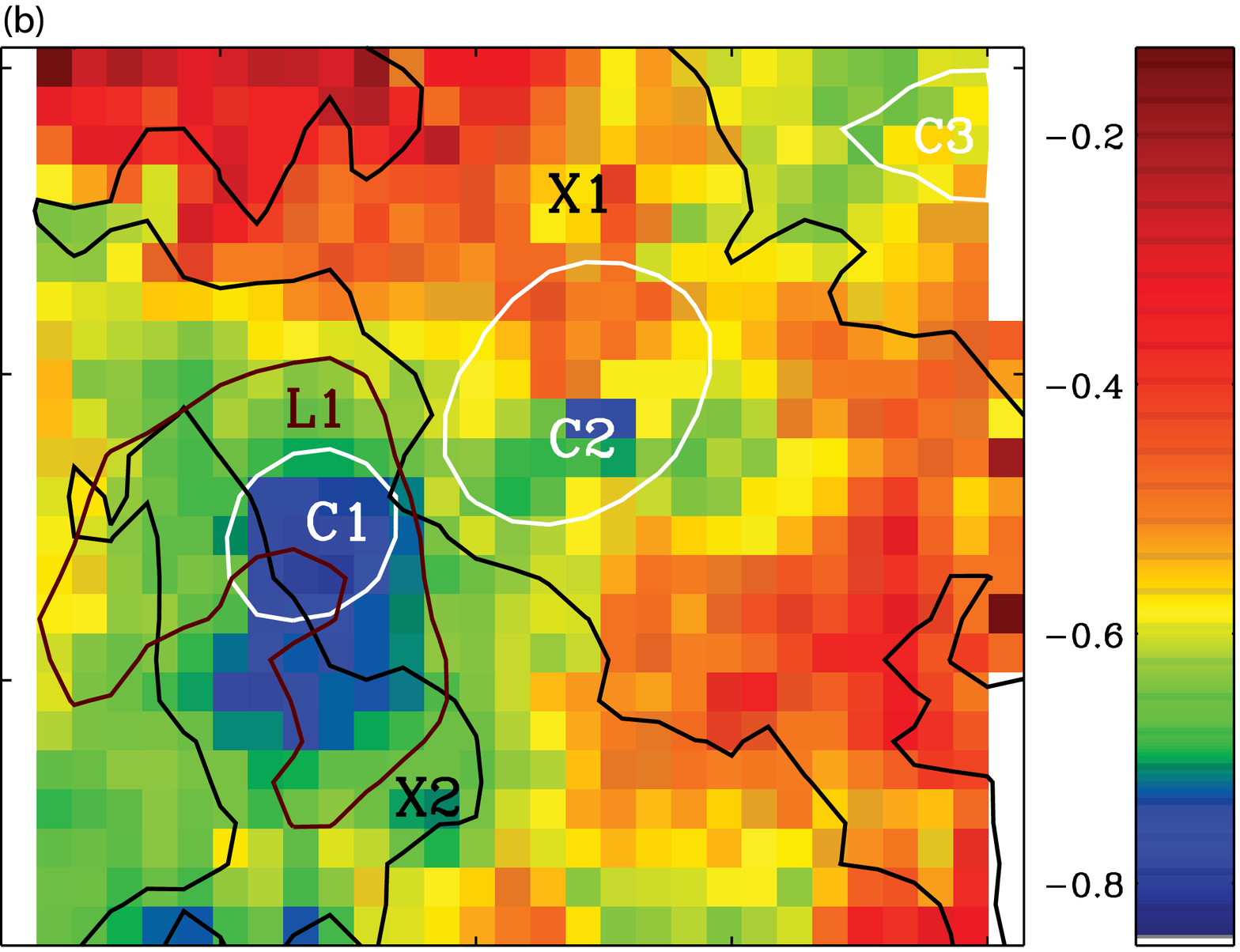}}
\caption{Non-thermal indicator maps. In (a), $\log($\nii\,$\lambda$\,6583/\ha) and in (b), $\log($\sii\,$\lambda$\,6716+6730/\ha). Contour lines are as in Fig.~\ref{conti} and \ref{ebv}.}
\label{gas_param}
\end{figure}

Fig.~\ref{gas_param} presents maps of the emission line ratios $\log$(\nii\,$\lambda$\,6583/\ha) and $\log$(\sii\,$\lambda$\,6716+6730/\ha). In general, both maps show typical ratios for \hii\ regions, i.e.  where \nii\ and \sii\ are weak compared to \ha. The dust lane (region X1) displays ratios which are closer to zero, but this is where the emission lines are weaker and, particularly in the North-East image corner, the signal is also lower. On the other hand, the region C2 has good signal and it shows a ratio of $\log$(\nii\,$\lambda$\,6583/\ha)$>$$-$0.2 which is high for an \hii\ region and more typical for an AGN according to \citet{2006MNRAS.371..972S}. However the value of $\log$(\sii\,$\lambda$\,6716+6730/\ha) in  C2 remains typical of an \hii\ region.

\begin{figure}
\centering
\subfigure{\label{bpt_a}\includegraphics[width=0.46\textwidth]{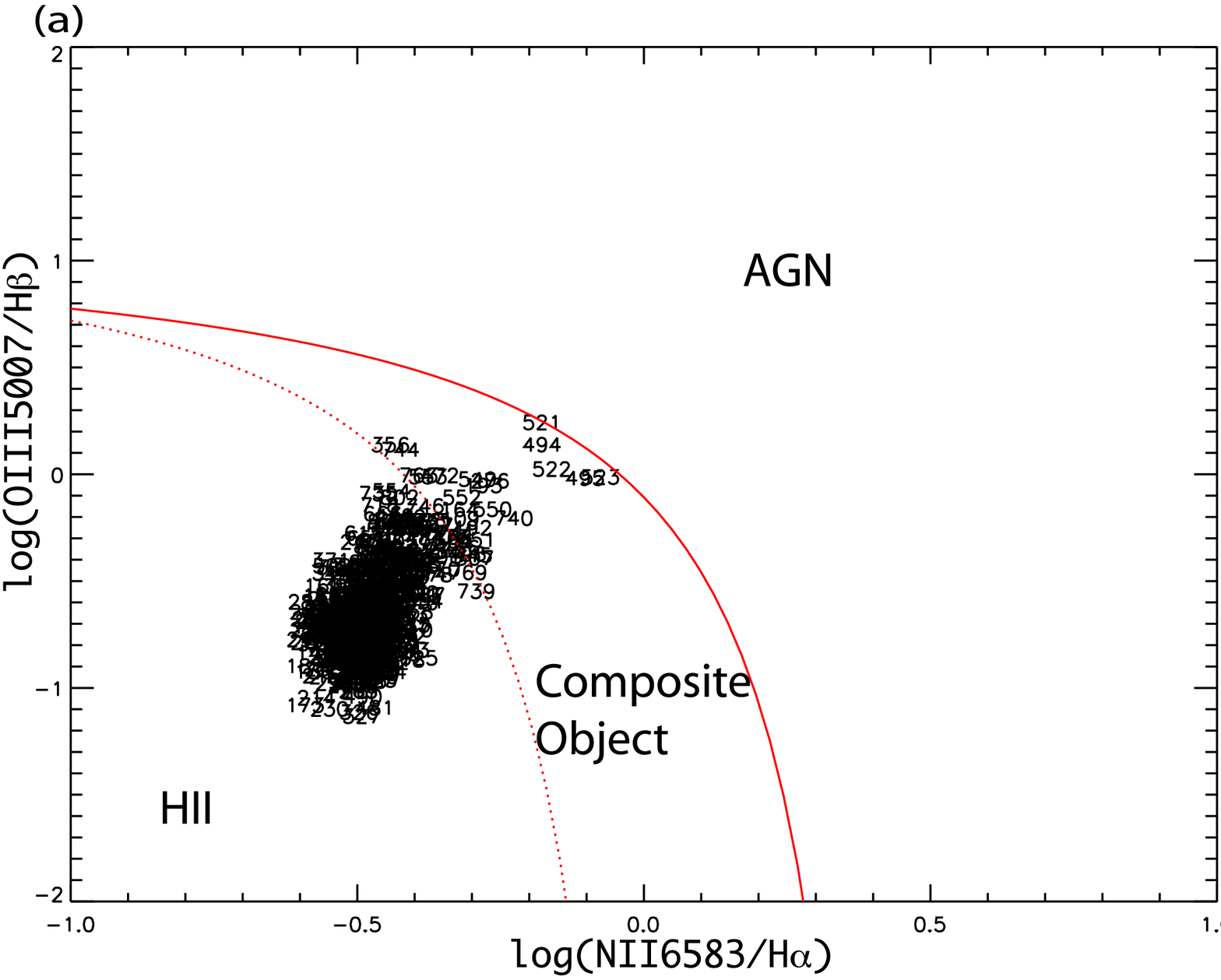}}
\subfigure{\label{bpt_b}\includegraphics[width=0.46\textwidth]{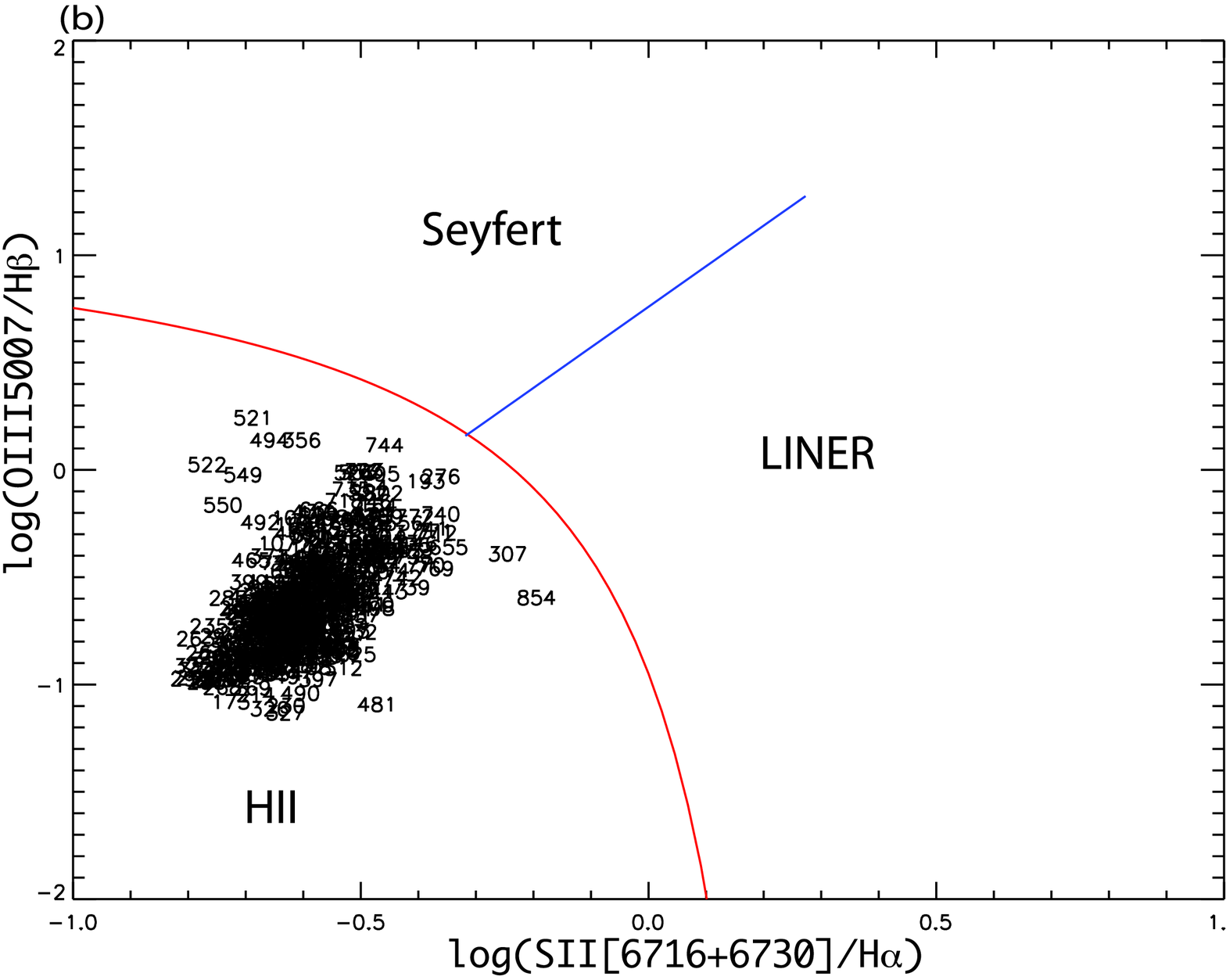}}
\subfigure{\label{bpt_c}\includegraphics[width=0.46\textwidth]{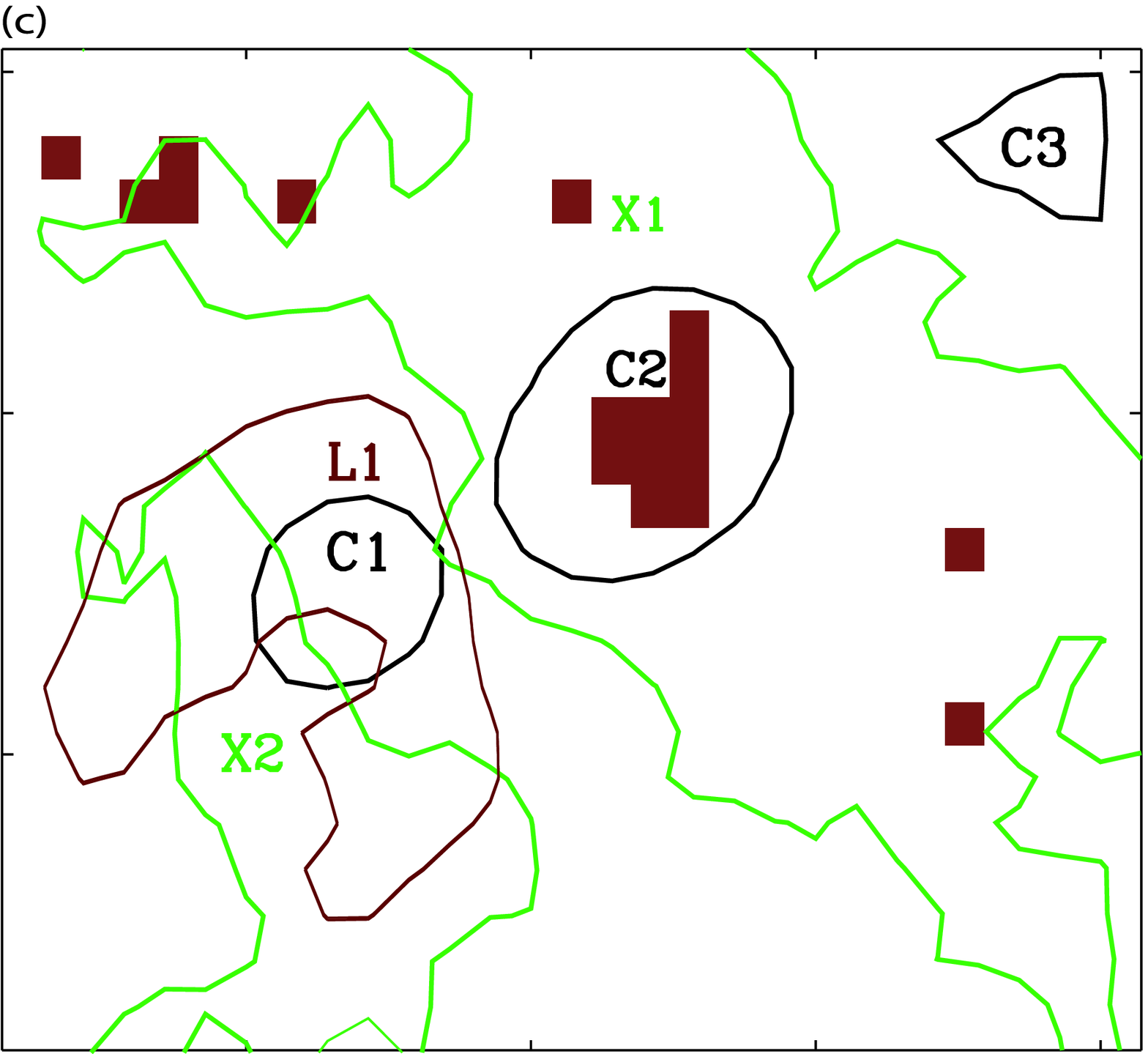}}
\caption{Activity indicators.  The BPT diagrams of (a) $\log$(\oiii\,$\lambda$\,5007/\hb) {\sl vs} $\log$(\nii\,$\lambda$\,6583/\ha) and (b) $\log$(\oiii\,$\lambda$\,5007/\hb) {\sl vs} $\log($\sii\,$\lambda$\,6716+6730/\ha) are shown. Lenslet \# are given on those plot. The map in (c) indicates the location of  pixels where an emission associated with a composite/transition region is observed according to the diagram in (a). Contour lines are as in Fig.~\ref{conti} and~\ref{ebv}.}

\label{bpt}
\end{figure}

The BPT diagrams, generally used to identify the type of the activity present in a galaxy, confirms the absence of a strong non-thermal activity. In Fig.~\ref{bpt} the red solid line the top panels indicate the limit between \hii\ regions and Seyfert/LINER activities according to \citet{2001ApJ...556..121K}. The dotted line in panel (a) defines the limit where \hii\ regions are observed according to \citet{2003MNRAS.346.1055K}.  The blue line in panel (b) separates Seyferts and LINERs \citep{2006MNRAS.372..961K}. No spectrum is lying above the AGN, Seyfert, or LINER limit.  Nevertheless, we see a significant number of spectra falling into the composite/transition region, i.e.  between the solid and dotted lines of the $\log$(\oiii\,$\lambda$\,5007/\hb) {\sl vs} $\log$(\nii\,$\lambda$\,6583/\ha) diagram. Fig.~\ref{bpt_c} shows the location of these spectra on the image. Most of them are near the galaxy center in C2. A few isolated spectra are located along the dust lane, where lower signal have been seen in the North-East corner. We therefore conclude that \gq\, hosts a composite/transition nucleus, i.e. a nucleus where both star formation and an AGN are present \citep[as defined by][]{2008ApJ...679...86W,2008ARA&A..46..475H}.

%%%%%%%%%%%%%%%%%%%%%%%%%%%%%%%%%%%%%%%%%%%%%%%%
\section{Discussion}

\subsection{The Iteration Technique}

The iteration technique has allowed us to refine the evaluation of the stellar population age, metallicity, and mass within a two population scenario. The age and mass of the old stellar population, based on the absorption lines, are the parameters that have improved the most. With the iterative process, the nebular emission in \hb\ can be removed, allowing us to use the \hb\ absorption component as an additional diagnostic and to reduce considerably  the error bar of the old population age.  By itself, the age value has little influence on the final mass of the old population. On the other hand, the presence of an important young stellar population affects the continuum level which also changes significantly  the determination of the old population mass and age. The young stars continuum emission, and its associated nebular continuum, dilute the absorption lines of the old stellar population.  In the region C2, for example, we find a typical age for the old population around 750$\pm$250\,Myr before the iterative process, while after, the age increases to 940$\pm$120\,Myr. The change is less spectacular for the young population, since the age and its uncertainty are, in this case, smaller (typically by 0.1-0.2\,Myr).  Nevertheless the young population age is also improved as a result of the better treatment of the underlying absorption in the nebular emission lines and the continuum level. This implies that emission line intensities for \hb\ and \ha\ are stronger than before and, in consequence, the age estimated from these indicators is younger.

The stellar metallicity has not changed significantly after the iterative process. It is strongly dependent on the \fei\ absorption lines, but mainly, it depends on statistics based on only a few large metallicity bins in the models. The oxygen abundance, however, is estimated from line intensity ratios,  involving the \hb\ and \ha\ lines which are better constrained as the iteration process converge.  The influence is keenly seen in the \oiii\,$\lambda$\,4959+5007/\hb\ ratio, although this is only a small part of the calculation, slightly increasing  the  oxygen abundance.

It is also important to realize that with the use of an integral field unit like OASIS we are able to resolve smaller star forming regions in nearby galaxies. The stellar population in one lenslet may become so small in mass that we could be losing the statistical aspect of the initial mass function. According to \citet{2003MNRAS.338..481C} a mass below 10$^5$\,M$_\odot$,  particularly for a young population, is becoming critical. On the other hand, we think that the statistical study of characteristics found over many lenslets is compensating for this.

The iterative technique  allows the identification of a weak peculiar activity in the central region of the galaxy. Before the iterative process, the underestimation of the \hb\ absorption flux tended to assign a non-thermal activity to many lenslets, even in some located far away from the galaxy center.  With the iterative process we have refined the line ratios to a point where most values are typical of \hii\ regions while only a few spectra, mostly those concentrated in the galaxy central region, belong to the composite/transition region of the BPT diagram. In the case of \gq, this non-thermal activity is very weak and did not cause difficulty for the analysis.  
A stronger non-thermal component is expected to affect the continuum level, the absorption lines, and also the extinction estimation. The presence of a strong non-thermal component will introduce an uncertainty for the population parameters with an iterative technique that only takes into account the superposition of stellar populations.

Two aspects that have been neglected in our iterative technique, and deserve a better treatment in the future when possible, concern the young stellar population contribution to the \hb, \mg, and \fei\ lines and the determination of the extinction. If the metallicity of the young stellar population is below solar and the old population is older than 1\,Gyr, it is not really a problem to neglect the absorption lines created by the young population since the old population will completely dominate these structures. However, a young population with an age between 10-30\,Myr and an abundance near or higher than solar may contribute to these lines, as can be seen in Fig.~\ref{evindsp}, and influence the characteristics found for both stellar populations.  For the moment, the models we are using do not include the absorption lines in all young stellar populations and therefore we could not consider this effect in our study.  The extinction was estimated in a simplified way in our study. It was done in the early phase of the analysis, prior to the iteration process, using the overall \ha\ line (without any correction for an absorption component) and our best measurement for the emission component in \hb\ (at the very base of an absorption component when present).  A specific extinction was calculated for each spectrum, allowing us to draw an extinction map, which is certainly a big advantage for our study. Another improvement would be to add a treatment of the extinction into the iterative technique and see its effect on the stellar population parameters. As the emission lines get better defined along the iteration process (i.e. as the underlying stellar absorption components are better known), it will modify the extinction, the slope of the continuum, and also influence the choice of the best model for the stellar populations. A preliminary test indicated, beside a much longer iteration time, a modification of the mass of the stellar populations (but not more than by a factor of 2) with a negligible effect on the age,  gas oxygen abundance, and  stellar metallicity.

%%%%%%%%%%%%%%%%%%%%%%%%%%%%%%%%%%
\subsection{An Evolutionary Scenario}

The information collected provides clues about the history of the central region of \gq. With these clues we propose here a simple evolutionary scenario leading to the structures and populations observed.  But first, let us summarize and further analyze these clues. We find a few young bursts, with an age between 5.5 and 6.5\,Myr, located along the galaxy large scale bar. We see an extended region of about 8\,Myr superposed to a dust lane which is oriented perpendicular to the galaxy large scale bar. These young populations are superposed to a longer episode of star formation with an average age near 850\,Myr. In the galaxy nucleus, the age of the old population reaches 940\,Myr and there is a composite/transition region activity. Furthermore there is a clear distinction between the gas metallicity,  2\,Z$_\odot$ on average (and higher near one of the younger bursts), and the old stellar population sub-solar average metallicity.

\begin{figure}
\centering
\subfigure{\label{SFH_a}\includegraphics[width=0.47\textwidth]{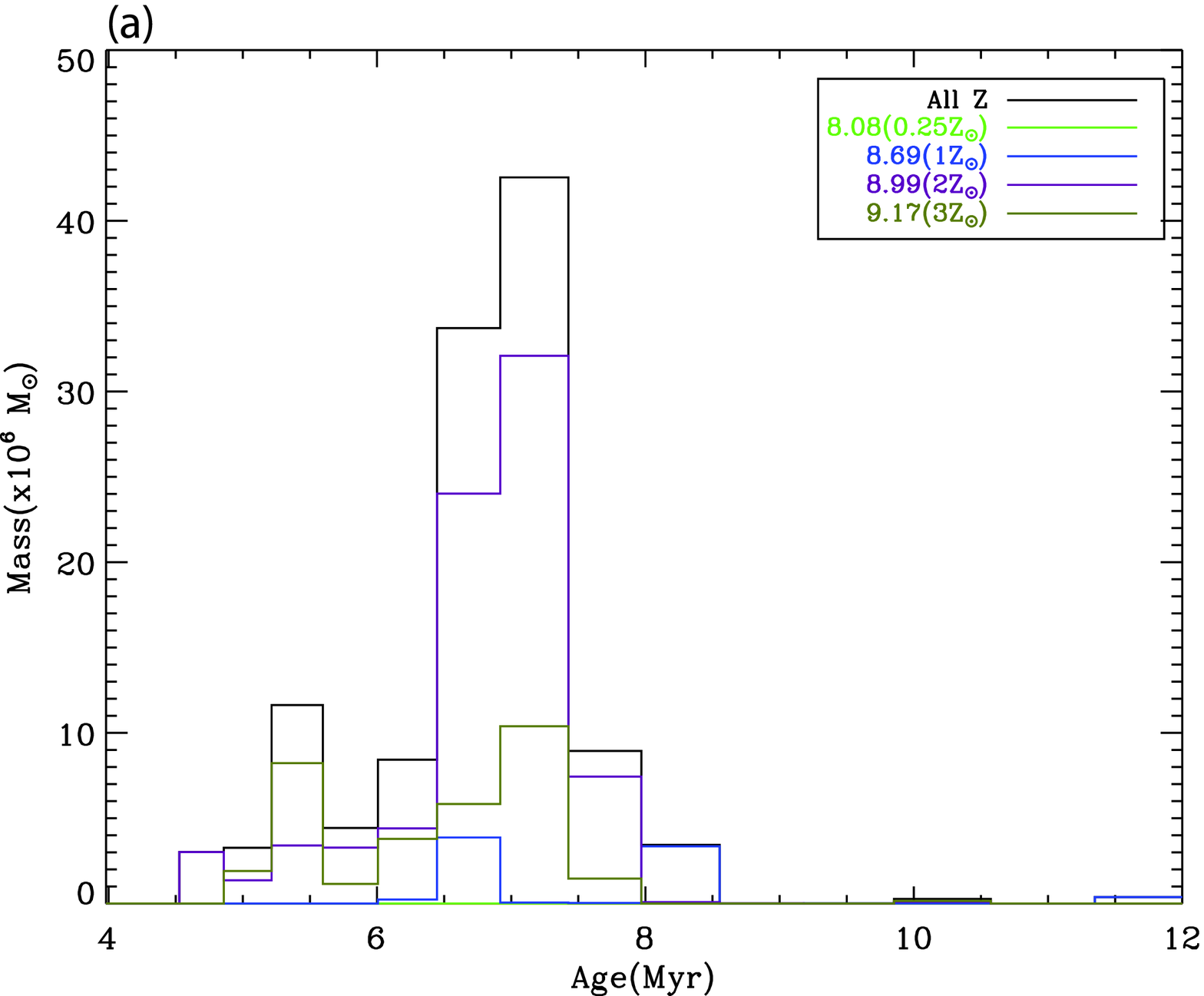}}
\subfigure{\label{SFH_b}\includegraphics[width=0.47\textwidth]{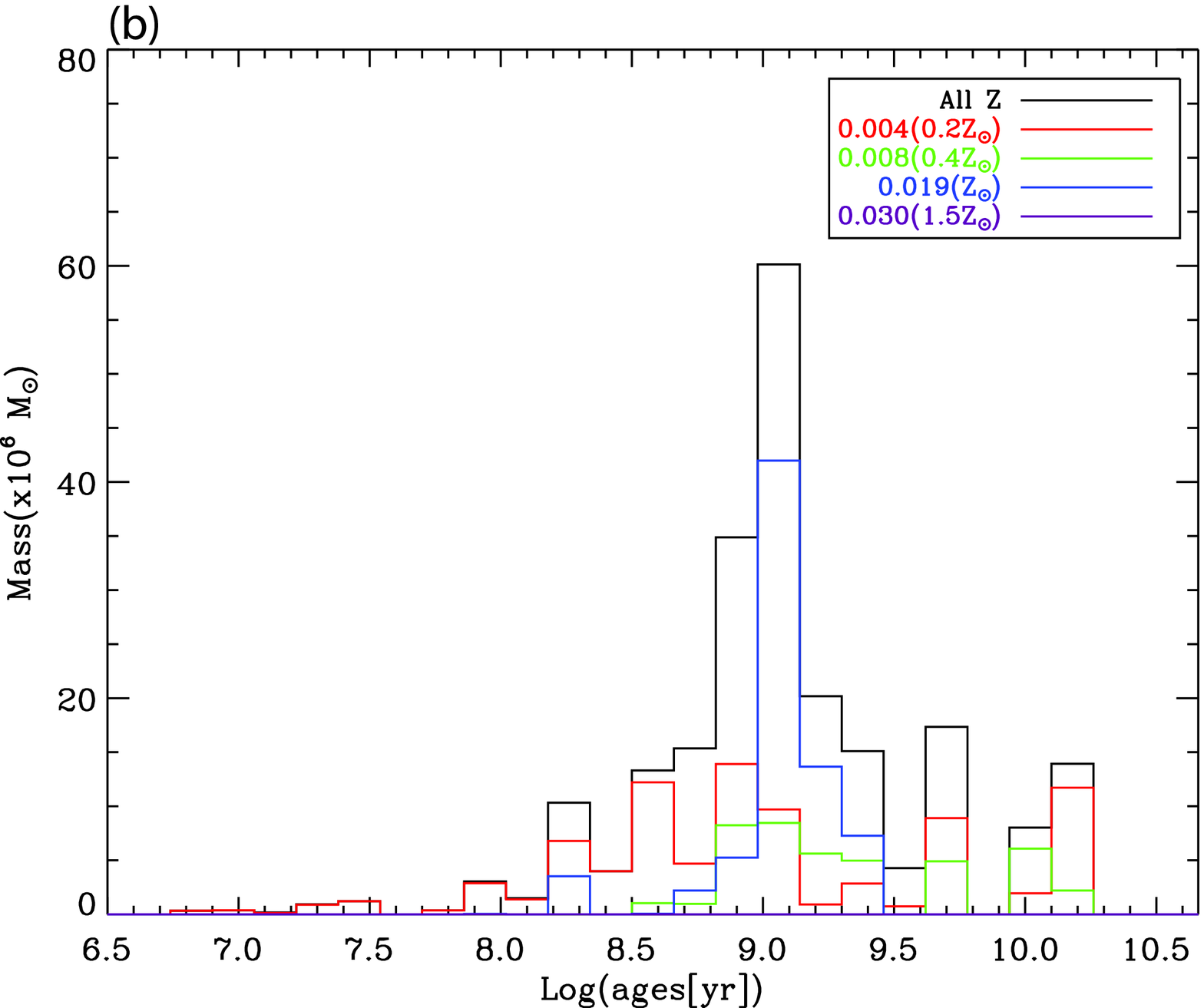}}
\caption{Histograms of the stellar population mass as a function of  age. The initial mass is represented here. In a), for the young populations based on the emission lines and in b), for the old populations based on the absorption lines. Different colors are used to show the mass in each age bin as a function of the stellar population metallicity (as in Fig.\ref{met_rev}).}
\label{SFH}
\end{figure}

\begin{figure}
\centering
\subfigure{\includegraphics[width=0.47\textwidth]{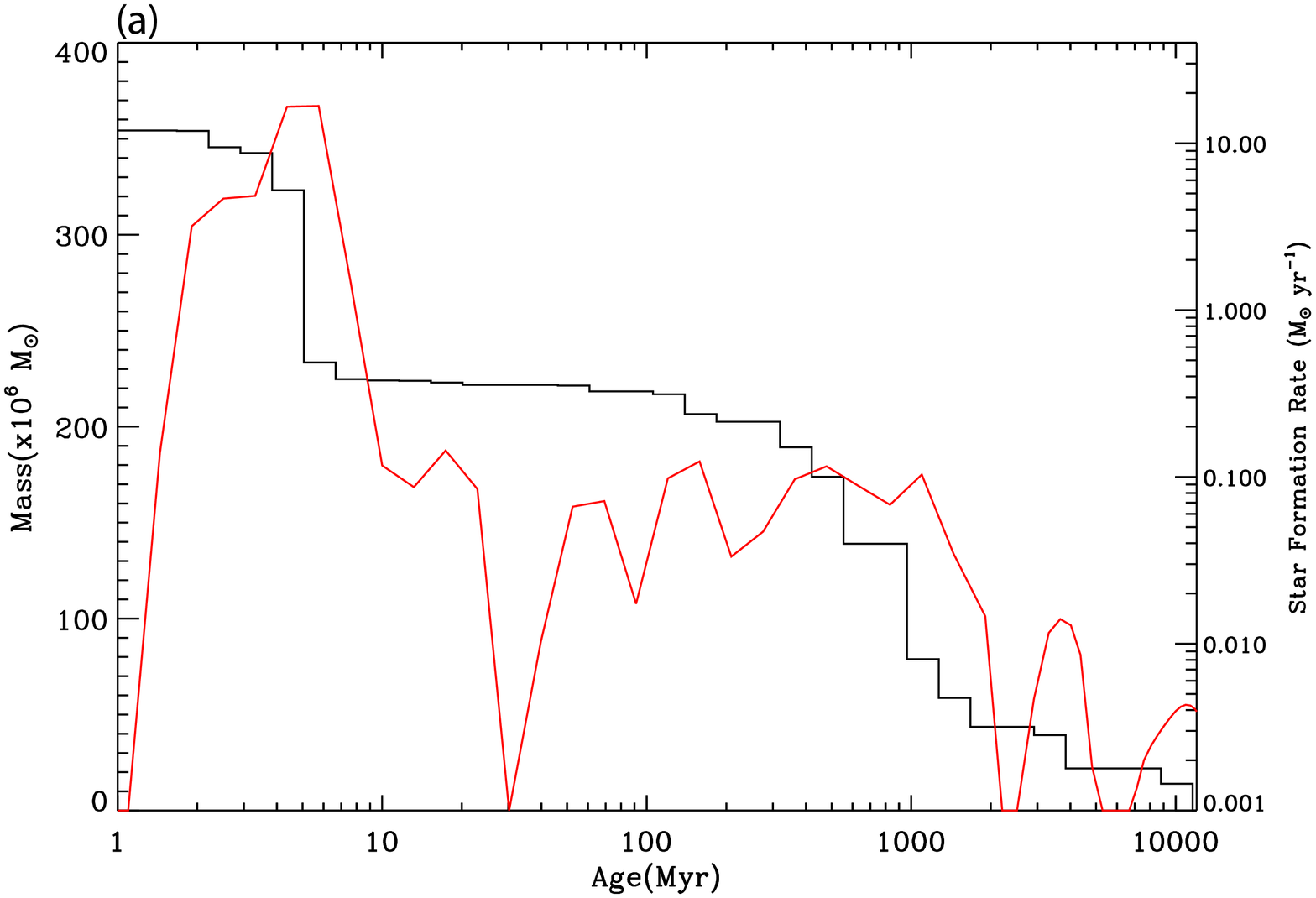}}
\subfigure{\includegraphics[width=0.47\textwidth]{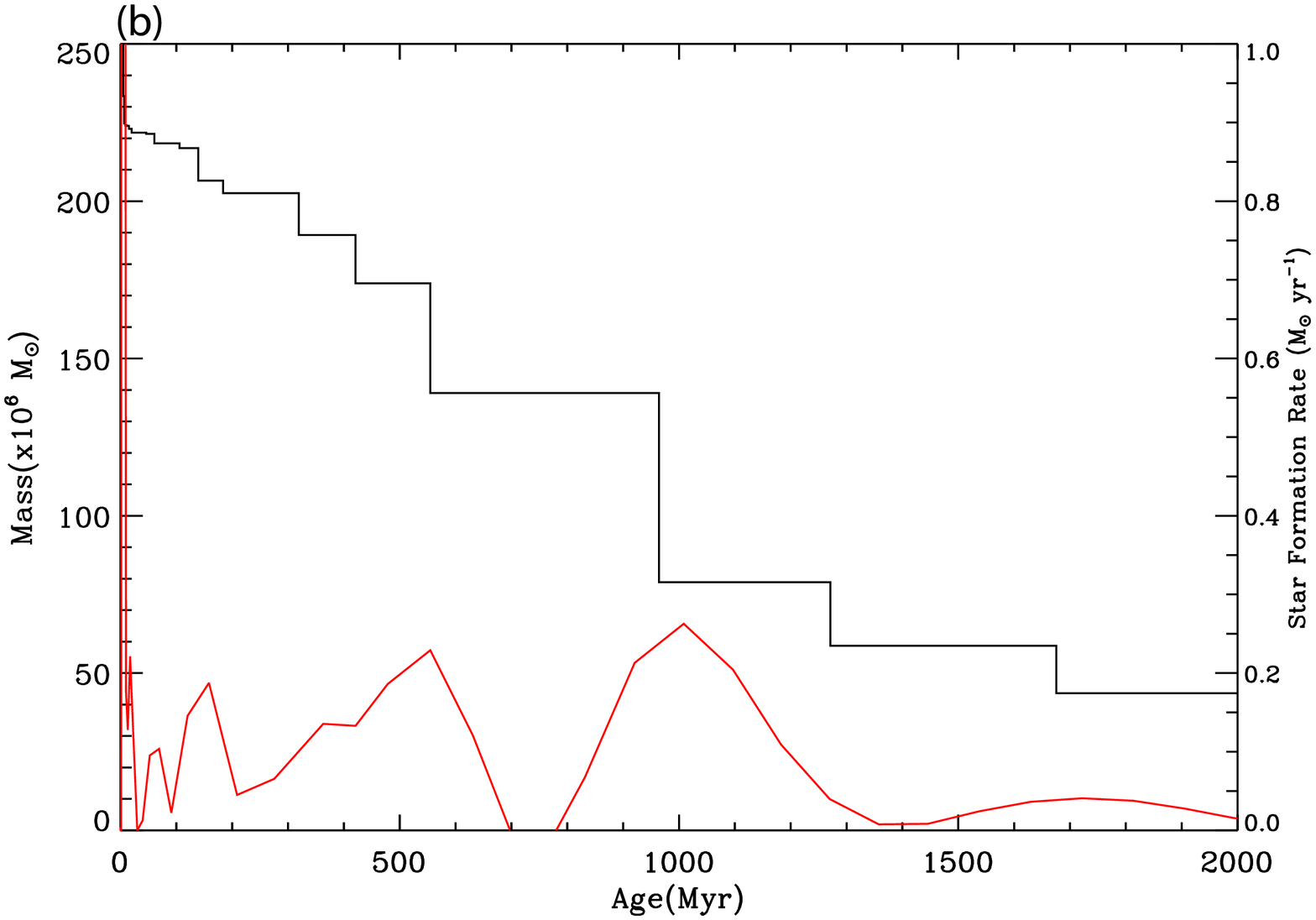}}
\subfigure{\includegraphics[width=0.47\textwidth]{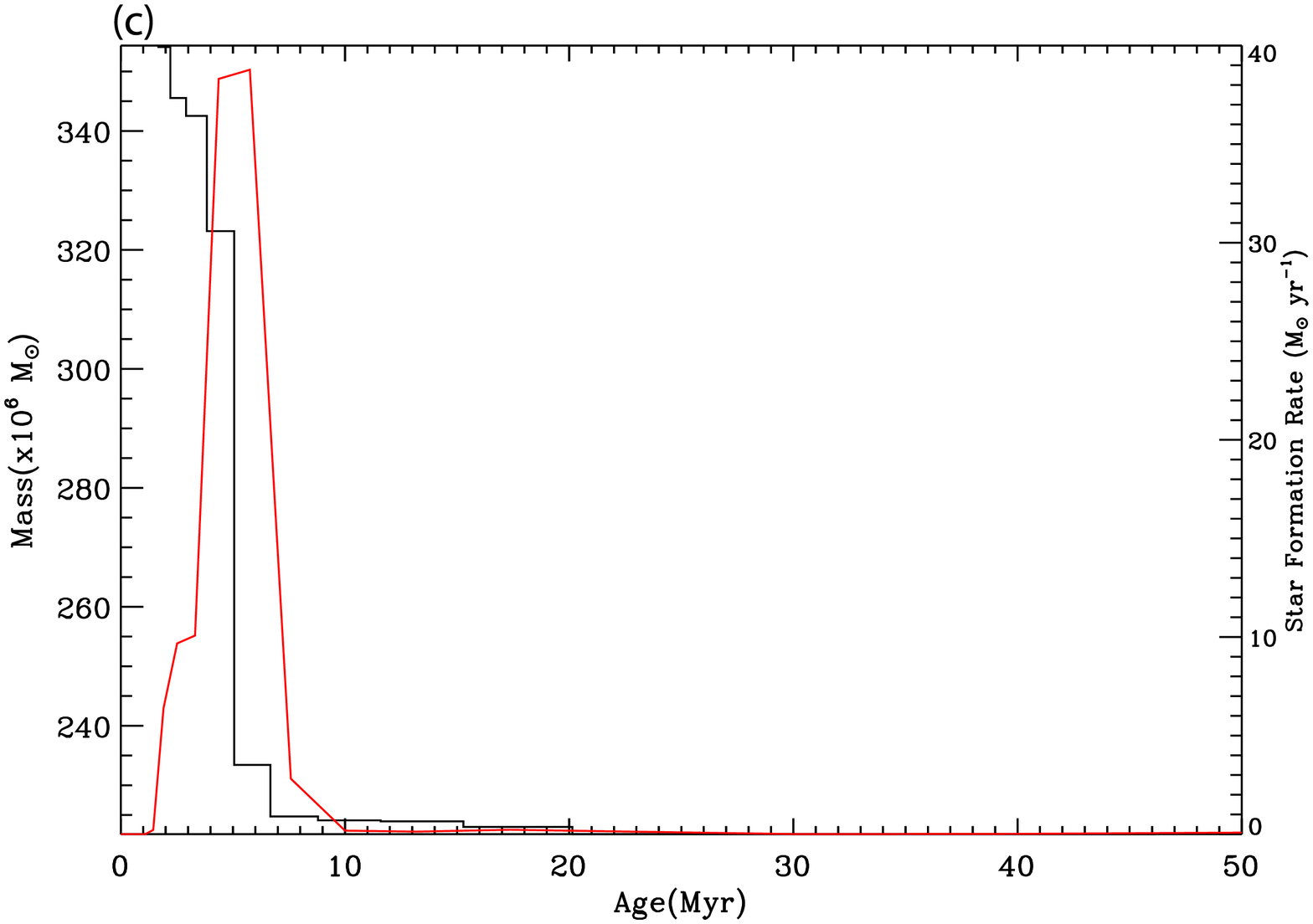}}
\caption{Historical star formation rate. In black, the total mass formed before an age, and in red,  the star formation rate for this age. 
In a) for the whole age range studied here (from 4 Myr to 12 Gyr), b) a zoom for the period covering the last 2 Gyr, and c) the most 
recent period between 4 and 50 Myr.}
\label{star_form_history}
\end{figure}

The stellar mass associated to the episodes of star formation shows an interesting distribution. Fig.~\ref{SFH} shows histograms of the initial mass for each populations as a function of age. As found for the current mass, nearly 40 per cent of the total mass converted into stars took place during the most recent bursts. This star formation was also very efficient, i.e. 83 per cent of the stellar mass in these bursts was created between 6 and 8\,Myr. Most of these young stars are showing a metallicity that is twice solar, while the remaining young stars have a higher metallicity (3\,Z$_\odot$) and dominate the mass budget bellow 6\,Myr.  The old star-forming episode, responsible for about 60 per cent of the inital stellar mass, looks very different. We interpret this episode as a long series of multiple bursts which took place in the past in the central region of the galaxy. The amplitude of the star formation was variable with time, reaching a maximum about 1\,Gyr ago. Furtheremore, the stellar metallicity was changing with time for these old stellar generations as well. At the 1\,Gyr  peak, an important part of the old populations had reached the solar metallicity while before, the metallicity was lower. It is interesting to see that among the more recent old generations (after the 1\,Gyr peak),  the metallicity is lower than solar. 

Using the mass histograms, we also built a plot of the total stellar mass created before a certain age. The first derivative of this mass may be interpreted as the star formation rate at a particular moment in the history of the galaxy. Looking at Fig.~\ref{star_form_history}, we find that the historical star formation rate was small (below 0.3\,M$_\odot$ yr$^{-1}$) until very recently, when the most recent star-forming episode shows a rate of almost 40\,M$_\odot$ yr$^{-1}$.  This rate is much higher then what would be found using the H$\alpha$ luminosity of the integrated spectrum \citep[0.21\,M$_\odot$ yr$^{-1}$ with the equation of ][]{1998ARA&A..36..189K}. Before the recent 6-8\,Myr burst, we see 3 bumps in the historical star formation rate:  around 150\,Myr,  near 500\,Myr, and at 1\,Gyr  corresponding to the maximum in the mass distribution of the old populations. We must point out that this method probably underestimate the  historical star formation rate since the width of the age bin increases with time. If we consider that the width of the most recent burst ($\sim$3\,Myr)  is typical for an episode of star formation, the historical star formation rate for the older bumps would increase to 7\,M$_\odot$ yr$^{-1}$, 16\,M$_\odot$ yr$^{-1}$, and 40\,M$_\odot$ yr$^{-1}$, respectively. This last number is comparable to the SFR of the most recent star forming event.

The narrowness of the histograms for the young stellar population mass (Fig.~\ref{SFH_a}) or  the number of spectra (Fig.~\ref{age_histo_a}) as a function of age, obtained based on the emission lines, may suggests that the young bursts have been triggered by a single specific event (internal or external) in the recent past. This event  had an influence, to some level, on the galaxy bar which then became more active in compressing the interstellar material and triggering or fuelling new star formation, as proposed by different simulations  \citep[see for example][]{1992MNRAS.259..345A,1996ApJ...462..114N,2004ApJ...600..595R,2009ApJ...692.1623H}. The fact that we see young bursts located along the galaxy bar supports the importance of the bar. Enriched interstellar material may be the result of past  episodes of star formation seen all over the galaxy central region. However,  the fact that we see a higher oxygen abundance for the gas near one of the recent burst (in the region L1) and the fact that we find a large amount of material in this region support a flow of enriched material toward the central region. 
We are not seeing star formation in a nuclear ring, but its radius could be over the limit of our field of view \citep[which has an average diameter of $\sim$\,720\,pc compared to a typical ring diameter of 1\,kpc;][]{1993AJ....105.1344B}. On the other hand, the extinction map shows clearly a dust lane oriented perpendicular to the galaxy large scale bar, which may be interpreted as a dusty nuclear bar. A nuclear bar is also often created in large scale bar simulations \citep{2004ApJ...617L.115E}. Nuclear bars and star forming knots are signatures of secular evolution \citep{2004ARA&A..42..603K} which is the mechanism used to explain the formation of pseudobulge in late spirals. 

\begin{figure}
\centering \includegraphics[width=0.47\textwidth]{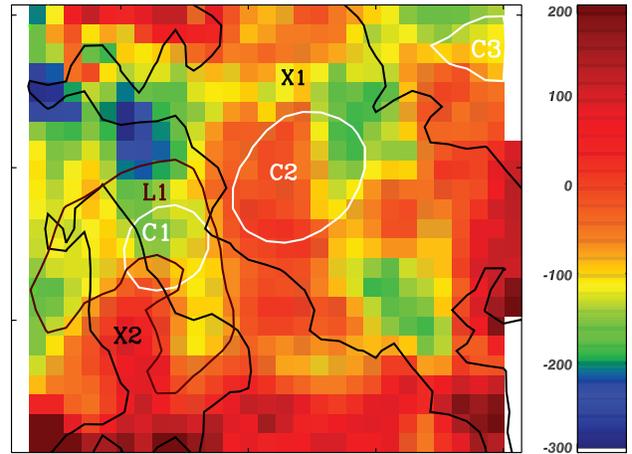}
\caption{Velocity map ($v\sin i$ in km s$^{-1}$) based on the \ha\ emission line center. Contour lines are as in Fig.~\ref{conti} and \ref{ebv}.}
\label{havel}
\end{figure}

The velocity map drawn using the centroid of the stronger emission line of \ha\ is not particularly clear with respect to the idea of gas from the disk flowing toward the galaxy center, but it surely indicates important motions.  This map (Fig.~\ref{havel}) shows variations in velocity ($v \sin i$, where the galaxy inclination is of the order of 20$^\circ$) from $-$300 to 200\,km\,s$^{-1}$ with some clear patterns: the central region has a velocity close to zero, the South-West edge of the image displays a redshifted velocity, while near and along the large scale bar we see blueshifted regions not necessarily superposed to the young star forming knots. One may see on the East side of the dust lane a redshifted velocity, while the West side displays mostly a blueshifted velocity. But overall this map is not easy to interpret.  Velocity maps for the other lines (e.g. \hb\ in emission and absorption, and \mg; not shown here) are significantly more noisy because of the line weakness.  No rotation curves have been published for \gq, but \citet{1988A&AS...72...57W} gives a maximum velocity of 192\,km\,s$^{-1}$, based on the H\,{\sc i}\ observations, which describe the motion in a typical disk, far away from the galaxy center. 

The old star formation episode appears to be more or less uniformly distributed over the central region with older and more metal rich objects closer to the nucleus.  This event is distributed over a long period of time; it took place between 100\,Myr and 3\,Gyr ago. It may indicate again the past signature of slow secular evolution which contributed to an initial phase of the pseudobulge formation. Was it purely internal? The amplitude of this old star forming episode, larger that the more recent $\sim$\,7\,Myr episode, may suggest that it has a different origin although the star formation rate are possibly comparable. Was it related to an interaction with another member of the Virgo cluster \citep[see for exemple:][]{1999NewA....4..101E,2006PASP..118..517B}, or did it involved a larger reservoir of gas already available at the time in the disk of the galaxy? In a simple scenario, all episodes of star formation could just be the consequence of the galaxy large scale bar evolution. If process such as  bar buckling \citep[i.e vertical instability that thickens and weakens the system bar][]{2004ApJ...604L..93D} combined with the triggering of star formation from infall from the bar \citep[][]{2007A&A...465L...1W}, could be repeated it could produce many star formation episode in the galaxy center reproducing the populations and structures observed here. To verify this hypothesis, it would be useful to apply our technique to more regions on the galaxy bar and disk at different radius, to look, for example, for variation in metallicity between the bar and the disk, to do a comparison of HII region properties, and to compare the old population of the disk with the pseudobulge. Adding more stellar indicators, like  the CaII triplet or the NaD, would be a good test for the intermediate 
(between 10 and 100\,Myr) stellar populations. A detailed dynamical analysis of the galaxy, to directly follow the material in the disk and along the bar would be important.  This study is also being repeated for more galaxies with different proprieties, Cantin et al. in preparation.

%%%%%%%%%%%%%%%%%%%%%%%%%%%%%%%%%%%%%%%%%%%%
\section{Conclusions}

The integrated spectrum covering the OASIS field of view on the galaxy central 780\,pc$\times$680\,pc shows strong nebular emission as seen in \hii\ regions as well as absorption lines of \mg\ and \fei\ possibly typical of an old stellar population. From this, we have performed an analysis using synthesis codes and an iterative technique to study, in a consistent way, two different stellar populations in all the individual spectra obtained. This technique takes into account the contribution to the flux from both populations and uses Bayesian statistics to get reliable uncertainties on the large number of parameters and spectra. We conclude that:

\begin{itemize}

\item The instrumental high spatial resolution allows us to separate and identify the position of the young star-forming populations. They are distributed in a few knots on each side of the galaxy center and oriented along the galaxy large scale bar. A dusty nuclear bar  is seen perpendicularly to the galaxy large scale bar. The old stellar population covers in a quasi uniform way the whole field of view.

\item The young bursts of star formation have an age between 5.5 and 8\,Myr with a metallicity near 2\,\zsun. They represents nearly 40 per cent of the stellar mass in the OASIS field of view. The old population age is distributed between 100\,Myr and 3\,Gyr with a peak around 1\,Gyr. Its metallicity is 0.2\,\zsun\ on average, a factor of 10 lower than the gas metallicity.

\item We speculate about a simple scenario to explain this activity which involve slow secular evolution to build a pseudobulge for the galaxy. In this scenario the galaxy large scale bar, as well as the star forming knots and the dusty nuclear bar are playing a major role.

\item An important advantage of the iteration technique is to decrease the uncertainty in the determination of age and mass of the stellar populations. There is a clear contribution to the continuum due to the younger population which needs to be taken into consideration, as we did in the iteration technique. It is also important to account for the spatial distribution into multiple knots of the younger population. For \gq, the age of the old stellar population obtained after the final iteration is larger than the one estimated from iteration 0. On the contrary, the young stellar population is at the end younger than obtained from the first analysis. 

\item The iteration technique has limits. It assumed a mixture of only two populations. To some extent, if the old stellar population is too small relative to the young one, it becomes impossible to study the absorption lines and characterize the old stellar population.  An important non-thermal contribution to the spectrum would not be treated properly with the actual technique. The technique could also be improved by considering a more detailed treatment of the extinction and more models for the absorption lines of \hb, \mg, and \fei\ in young and metal rich stellar populations.

\item Finally, the high spatial and spectral resolution of OASIS combined to spectral synthesis codes and our iteration technique are extremely powerful to extract information about the morphology, content, and history of galaxies.

\end{itemize} 

%%%%%%%%%%%%%%%%%%%%%%%%%%%%%%%%%%%%%%%%%%%%%%%%
\section*{Acknowledgments}

We would like to thank the anonymous referee for very useful comments and suggestions. We thank Pierre Martin for his help during the observations at the CFHT. We are also grateful to Pierre Ferruit and Eric Emsellem for making the reduction code XOASIS available to us and also directing us during the data processing.  This work was supported by the Natural Sciences and Engineering Research Council of Canada, by the Fonds Qu\'eb\'ecois de la Recherche sur la Nature et les Technologies of the Government of Qu\'ebec, and by the Spanish PNAYA projects AYA2004-02860-C03.

%%%%%%%%%%%%%%%%%%%%%%%%%%%%%%%%%%%%%%%%%%%%%%%%
\bibliographystyle{mn}
\bibliography{cantin_ngc4900}
%%%%%%%%%%%%%%%%%%%%%%%%%%%%%%%%%%%%%%%%%%%%%%%%
\appendix

\section{Bayesian Statistics}

For the detailed analysis of \S~5, we used a probability program based on Bayesian statistics \citep{2005blda.book.....G} to extract: 1) a single age-metallicity solution for a stellar population when comparing its observational set of indicators (absorption or emission lines) with  models, and 2) the individual mass of two populations when comparing the observational spectrum with  synthetic spectra. One main advantage of this  program is to quickly give reliable uncertainties for a large number of spectra.  We give here a brief description of the probability program used.

\begin{figure}
\centering
\subfigure{\includegraphics[width=0.47\textwidth]{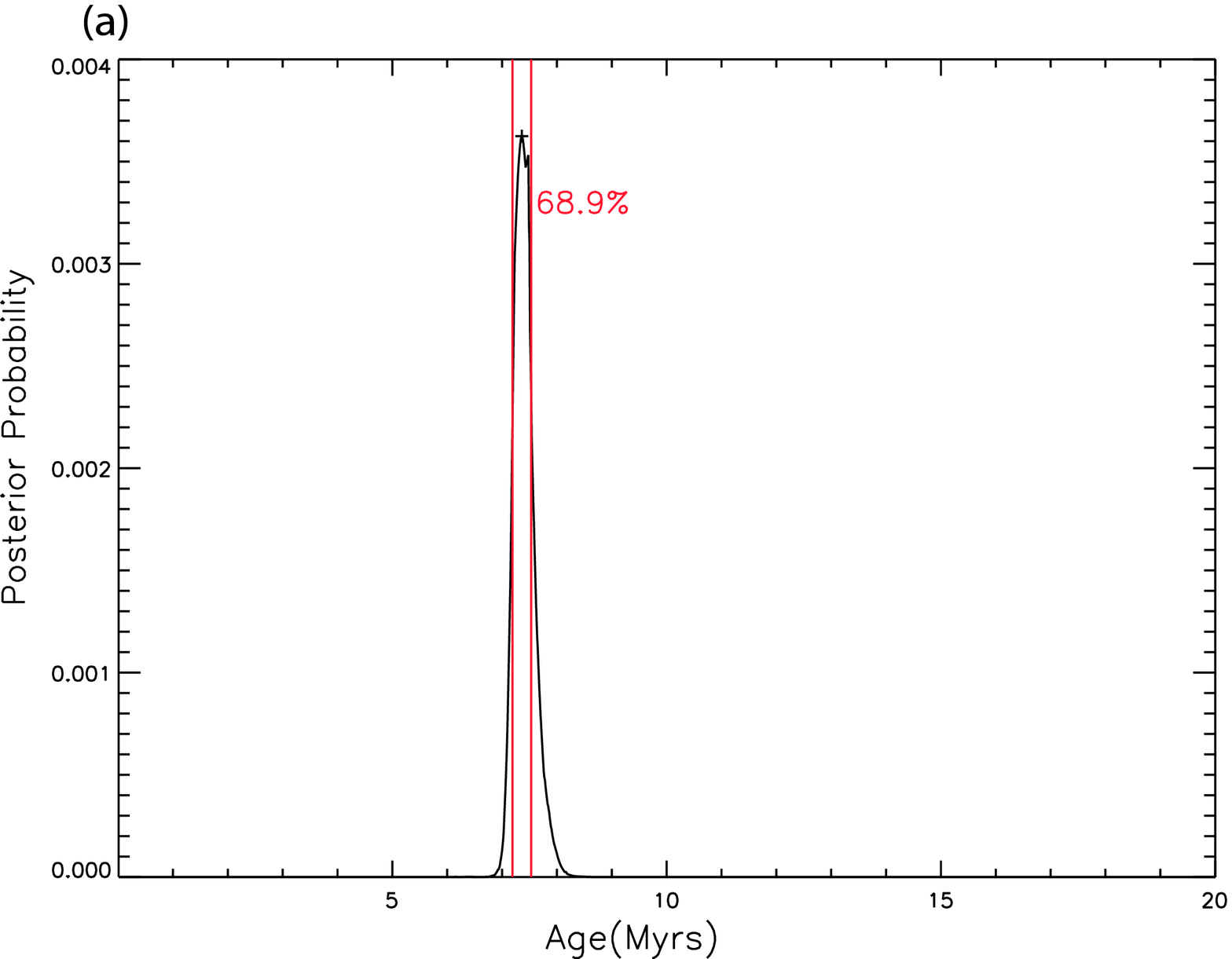}}
\subfigure{\includegraphics[width=0.47\textwidth]{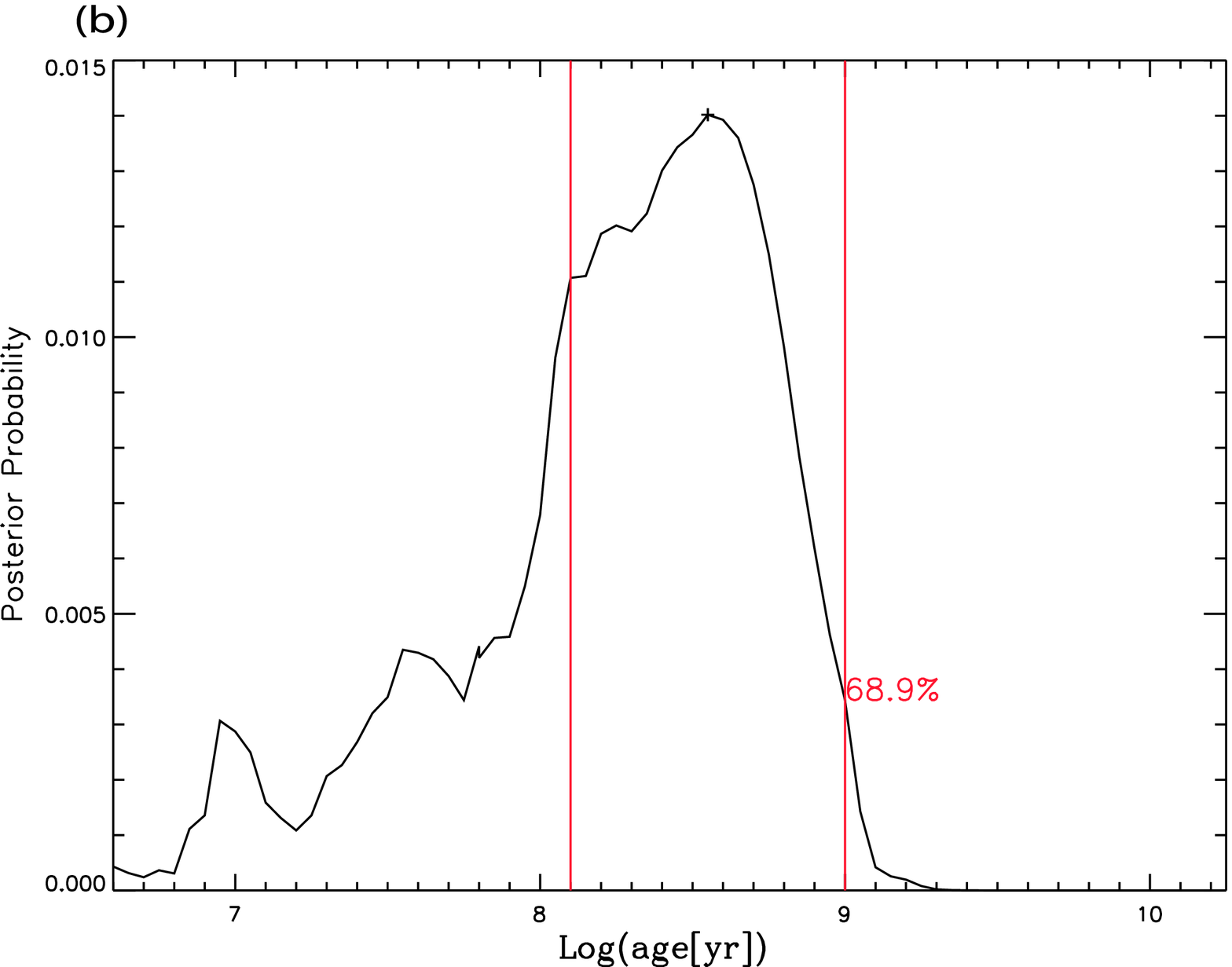}}
\caption{The marginalized probability distribution for the age of the stellar population in the spectrum \#160 in the region L1. These plots are the results of the Bayesian analysis: (a) for the young stellar population,  and (b) for the old stellar population.}
\end{figure}

The first step is to define the prior information $W(p_1,p_2 | I)$ based on the previous iteration $I$. The parameters, $p_1$ and $p_2$,  are either: 1)  the  age and the metallicity of a population, or 2) the mass of the old population and  the mass of the young population. We then compute the likelihood $\mathcal{L}(D | p_1,p_2, I)$ of the models, which is the probability of obtaining the data $D$ given that $p_1$, $p_2$ and $I$ are true. Each model gives many indicators $L_i[p_1,p_2]$, which are related to measurements of the lines or the continuum level. The likelihood is calculated for $N$ indicators using the following equation:

\begin{equation}
\label{likelyhood}
\mathcal{L}(D | p_1,p_2, I) = (2\pi)^{N \over 2}~\bar{\sigma}^{N \over 2}\exp^{ -\sum_{i=1}^N
\big(\frac{(O_i-L_i[p_1, p_2])^2}{2\sigma_{i}^2} \big)},
\end{equation}

\noindent where $\sigma_i$ is the uncertainty measured for the indicator $O_{i}$ and $\bar{\sigma}$  is the product of the uncertainty for all indicators. 

A posterior probability $P[p_1,p_2 | D, I]$ is then calculated by applying the prior $W[p_1,p_2 | I]$ to each likelihood found for all the models considered and by normalizing according to the equation:

\begin{equation}
\label{bayes_age}
P[p_1,p_2 | D,I]= {  {W[p_1, p_2 | I]  * \mathcal{L}[D | p_1, p_2, I]}  \over 
{\sum_{[p_1,p_2]} W[p_1,p_2 | I] * \mathcal{L}[D | p_1,p_2, I]}}.
\end{equation}

The highest probability, i.e. closest to 1,  indicates the best solution. To compute the incertainties we marginalize the posterior probability over one parameter according to equation: 

\begin{equation}
\label{marg_eq}
P[p_1 | D,I]= {\sum_{p_2}}P[p_1,p_2 | D,I].
\end{equation}

\begin{figure}
\centering
\includegraphics[width=0.47\textwidth]{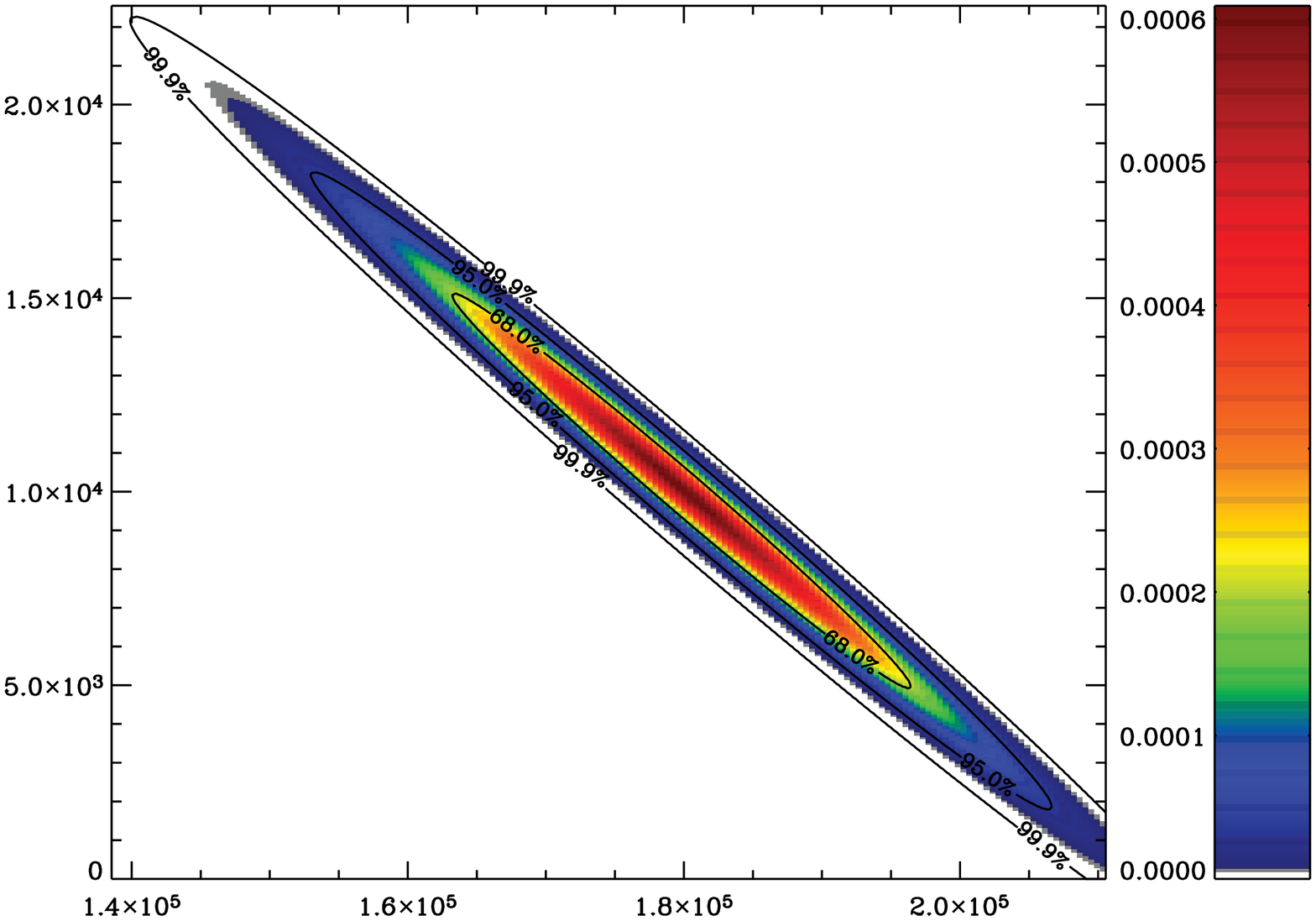}
\caption{The probability density plot for the mass of the young and old stellar population in the spectrum \#160 in the region L1 based on the Bayesian analysis.}
\end{figure}

\begin{figure}
\centering
\subfigure{\includegraphics[width=0.47\textwidth]{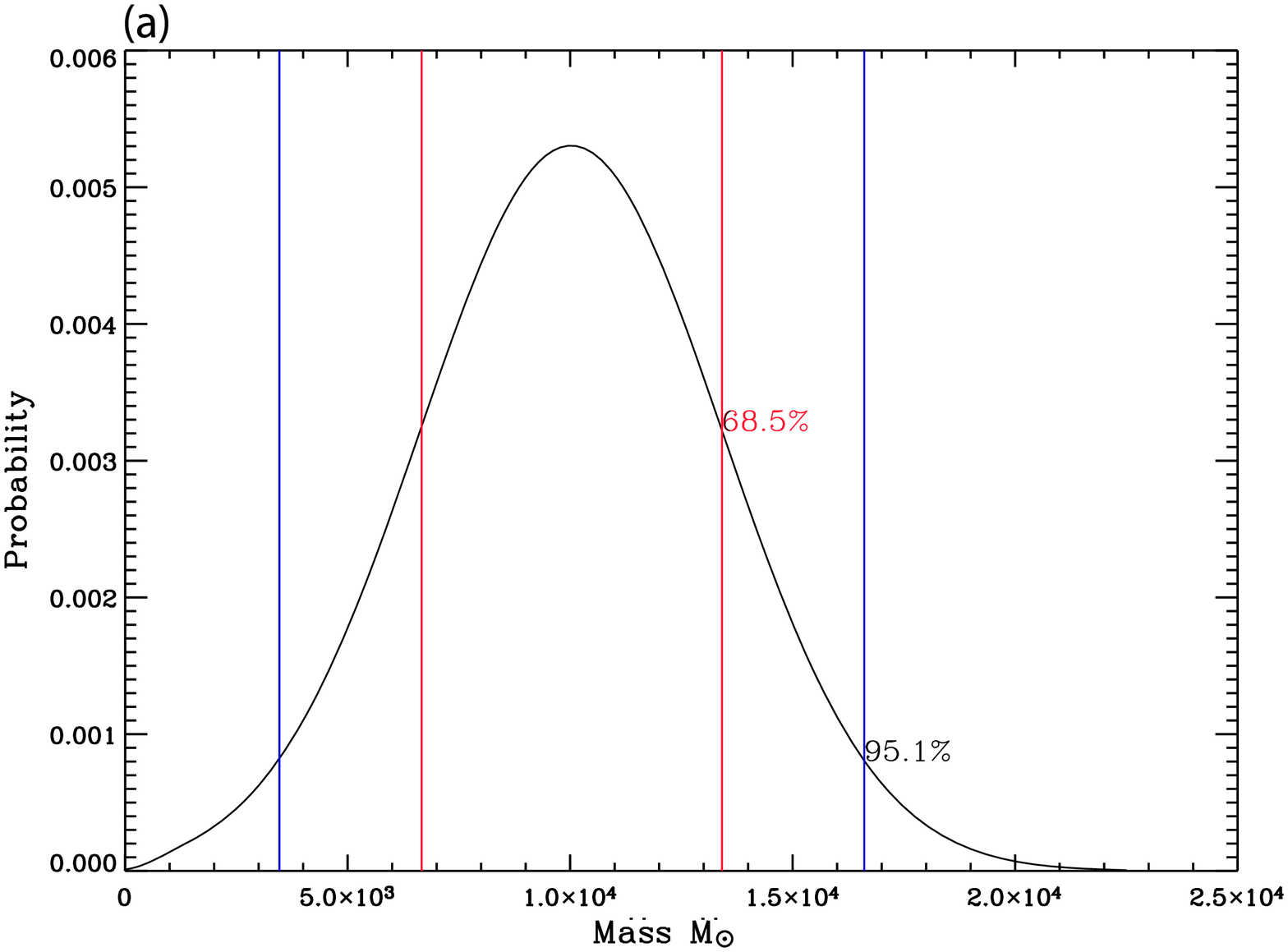}}
\subfigure{\includegraphics[width=0.47\textwidth]{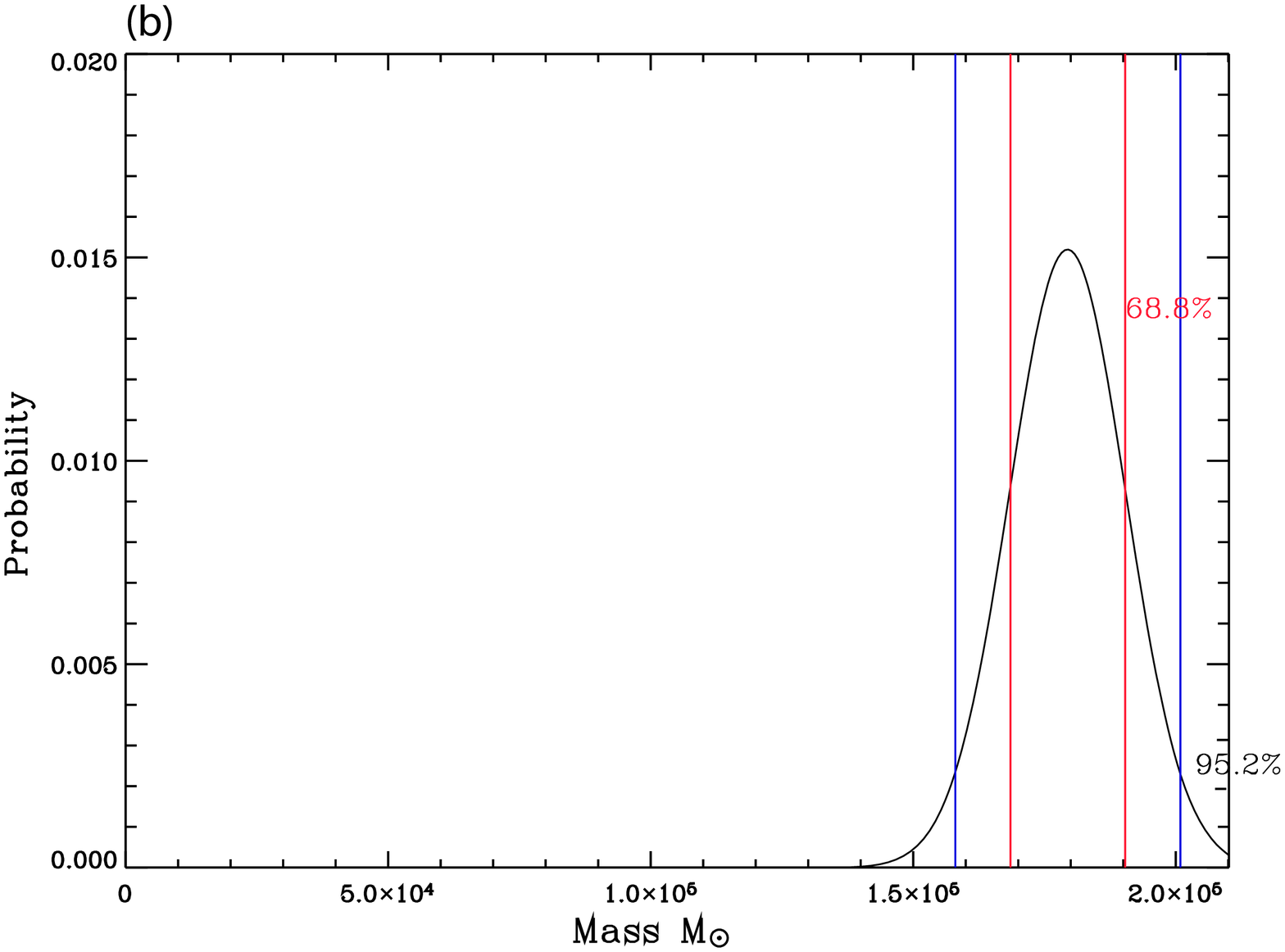}}
\caption{The marginalized probability distribution for the mass of the stellar population in the spectrum \#160 in the region L1.  These plots are the results of the Bayesian analysis: (a) for the young stellar population,  and (b) for the old stellar population.}
\end{figure}

\noindent When searching for the age-metallicity pair solution for a young stellar population, we use a semi-flat prior for the first iteration. In this case, all ages are given the same weight but the models with the metallicity closer to the one measured for the gas is given twice as much weight as the other three  possibilities (there are four possible metallicities with {\sc lavalsb}). Subsequent iterations will adopt the posterior probability of the previous iteration as the new prior.  Fig.~A1 shows, for example, the marginalized probability plot obtained for the age of the young stellar population in one spectrum (\#160) in the region L1 after the last iteration (as described in \S~5, the iteration technique takes into account the underlying older stellar population). It was obtained by comparing two indicators, the equivalent widths of \hb\ and \ha, with those produced by {\sc lavalsb}. This plot shows the probability marginalized over the metallicities, i.e. the probability at a given age summed over all four possible metallicities. The marginalized probability is used here to visualize the  solution and calculate an uncertainty for the best model parameters. Vertical lines on this plot show the region containing 68 per cent of the total probability distribution, i.e the 1 sigma deviation for the best parameter pair. We see here  a  narrow Gaussian indicating an age of 7.3$\pm$0.2\,Myr for the young population.

When searching for the age-metallicity pair solution for an old stellar population, the first iteration considers a flat prior, which gives the same weight to all models (there are  more indicators in this case).  Subsequent iterations are computed using the posterior probability found in the previous iteration. Three to four indicators are available in the spectrum for the old stellar population, the \fei\,$\lambda$\,5270,  \fei\,$\lambda$\,5335,  \mg\ indice,  and the \hb\ line when not contaminated by emission (i.e after  the first iteration).   Fig.~A1 shows, for example, the marginalized probability plot obtained for the age of the old stellar  population in the same spectrum (\#160). We obtain $\log(\rm{age~[yr]})$=8.4$\pm$0.5.

When the method is used to find the mass of the young and the old stellar population,  a flat prior is considered for the first iteration. It seems more efficient to keep a flat prior for the subsequent iterations.  In this case, the MR1 and the MR2 spectral range are considered as two indicators (i.e. the sum of all  $L_i - L_i[young,old]$  is first calculated for each pixel of MR1 and then MR2 and treated as two indicators in eq.~A1). Fig.~A2 shows the probability density plot obtained after the last iteration of the same spectrum presented in Fig.~A1. Such a plot is possible since the mass parameter space  is much larger than the metallicity parameter space. Ellipses in this plot are guiding lines for the mass uncertainty which is again determined using the marginalized probability as shown in Fig.~A3 for the young and old stellar population, respectively. 

\label{lastpage}
\end{document}